\documentclass[trackchange,twocolumn]{aastex701}

\usepackage{adjustbox}

\newcommand{\iras}{{\it IRAS}}

\newcommand{\arii}{\hbox{[Ar$\,${\scriptsize II}]}}
\newcommand{\ariii}{\hbox{[Ar$\,${\scriptsize III}]}}

\newcommand{\niII}{\hbox{[Ni$\,${\scriptsize II}]}}

\newcommand{\neii}{\hbox{[Ne$\,${\scriptsize II}]}}
\newcommand{\neiii}{\hbox{[Ne$\,${\scriptsize III}]}}

\newcommand{\nev}{\hbox{[Ne$\,${\scriptsize V}]}}
\newcommand{\nevi}{\hbox{[Ne$\,${\scriptsize VI}]}}

\newcommand{\feii}{\hbox{[Fe$\,${\scriptsize II}]}}
\newcommand{\feiii}{\hbox{[Fe$\,${\scriptsize III}]}}

\newcommand{\siii}{\hbox{[S$\,${\scriptsize III}]}}
\newcommand{\siv}{\hbox{[S$\,${\scriptsize IV}]}}
\newcommand{\piii}{\hbox{[P$\,${\scriptsize III}]}}
\newcommand{\clii}{\hbox{[Cl$\,${\scriptsize II}]}}

\newcommand{\nii}{\hbox{[N$\,${\scriptsize II}]}}

\newcommand{\kms}{km\,s$^{-1}$} 
 
\newcommand{\um}{$\mu$m}

\newcommand{\ergs}{erg\,s$^{-1}$}
\newcommand{\ergscm}{erg\,s$^{-1}$\,cm$^{-2}$}

\newcommand\jwst{\emph{JWST}}
\newcommand\spitzer{\emph{Spitzer}}

\newcommand\ifsfit{\texttt{IFSFIT}}
\newcommand\lmfit{\texttt{lmfit}}
\newcommand\cafe{\texttt{CAFE}}

\newcommand\qtdfit{\texttt{q3dfit}}
\newcommand\astropy{\texttt{Astropy}}
\newcommand\numpy{\texttt{NumPy}}
\newcommand\scipy{\texttt{SciPy}}
\newcommand\pyneb{\texttt{PyNeb}}
\newcommand\matplotlib{\texttt{matplotlib}}

\newcommand\fonezero{F10565$+$2448}

\shorttitle{JWST Data of F10565+2448}
\shortauthors{Dan, et al.}
\submitjournal{ApJ}

\begin{document}

\title{Resolving the Multiphase Outflow, Shock Signatures, and PAHs in the AGN-Starburst Composite ULIRG F10565+2448 with \jwst\ MIRI/MRS} 

\author[orcid=0000-0001-5894-4651]{Kylie Yui Dan}
\affiliation{Department of Astronomy, University of Maryland, College Park, MD 20742, USA}
\email[show]{kydan@umd.edu}

\author[orcid=0000-0002-4014-9067]{Jerome Seebeck}
\affiliation{Department of Astronomy, University of Maryland, College Park, MD 20742, USA}
\email{jseebeck@umd.edu}

\author[orcid=0000-0002-3158-6820]{Sylvain Veilleux}
\affiliation{Department of Astronomy, University of Maryland, College Park, MD 20742, USA}
\affiliation{Joint Space-Science Institute, Department of Astronomy, University of Maryland, College Park, MD 20742, USA}
\email{veilleux@umd.edu}

\author[orcid=0000-0002-1608-7564]{David Rupke}
\affiliation{Department of Physics, Rhodes College, Memphis, TN 38112, USA}
\email{rupked@rhodes.edu}

\author[orcid=0000-0001-5285-8517]{Eduardo Gonzalez-Alfonso}
\affiliation{Universidad de Alcalá, Departamento de Física y Matemáticas, Campus Universitario, 28871 Alcalá de Henares, Madrid, Spain}
\email{eduardo.gonzalez@uah.es}

\author[orcid=0000-0002-9627-5281]{Ismael Garcia-Bernete}
\affiliation{Centro de Astrobiolog\'ia (CAB), CSIC-INTA, Camino Bajo del 497 Castillo s/n, E-28692 Villanueva de la Ca\~nada, Madrid, Spain}
\email{igbernete@gmail.com}

\author[orcid=0000-0003-3762-7344]{Weizhe Liu}
\affiliation{Department of Astronomy, Steward Observatory, University of Arizona, Tucson, AZ 85719, USA}
\email{oscarlwz@gmail.com}

\author[orcid=0000-0003-0291-9582]{Dieter Lutz}
\affiliation{Max Planck Institute for Extraterrestrial Physics, Giessenbachstraße 1, 85748 Garching, Germany}
\email{lutz@mpe.mpg.de}

\author[orcid=0000-0001-8485-0325]{Marcio Melendez}
\affiliation{Space Telescope Science Institute, 3700 San Martin Drive, Baltimore, MD 21218, USA}
\email{melendez@stsci.edu}

\author[orcid=0000-0002-4005-9619]{Miguel Pereira Santaella}
\affiliation{Instituto de Física Fundamental (IFF), CSIC, Serrano 123, E-28006 Madrid, Spain}
\email{miguel.pereira@iff.csic.es}

\author[orcid=0000-0002-0018-3666]{Eckhard Sturm}
\affiliation{Max Planck Institute for Extraterrestrial Physics, Giessenbachstraße 1, 85748 Garching, Germany}
\email{sturm@mpe.mpg.de}

\author[orcid=0000-0002-6562-8654]{Francesco Tombesi}
\affiliation{Physics Department, Tor Vergata University of Rome, Via della Ricerca Scientifica 1, 00133 Rome, Italy}
\affiliation{INAF – Astronomical Observatory of Rome, Via Frascati 33, 00040 Monte Porzio Catone, Italy}
\affiliation{INFN - Rome Tor Vergata, Via della Ricerca Scientifica 1, 00133 Rome, Italy}
\email{francesco.tombesi@roma2.infn.it}


\begin{abstract}
We present new {\em James Webb Space Telescope} Mid-Infrared Instrument (MIRI) Medium-Resolution Spectrometer (MRS) observations of the nearby ultra-luminous infrared galaxy F10565+2448. These integral field spectroscopic data reveal an unresolved nuclear outflow in both warm-ionized and warm-molecular gas phases as well as a resolved blueshifted kpc-scale warm-molecular outflow. The unresolved warm-ionized outflow has a mean projected velocity up to $-520$ \kms,  while the unresolved warm-molecular outflow is slower at $-150$ \kms. For the resolved warm-molecular outflow, the projected mean velocity ($-280 < v_{50} < -110$ \kms) is only slightly faster than the velocity of the disk ($-70 < v_{50} < 120$ \kms) and as such likely does not exceed the estimated escape velocity of $\gtrsim 300$ \kms. The warm-molecular outflow is slightly hotter ($507 \pm 25$K) than the disk ($329 \pm 5$K), and displays areas of higher temperature and lower column density that may indicate a shock front, which we explore using the \feii\ 5.34 $\mu$m/Pf$\alpha$ shock diagnostic. Analysis of the polycyclic aromatic hydrocarbon features reveal trends of ionization and grain size that first decrease with radius up to 1 kpc before increasing up to 3 kpc. These results bolster the picture of F10565+2448 being an AGN-starburst composite where both star formation and AGN-powered phenomena are required to explain the outflow energetics. 

\end{abstract}


\section{Introduction} 
Our current picture of galaxy evolution requires some form of feedback mechanism to regulate star formation. Models without feedback consistently over-predict the abundance of galaxies at the highest and lowest mass ends when compared to observations \citep[e.g.,][]{dekel1986, silk1998, hopkins2014, somerville2015}. Galactic winds (or ``outflows") have proven to be an essential part of feedback, enriching the intergalactic medium with metals and suppressing star formation and growth of supermassive black holes \citep[e.g.,][]{king2003, murray2005, veilleux2005, heckman2017, veilleux2020}. Outflows are commonly driven either by starbursts, active galactic nuclei (AGN), or a combination of both.

A key population for studying outflows is the ultraluminous infrared galaxies (ULIRGs) with infrared (8 $-$ 1000 \um) luminosities above 10$^{12}$ $L_\odot$. These objects are often galaxy mergers, creating a compelling theoretical picture: during a gas-rich merger, gas is funneled into the circumnuclear region of the merger remnant, leading to dust enshrouded star formation and supermassive black hole (SMBH) accretion. The resulting galactic winds may provide negative feedback through the heating and expulsion of gas and dust, limiting further star formation \citep{dimatteo2005, king2005, hopkins2012}, or positive feedback through compression of the interstellar medium, leading to increased star formation \citep{Elmegreen1977, krumholz2009, silk2013, zinn2013, Maiolino2017}. Exactly how outflows affect their host galaxies and lead to feedback is being actively explored in the field. 
Local ULIRGs can provide opportunities to study feedback in exquisite spatial detail, which may help us understand exactly how outflows can affect their hosts \citep[e.g.,][]{Westmoquette2012, Rupke2013, Rupke2017, Pereira2018, Spence2018}. 

Outflows are often observed in several different gas phases, from hot ionized gas to cold molecular gas \citep{Rupke2005, veilleux2005, gonzalez2017, lutz2020, veilleux2020, lamperti2022}. It is crucial to observe them across the wavelength spectrum if we want to derive the total outflow energy and thus obtain a more complete picture of the impacts of feedback. With the advent of the James Webb Space Telescope (\jwst), it is now possible to spatially resolve these objects in the mid-infrared for the first time, leading to precise measurements of the warm molecular and warm ionized gas phases. 

We present results from the cycle 2 \jwst\ program (PID 3869, PI Veilleux) aimed at studying ULIRGs with known fast and powerful molecular outflows. The target here is the primary galaxy of \fonezero, a nearby \citep[$z = 0.0431$;][]{Rupke2013} merger made up of three distinct galaxies, with the secondary galaxy being 20 kpc northeast of the primary galaxy and the tertiary galaxy being 6.2 kpc southeast of the primary \citep{downes1998}. The primary galaxy contributes most of the total IR luminosity of the system $L_{\mathrm{IR, total}(8-1000\mu\mathrm{m})} =
1.1 \times 10^{12} L_\odot$ \citep{sanders2003, Rupke2013} and is the only galaxy of the three within the field of view of our data. \fonezero\ is likely a starburst-AGN composite galaxy, with an AGN bolometric fraction of 17\% \citep{veilleux2009_agn_fraction} and a star formation rate (SFR) of 150 M$_\odot$ yr$^{-1}$ \citep{Rupke2013}. A relatively low-velocity, blueshifted outflow to the southwest has been detected in the neutral, ionized, and cold molecular gas phases.
Arecibo and Giant Metrewave Radio Telescope HI observations reveal a blueshifted outflow in the neutral gas phase up to $\sim -250$ km s$^{-1}$ \citep{mirabel1988, su2023}. 
Gemini IFU data of the Na $\mathrm{I}$ D absorption doublet and \nii\ and H$\alpha$ emission lines show a blueshifted outflow in the neutral and ionized gas phases with average velocities of $-$218 km s$^{-1}$ and $-$133 km s$^{-1}$, respectively \citep{shih2010, Rupke2013}. 
\fonezero\ has been observed in the CO(2-1) and CO(3-2) molecular transitions with the Caltech Submillimeter Observatory \citep{glenn2001} and the Submillimeter Array \citep{wilson2008}, although neither detected a molecular outflow. IRAM Plateau de Bure Interferometer data reveal broad CO(1–0) wings of up to $\pm 600$ km s$^{-1}$ \citep{cicone2014}. F10565+2448 also shows
prominent OH119 and OH79 P Cygni profiles and weak blueshifted OH65 and OH84 absorption \citep{veilleux2013, gonzalez2017}. 

In this work, we complement the multi-wavelength analysis of \fonezero\ with \jwst\ MIRI/MRS integral field spectroscopic (IFS) observations of the warm molecular and warm ionized gas phases, as well as the polycyclic aromatic hydrocarbon (PAH) emission. In Section \ref{sec:reduction}, we go over the observations and data reduction process. We describe our analysis methods in Section \ref{sec:analysis} and present the results in Section \ref{sec:results}. In Section \ref{sec:discussion}, we discuss the mass and energetics of the outflow, trends in the properties of the PAH with distance from the center, and a mid-IR shock diagnostic. We summarize our conclusions in Section \ref{sec:conclusion}. We adopt a cosmology of $H_0 = 69.6$ \kms Mpc$^{-1}$, $\Omega_M = 0.286$, and $\Omega_L = 0.714$, which implies a scale of 0.855 kpc arcsec$^{-1}$ and a luminosity distance of 191.8 Mpc \citep{Bennett2014, Wright2006}.

\section{Observations and Data Reduction}
\label{sec:reduction}

\fonezero\ was observed with \jwst\ on 2024-05-27 using the Medium-Resolution Spectrometer (MRS) mode of the Mid-InfraRed Instrument \citep[MIRI;][PID 3869, PI Veilleux]{wright2023, argyriou2023}. All of the grating settings - Short, Medium, and Long - were used to cover the entire wavelength range of MIRI/MRS (4.9 to 27.9 $\mu$m). A 4-point extended dither pattern was used to help remove background contamination and reduce undersampling, resulting in a total exposure time of 1354 s per channel. A background cube was obtained immediately after the on-source cube using the same settings, with the exception of using the 2-point dither. The data used in this paper can be found in MAST: \dataset[10.17909/0am6-8907]{http://dx.doi.org/10.17909/0am6-8907}. We reduced these data with \jwst\ Calibration Pipeline v1.19.1 \citep{Bus2022, Bus2024}. 
The reduction was completed with the MIRI pipeline sample notebook publicly available on the Space Telescope GitHub (\href{https://github.com/spacetelescope/jwst-pipeline-notebooks/}{JWPipeNB-MIRI-MRS.ipynb}), using the master-background subtraction setting and 2D residual fringing step turned on. 

\begin{figure}
    \centering
    \includegraphics[width=0.9\columnwidth]{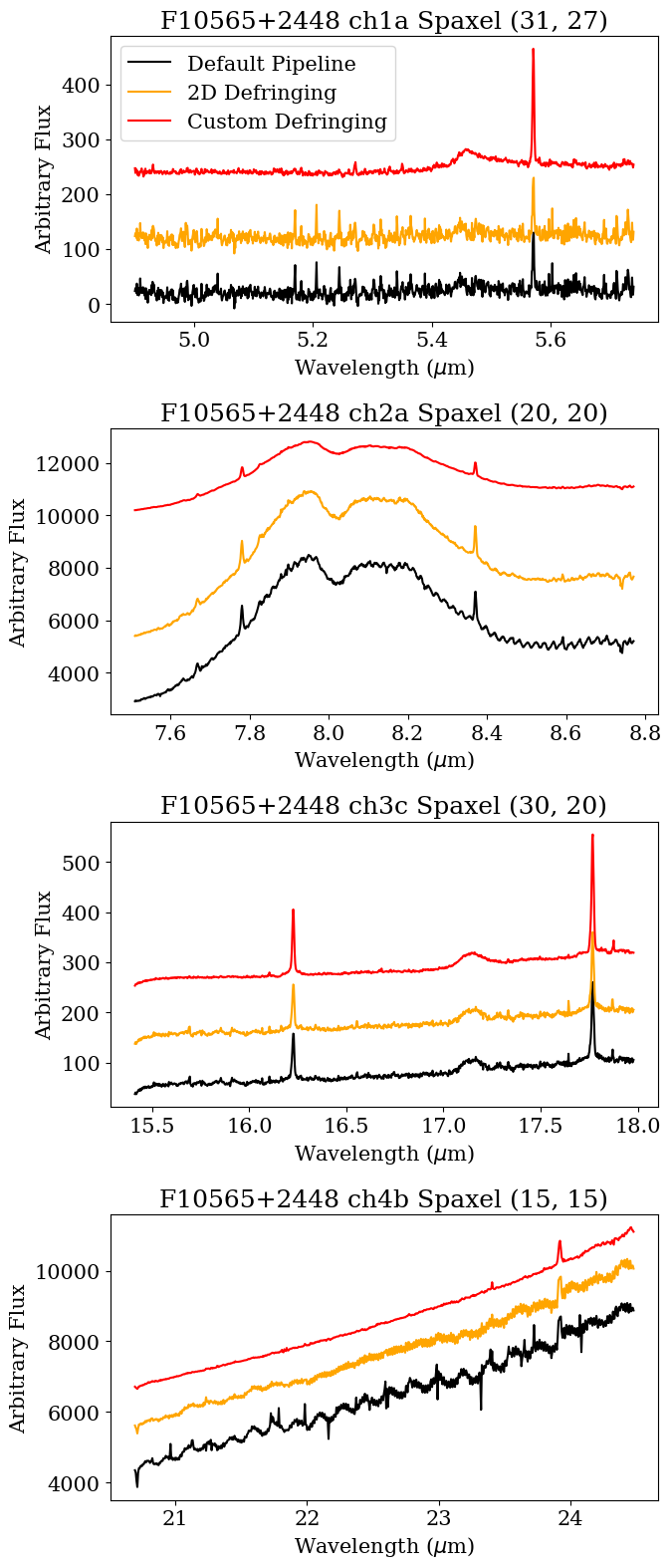}
    \caption{Example spectra for each MIRI channel, showing the defringing process for the F10565+2448 data. Each subplot shows the data reduced by the \jwst\ pipeline \textit{without} 2D defringing (black), the data reduced by the \jwst\ pipeline \textit{with} 2D defringing (orange), and the final post-processed spectra which have been smoothed and defringed again using the \jwst\ pipeline function \texttt{fit\_residual\_fringes\_1d} (red). The spectra are arbitrarily offset in flux for visibility.
    \label{fig:spec_comp}}
\end{figure}

To further correct for fringing seen across all 4 channels that increases with higher frequencies, we first smooth the cube with a running circular average using a radius of 1.5 pixels to eliminate the fringes caused by spatial undersampling and increase the S/N ratio of the higher frequency fringes. Then we apply the \jwst\ pipeline function \texttt{fit\_residual\_fringes\_1d}  spaxel-by-spaxel to the reduced cubes. We find that the fringing was eliminated or significantly reduced in a majority of spaxels, allowing for improved analysis of line kinematics. For detailed comparisons of the before and after of this process, see Figure \ref{fig:spec_comp}. Additionally, we correct the World Coordinate System (WCS) of each cube to make the location of the brightest spaxel in integrated light equal to the center of \fonezero\ \citep[RA 10 59 18.128, DEC +24 32 34.74;][]{alam2015}.

For spatial comparisons of the spectrally distant H$_2$ emission lines, we create a set of cubes with unified effective point spread function (PSF) sizes, hereafter referred to as the ``PSF-matched'' cubes. To do this, we spatially smooth a set of the final reduced cubes to the effective full-width-at-half-maximum (FWHM) of the PSF at the observed wavelength of the longest wavelength line: H$_2$ 0$-$0 S(1) 17.03 $\mu$m. We find the PSF FWHM of H$_2$ 0$-$0 S(1) from interpolating the wavelength-dependent PSF FWHM of MIRI/MRS which was estimated from the values observed in 3D drizzled cubes (Figure 12 of \citealt{Arg2023}). At every wavelength, we calculate the quadrature difference ($\sigma_{convolve} = \sqrt{\sigma^2_{17.03}-\sigma_{\lambda}^2}$) to the PSF FWHM of H$_2$ 0$-$0 S(1) and convolve the slice with a 2D Gaussian. This corresponds to a final PSF FWHM of $\sim$0.75\arcsec\ for F10565+2448. Since the PSF unification process results in lower spatial resolution for the shorter wavelength lines, the PSF-matched cubes are only used in the specific cases when direct spaxel-to-spaxel comparisons need to be made.

For the one-dimensional nuclear spectral analysis, we adopt a procedure similar to that described by \cite{See2024}. The nuclear spectrum is extracted directly from the pipeline-processed data cubes using an aperture diameter defined by a wavelength-dependent FWHM: FWHM$(\lambda) = 0.75$\arcsec\ for $\lambda < 8$ $\mu$m, and FWHM$(\lambda) = 0.75 \times [\lambda(\mu\mathrm{m})/8]$\arcsec\ for $\lambda \geq 8$ $\mu$m. Each spectral channel is first defringed using the \texttt{fit\_residual\_fringes\_1d} routine, after which the individual segments are concatenated into a continuous spectrum by interpolating across overlapping wavelength regions. The final spectrum is rebinned onto a uniform grid with a dispersion of $\Delta\lambda = 8$ \AA, corresponding to the spectral element width of Channel 1.

\section{Data Analysis}
\label{sec:analysis}

\subsection{Nuclear Emission}
To fit the nuclear spectrum, we utilize the Python version of the \cafe\ \citep[Continuum And Feature Extraction;][]{Diaz2025} \jwst\ IFU spectral fitting code, which can be accessed on the GOALS survey \href{https://github.com/GOALS-survey/CAFE}{GitHub repository}. \cafe\ performs a spectral decomposition of the continuum emission (stellar and/or dust), as well as emission and absorption features and broad emission from PAHs. 

We provide \cafe\ with the manually extracted nuclear spectrum and specify several user-defined settings.
We model the continuum using three of \cafe's fully characterized dust continuum emission components defined by their blackbody emissivity at the equilibrium temperature (cool, warm, and hot), which we find best fits the continuum. For the heating sources, we assign the ISRF (representing the average interstellar radiation field) to the cool and warm dust, and an AGN SED to the hot component. Additionally, we include the 6.1 $\mu$m optical depth from water ice.

We utilize a PAH template based on the mean values from Table B.1 in \citet{putte2025} while including the two longest wavelength PAH components in the standard \cafe\ PAH template (at 17.87 and 18.92 $\mu$m), as the \citet{putte2025} table only goes up to 17.76 $\mu$m. The PAHs' central wavelength and gamma (FWHM $=$ rest wavelength $\times$ gamma) are allowed to vary within 500 km s$^{-1}$ and 10\% respectively. The PAHs' attenuation is set to be the same as the warm dust component.

Although we allow \cafe\ to fit emission lines, we do not use the results for our analysis. \cafe\ sacrifices accuracy over narrow wavelength regions in favor of an effective broad continuum fit, resulting in line fits that are less accurate. For more accurate line fits, we subtract the continuum provided by the \cafe\ fits and then perform another manual polynomial continuum subtraction before fitting one or two Gaussian components to the residual emission line flux. To correct for instrumental broadening, we utilize the resolution vs. wavelength relation for MIRI/MRS from \citet{jones2023}. 

\subsection{Extranuclear Emission}
We fit the data cube spaxel-by-spaxel with the software package \qtdfit\ \citep{Rupke2023}, inspired by \ifsfit\ \citep{q3d2014}. \qtdfit\ is designed specifically for the analysis of IFS data on QSOs, as it removes the bright PSF caused by the central compact AGN emission, revealing the much fainter emission from the host galaxy without contamination from the bright PSF. 

\qtdfit\ extracts a spectrum to use as a PSF template from either the brightest spaxel in the data, a defined radius around that bright spaxel, or a manually set spectrum. In an initial fit, the template, along with emission lines, is fit to the data. For each spaxel, the template is scaled up or down with a series of exponentials to match the continuum and remove any signal that resembles the nuclear spectrum. The process loops, refitting the emission lines and continuum until certain residual levels are met. Emission lines are fit with a specified number of Gaussian components to the spectrum with nuclear PSF emission removed. With each iteration, if a line fit does not pass a specified significance cut, it is removed and fit with fewer components or not fit at all. We require the Gaussian components to be detected at the 3-$\sigma$ level, except in a few spaxels where a second component below the 3-$\sigma$ threshold is allowed after visual inspection of the data. This entire process is done for every spaxel in the data and can be accelerated with multicore processing. More details on the \qtdfit\ procedure and its use in the analysis of \jwst\ IFS data can be found in \citet{Wylezalek2022, Rupke2023, Veilleux2023, Vayner2023}. As MRS bands can often have spectral jumps due to discontinuities between sub-bands, we separately run \qtdfit\ on specific sub-bands where each line is observed. 

We fit the same lines specified in the nuclear spectrum fits with \cafe. However, broad components are not seen in the extended fine structure line emission, so multiple component fits are only utilized in the H$_2$ rotational lines S(1) through S(5). 

As \qtdfit\ currently lacks the capability to fit PAH emission, we adopt the method from \citet{draine2021} for the extranuclear PAH analysis. We obtain the ``clipped" flux $F_{\mathrm{clip}}$ of each feature by specifying points $\lambda_1$ and $\lambda_2$ bracketing the feature where the strength of the feature is taken as zero and defining a ``clip-line" $\lambda F_\lambda^{\mathrm{c.l.}}$ between $\lambda_1$ and $\lambda_2$ to be a linear function of $\log\lambda$ connecting $\lambda F_\lambda$ at the clip points. $F_{\mathrm{clip}}$ is defined by Equation 19 in \citet{draine2021}: 

\begin{equation}
    F_{\mathrm{clip}} = \int_{\lambda_1}^{\lambda_2}(F_\lambda - F_\lambda^{\mathrm{c.l.}})\ \mathrm{d}\lambda
\end{equation}

As the broad PAH emission spans all channels of MIRI/MRS, we utilize the``PSF-matched" cubes for this analysis. When individual PAH complexes span two channels, the channels are aligned by applying a corrective ratio to raise the shorter wavelength channel to match the longer wavelength channel. 

\section{Results}
\label{sec:results}
\subsection{Nuclear Emission}

Our \cafe\ fit approximates the nuclear spectrum well with $\sim$ 5\% residuals, shown in Figure \ref{fig:CAFE}. The largest residual feature from 11.5 $\mu$m  to 11.9 $\mu$m could be indicative of an unknown PAH feature in the 11.3 $\mu$m complex. The equivalent widths, fluxes, and luminosities of the PAH complexes are listed in Table \ref{tab:PAH}.

\begin{figure*}
    \centering
    \includegraphics[width=0.8\textwidth]{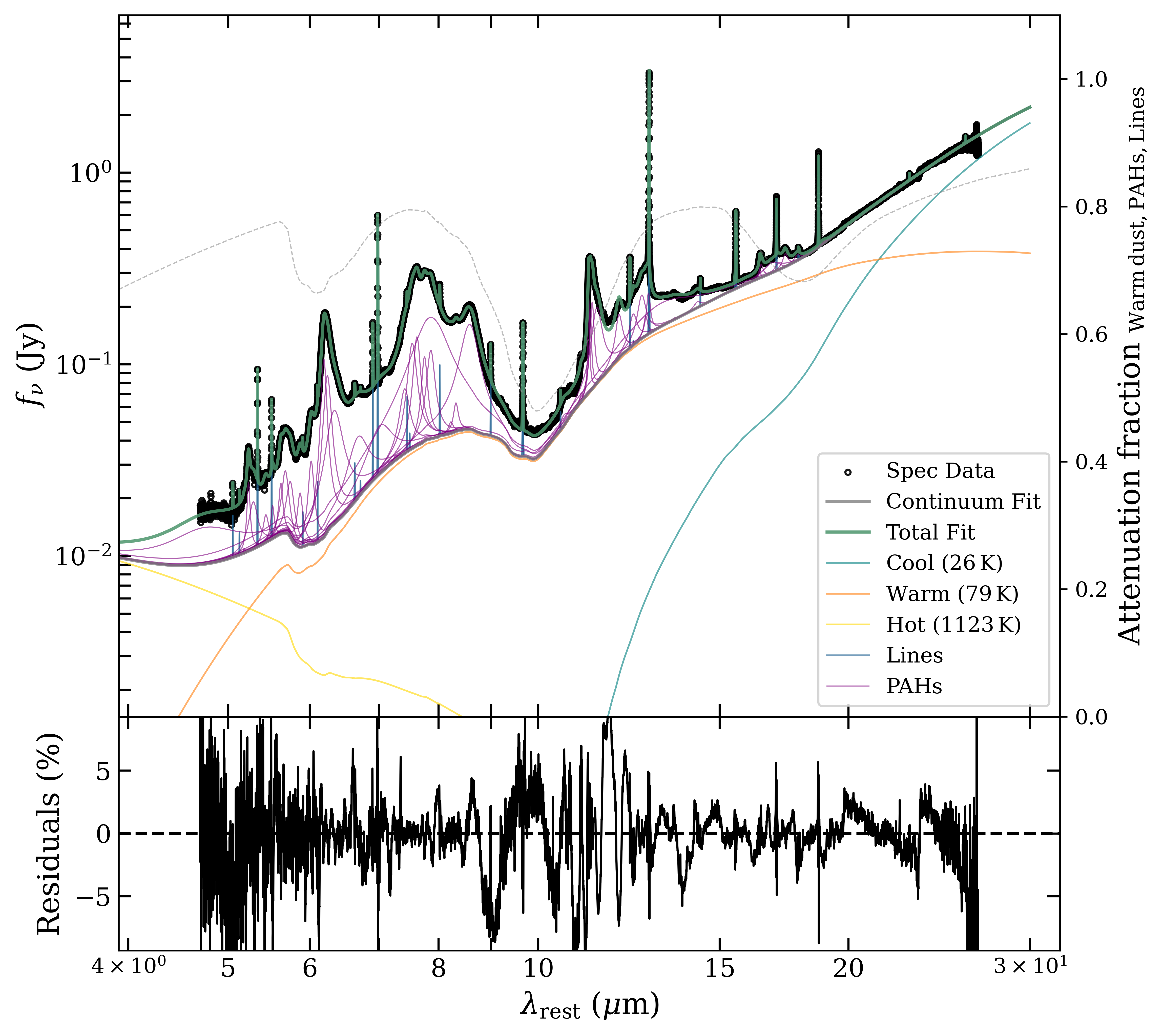}
    \caption{Top: \cafe\ fit of the nuclear extracted spectrum of F10565+2448. Bottom: Residuals of the fit. We included three continuum dust components (cool, warm, and hot), custom PAH features from Table B.1 in \citet{putte2025}, emission lines, and the 6.1 $\mu$m optical depth feature from water ice. The fitted attenuation at 9.6 $\mu$m of the cool, warm, and hot dust components as well as the water ice are respectively $\tau_{\mathrm{cool}} \sim 5.19$, $\tau_{\mathrm{warm}} \sim 1.45$, $\tau_{\mathrm{hot}} \sim 0.00$, and $\tau_{\mathrm{ice}} \sim 0.51$. Emission lines and PAH features are extincted by $\tau_{\mathrm{warm}}$, which is plotted as the dotted gray line (right y-axis). 
    \label{fig:CAFE}}
\end{figure*}

\begin{deluxetable}{l c c c c}
 \tablecaption{Nuclear PAH Measurements
 \label{tab:PAH}}
 \tablehead{\colhead{PAH Feature} & \colhead{6.2 \um} & \colhead{7.7 \um} & \colhead{11.3 \um} & \colhead{17.0 \um}}
 \startdata
    $\lambda$ range (\um) & 5.9 $-$ 6.5 & 6.9 $-$ 9.2 & 10.8 $-$ 11.7 & 15.7 $-$ 18.0\\
    EW\tablenotemark{a} & 1.6 & 6.1 & 1.0 & 0.4 \\
    Flux\tablenotemark{b} & 2.5 $\pm$ 0.4 & 8.8 $\pm$ 0.6 & 2.06 $\pm$ 0.06 & 1.0 $\pm$ 0.3\\
    Flux (corrected)\tablenotemark{b} & 3.7 $\pm$ 0.6 & 11.5 $\pm$ 0.7 & 3.4 $\pm$ 0.1 & 1.4 $\pm$ 0.5\\
    Luminosity\tablenotemark{c} & 4.3 $\pm$ 0.6 & 13.2 $\pm$ 0.9 & 3.9 $\pm$ 0.1 & 1.6 $\pm$ 0.6\\
 \enddata
 \tablenotetext{a}{In units of \um}
 \tablenotetext{b}{In units of $10^{-12}$ \ergscm}
 \tablenotetext{c}{In units of 10$^9$ $L_\odot$}
 \tablecomments{Rows: wavelength range of the selected Drude profiles, equivalent width, total flux of the complex, extinction corrected flux using the warm dust extinction from \cafe\ ($\tau_{9.6\mu \mathrm{m, warm}} \sim 1.45$), and luminosity calculated using the extinction corrected fluxes. The errors for the PAH fits are calculated by \cafe\ using the standard error on fitting parameters from \lmfit, which estimates those errors as the square root of the fit covariance matrix.}
\end{deluxetable}

The emission line fits are summarized in Table \ref{tab:nuc_lines}, which denotes the feature name, rest wavelength, IP (ionization potential needed to produce the ion; 0 in the case of H$_2$), flux, extinction-corrected flux, $v_{50}$, and $w_{80}$. $v_{50}$ is the 50th-percentile or median velocity, and $w_{80}$ is the 80th-percentile velocity width ($|v_{90} - v_{10}|$). Several of the lines are fit with two Gaussian components to represent both the low-velocity, narrow disk emission and the broader extended blue wing; in Table \ref{tab:nuc_lines}, the broad components are denoted in the rows below their respective narrow components. 

The $v_{50}$ and $w_{80}$ of the broad components of the ionized gas tracers are shown in Figure \ref{fig:IP} plotted against the IP and critical density \citep[gathered from \pyneb,]{Luridiana2015} of each tracer. The data show a significant increase in $|v_{50}|$ with increasing IP, driven by the high ionization lines \neiii\ and \nev.To illustrate this positive correlations, we showcase the sequence of strong mid-IR neon emission lines (\neii\ 12.81 $\mu$m, \neiii\ 15.56 $\mu$m, and \nev\ 14.32 $\mu$m) in Figure \ref{fig:nuclear_Ne}. \neii\ and \neiii\ show strong low-velocity narrow emission with extended broad blue wings. \nev\ lacks the narrow low-velocity emission and only shows a broad blueshifted component. The positive correlation with IP may suggest a decelerating outflow with increasing distance from the center, but we cannot rule out the possibility that it is a spatially stratified cone with faster and more highly ionized species near the center. The detection of \nev\ also suggests photoionization by the central AGN \citep{Spinoglio1992, Genzel1998}.

\begin{figure}
    \centering
    \includegraphics[width=\columnwidth]{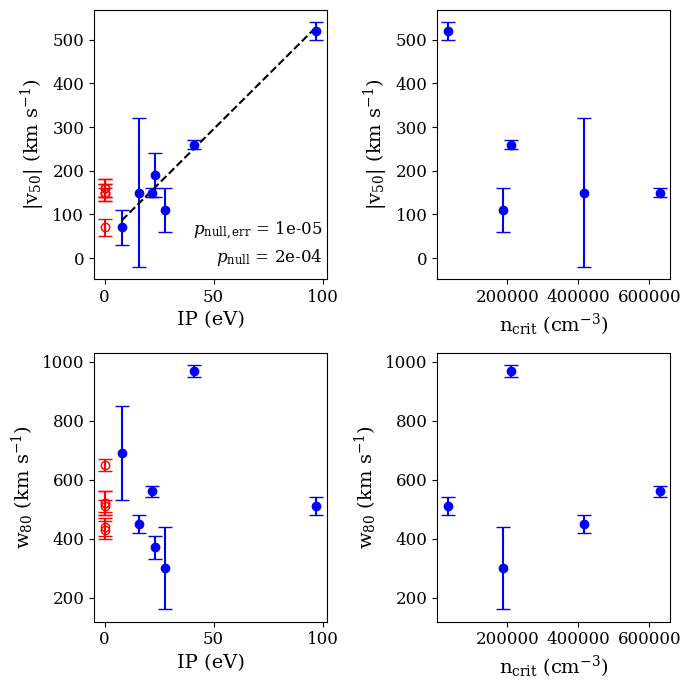}
    \caption{Nuclear results. Left column: plot of $v_{50}$ (top left) and $w_{80}$ (bottom left) vs ionization potential. Right column: plot of $v_{50}$ (top right) and $w_{80}$ (bottom right) vs critical density. Blue points represent ionized gas broad components; red points mark H$_2$ broad components. The red points are not included in the linear regression. Statistically significant linear fits (p[null] $< 0.05$) are displayed with their corresponding p[null] values calculated with and without the error bars. 
    \label{fig:IP}}
\end{figure}

\begin{figure}
    \centering
    \includegraphics[width=0.7\columnwidth]{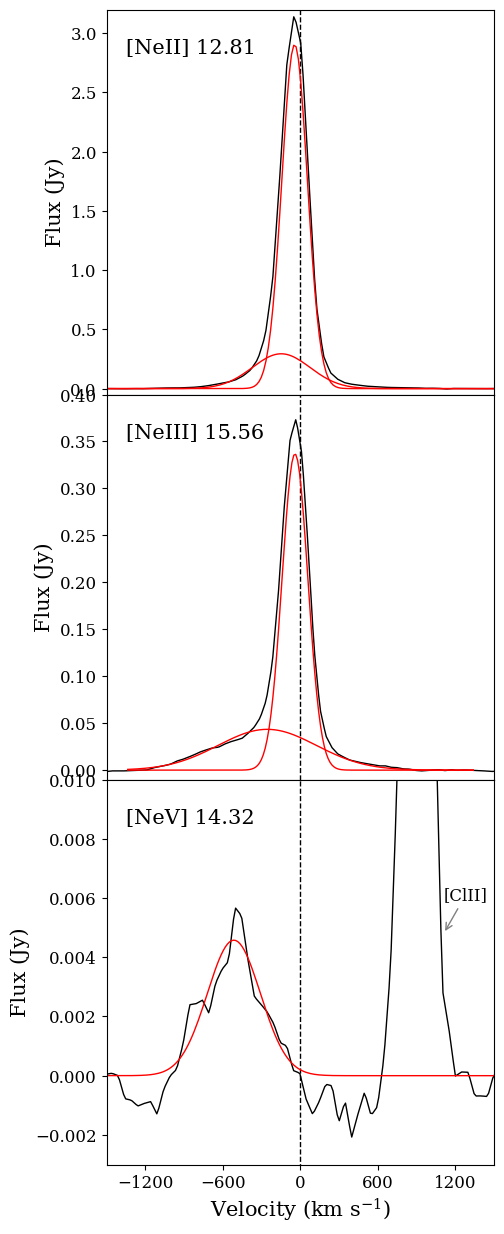}
    \caption{Nuclear fits of the detected neon lines within the MIRI/MRS wavelength range. The continuum-subtracted data are shown in black, while fits are marked in red. We detect a strong, low-velocity narrow component for \neii\ 12.81 $\mu$m (top panel) and \neiii\ 15.56 $\mu$m (middle panel). The \nev\ 14.32 $\mu$m profile (bottom) does not show a narrow component. All three lines display a blueshifted broad component with respective $v_{50}$ values of $-150 \pm 10$, $-260 \pm 10$, and $-520 \pm 10$ \kms\ that are indicative of a nuclear ionized outflow. Note: The narrow emission line in the bottom panel is \clii\ 14.37 $\mu$m. 
    \label{fig:nuclear_Ne}}
\end{figure}

Similarly, several of the H$_2$ rotational lines show blueshifted wings. The wing is detected in S(1), S(2), S(3), S(4), S(5), and S(7) but is strongest in S(3). The flux-weighted average velocity and $w_{80}$ for the broad component are $-$150 \kms\ and 480 \kms, respectively. 

\begin{deluxetable*}{l c c c c c c}
 \tablecaption{Nuclear Emission Line Measurements
 \label{tab:nuc_lines}}
 \tablehead{ 
 \colhead{Feature ID} & \colhead{$\lambda_{rest}$} & \colhead{IP} & \colhead{Flux} & \colhead{Flux (Corrected)} & \colhead{$v_{50}$} & \colhead{$w_{80}$} \\
 \colhead{} & \colhead{$\mu$m} & \colhead{eV}& \colhead{$10^{-14}$ \ergscm} & \colhead{$10^{-14}$ \ergscm} & \colhead{\kms} & \colhead{\kms} }
 \startdata
   H$_2$ $0-0$ S(8) & 5.05 &  --- & 0.29 $\pm$ 0.06 & 0.39 $\pm$ 0.01 & $-$40 $\pm$ 10 & 220 $\pm$ 20  \\
   Humphreys (10$-$6) & 5.13 & --- & 0.11 $\pm$ 0.03 & 0.15 $\pm$ 0.05 & $-$50 $\pm$ 10 & 170 $\pm$ 30 \\
   \feii\ & 5.34 & 7.9 & 2.87 $\pm$ 0.08 & 3.7 $\pm$ 0.1 & $-$40 $\pm$ 10 & 220 $\pm$ 20 \\
   &  &  & 0.4 $\pm$ 0.2 & 0.6 $\pm$ 0.3 & $-$70 $\pm$ 40 & 690 $\pm$ 160 \\
   H$_2$ $0-0$ S(7) & 5.51 &  --- & 1.3 $\pm$ 0.05 & 1.71 $\pm$ 0.09 & $-$60 $\pm$ 10 & 210 $\pm$ 20  \\
   &  &  & 0.6 $\pm$ 0.1 & 0.7 $\pm$ 0.2 & $-$160 $\pm$ 20 & 520 $\pm$ 40 \\
   Humphreys (9$-$6) & 5.91 & --- & 0.24 $\pm$ 0.03 & 0.34 $\pm$ 0.06 & 30 $\pm$ 10 & 220 $\pm$ 20 \\
   H$_2$ $0-0$ S(6) & 6.11 &  --- & 0.68 $\pm$ 0.03 & 0.99 $\pm$ 0.07 & $-$70 $\pm$ 10 & 260 $\pm$ 20 \\
   \niII & 6.64 &  14.5 & 0.43 $\pm$ 0.02 & 0.58 $\pm$ 0.05 & $-$30 $\pm$ 10  & 360 $\pm$ 20 \\
   \feii & 6.72 & 7.9 & 0.20 $\pm$ 0.04 & 0.27 $\pm$ 0.07 & $-$30 $\pm$ 20 & 250 $\pm$ 40 \\
   H$_2$ $0-0$ S(5) & 6.91 &  --- & 2.25 $\pm$ 0.08 & 2.9 $\pm$ 0.1 & $-$20 $\pm$ 10 & 190 $\pm$ 20 \\
   & & & 0.8 $\pm$ 0.2 & 1.0 $\pm$ 0.3 & $-$70 $\pm$ 20 & 430 $\pm$ 30 \\
   \arii\ & 6.99 & 15.8 & 15.8 $\pm$ 3 & 20 $\pm$ 5 & $-$30 $\pm$ 10 & 220 $\pm$ 20  \\
    &  &  & 2 $\pm$ 3 & 3 $\pm$ 5 & $-$150 $\pm$ 170 & 300 $\pm$ 140  \\
   Pfund (6$-$5) & 7.46 & --- & 1.16 $\pm$ 0.03 & 1.45 $\pm$ 0.04 & 40 $\pm$ 10 & 260 $\pm$ 20 \\
   Humphreys (8$-$6) & 7.50 & --- & 0.43 $\pm$ 0.06 & 0.5 $\pm$ 0.1 & 90 $\pm$ 20 & 500 $\pm$ 60 \\
   H$_2$ $0-0$ S(4) & 8.03 &  --- & 1.21 $\pm$ 0.04 & 1.56 $\pm$ 0.06 & $-$80 $\pm$ 10 & 190 $\pm$ 20 \\
   &  &  & 0.43 $\pm$ 0.08 & 0.6 $\pm$ 0.1 & $-$150 $\pm$ 20 & 520 $\pm$ 40 \\
   \ariii\ & 8.99 & 27.6 & 0.9 $\pm$ 0.2 & 1.4 $\pm$ 0.4 & $-$20 $\pm$ 10 & 200 $\pm$ 20 \\
   &  &  & 0.4 $\pm$ 0.2 & 0.6 $\pm$ 0.5 & $-$110 $\pm$ 50 & 300 $\pm$ 30 \\
   H$_2$ $0-0$ S(3) & 9.66 &  --- & 2.10 $\pm$ 0.04 & 4.2 $\pm$ 0.2 & $-$40 $\pm$ 30 & 190 $\pm$ 60 \\
   &  &  & 0.56 $\pm$ 0.09 & 1.1 $\pm$ 0.4 & $-$220 $\pm$ 40 & 330 $\pm$ 60 \\
   \siv & 10.51 & 34.9 & 0.23 $\pm$ 0.01 & 0.45 $\pm$ 0.04 & $-$30 $\pm$ 10 & 210 $\pm$ 20 \\
   H$_2$ $0-0$ S(2) & 12.28 &  --- & 2.17 $\pm$ 0.02 & 3.02 $\pm$ 0.05 & $-$70 $\pm$ 10 & 200 $\pm$ 20 \\
   &  &  & 0.67 $\pm$ 0.05 & 0.9 $\pm$ 0.1 & $-$150 $\pm$ 10 & 580 $\pm$ 20 \\
   Humphreys (7$-$6) & 12.37 & --- & 0.28 $\pm$ 0.01 & 0.39 $\pm$ 0.01 & 50 $\pm$ 10 & 220 $\pm$ 20 \\
   H8 (11$-$8) & 12.39 & --- & 0.042 $\pm$ 0.006 & 0.058 $\pm$ 0.01 & 3 $\pm$ 10 & 260 $\pm$ 30 \\
   \neii\ & 12.81 & 21.6 & 50.3 $\pm$ 0.6 & 66 $\pm$ 1 & $-$40 $\pm$ 10 & 230 $\pm$ 20  \\
    &  &  & 12 $\pm$ 1 & 17 $\pm$ 2 & $-$150 $\pm$ 10 & 560 $\pm$ 20  \\
   \nev\ & 14.32 & 97.2 & --- & --- & --- & ---  \\
   &  & & 0.16 $\pm$ 0.01 & 0.2 $\pm$ 0.02 & $-520$ $\pm$ 10 & 510 $\pm$ 30  \\
   \clii & 14.37 & 12.97 & 0.56 $\pm$ 0.01 & 0.70 $\pm$ 0.01 & $-$60 $\pm$ 10 & 210 $\pm$ 20  \\
   \neiii\ & 15.56 & 41.0 & 4.68 $\pm$ 0.03 & 6.06 $\pm$ 0.05 & $-$40 $\pm$ 10 & 220 $\pm$ 20  \\
    &  &  & 2.65 $\pm$ 0.06 & 3.4 $\pm$ 0.1 & $-$260 $\pm$ 10 & 970 $\pm$ 20  \\
   H$_2$ $0-0$ S(1) & 17.03 &  --- & 4.0 $\pm$ 0.1 & 5.7 $\pm$ 0.2 & $-$80 $\pm$ 10 & 220 $\pm$ 20 \\
   &  &  & 1.9 $\pm$ 0.2 & 2.7 $\pm$ 0.5 & $-$150 $\pm$ 20 & 510 $\pm$ 20 \\
   \piii & 17.89 & 19.77 & 0.32 $\pm$ 0.01 & 0.47 $\pm$ 0.02 & $-$60 $\pm$ 10 & 190 $\pm$ 20 \\
   \feii & 17.94 & 7.9 & 0.43 $\pm$ 0.01 & 0.64 $\pm$ 0.02 & $-$100 $\pm$ 10 & 310 $\pm$ 20 \\
   \siii\ & 18.71 & 23.3 & 8.7 $\pm$ 0.5 & 13 $\pm$ 1 & $-$70 $\pm$ 10 & 210 $\pm$ 20  \\
   &  &  & 1.7 $\pm$ 0.7 & 2 $\pm$ 1 & $-$190 $\pm$ 50 & 370 $\pm$ 40  \\
   \feiii\  & 22.93 & 16.19 & 0.85 $\pm$ 0.02 & 1.06 $\pm$ 0.03 & $-$3 $\pm$ 10 & 210 $\pm$ 20  \\
   \feii & 25.99 & 7.9 & 1.71 $\pm$ 0.07 & 2.05 $\pm$ 0.1 & $-$100 $\pm$ 10 & 260 $\pm$ 20 \\
 \enddata
 \tablecomments{Columns: Name of the emission line, rest wavelength, ionization potential, flux of the fit, extinction corrected flux using the warm dust extinction from \cafe\ ($\tau_{9.6\mu \mathrm{m, warm}} \sim 1.45$), 50th-percentile velocity, and 80th-percentile width.  Uncertainties were calculated as the square root of the diagonal of the fit covariance matrix; however, in the case of $v_{50}$ and $w_{80}$, we applied floors of 10 and 20 \kms\ respectively. When two Gaussian components are used, the broad component is listed as the row below the narrow component.}
\end{deluxetable*}

Although we fit the Hydrogen recombination lines (Hu $\alpha$, Hu $\beta$, Hu $\gamma$, Hu $\delta$, Pf $\alpha$, and H8 (11-8)) in this object, the lack of difference in $A(\lambda)/A(V)$ of each line made estimating the extinction difficult. A plot of the mid-IR extinction curve from \cite{gordon2023} with the locations of our Hydrogen recombination lines overplotted is shown in Figure \ref{fig:ext}. To correct for extinction in the nuclear emission lines and PAH flux measurements, we used the warm dust extinction value from \cafe\ ($\tau_{9.6\mu \mathrm{m, warm}} \sim 1.45$).

\begin{figure}
    \centering
    \includegraphics[width=\columnwidth]{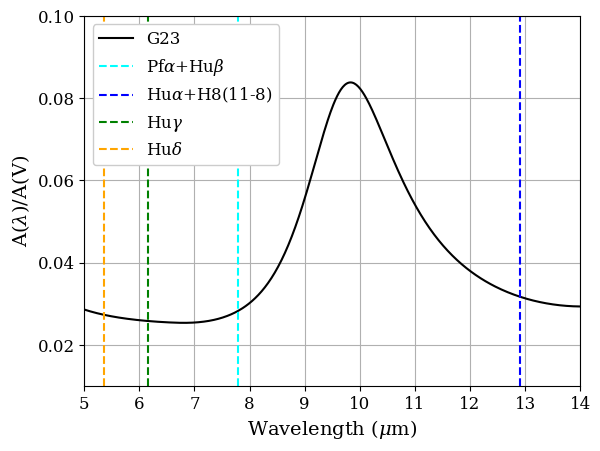}
    \caption{Extinction curve of \citet{gordon2023} with the positions of the detected Hydrogen recombination lines overplotted. 
    \label{fig:ext}}
\end{figure}

\subsection{Extranuclear Emission}

Blueshifted wings are also visible in the extranuclear H$_2$ rotational lines, sometimes even resolving into distinct split peaks. Figure \ref{fig:H2_maps} shows the full fitted PSF-matched cubes for H$_2$ 0$-$0 S(1) and S(3), the two strongest rotational transitions in these data. The flux map for each Gaussian component is shown in the left column, $v_{50}$ in the middle column, and $w_{80}$ in the right column. Component 1 traces a nearly face-on disk rotating with an asymmetric spread of velocities ($-70 < v_{50} < 120$ \kms) and narrow widths ($w_{80} < 150$ \kms). Component 2 reveals a slightly faster ($-300 < v_{50} < -110$ \kms), broader ($w_{80} < 260$ \kms), and exclusively blueshifted outflow to the southwest. Figure \ref{fig:all_H2} displays example fits to H$_2$ 0$-$0 S(1) through S(4) for an extracted region within the outflow (right panels) marked as a white circle in the central H$_2$ 0$-$0 S(1) $v_{50}$ map, as well as fits to the nuclear extracted region (left panels). Although not shown in Figure \ref{fig:all_H2}, the outflow is also weakly detected in S(5).

\begin{figure*}
    \centering
    \includegraphics[width=0.7\textwidth]{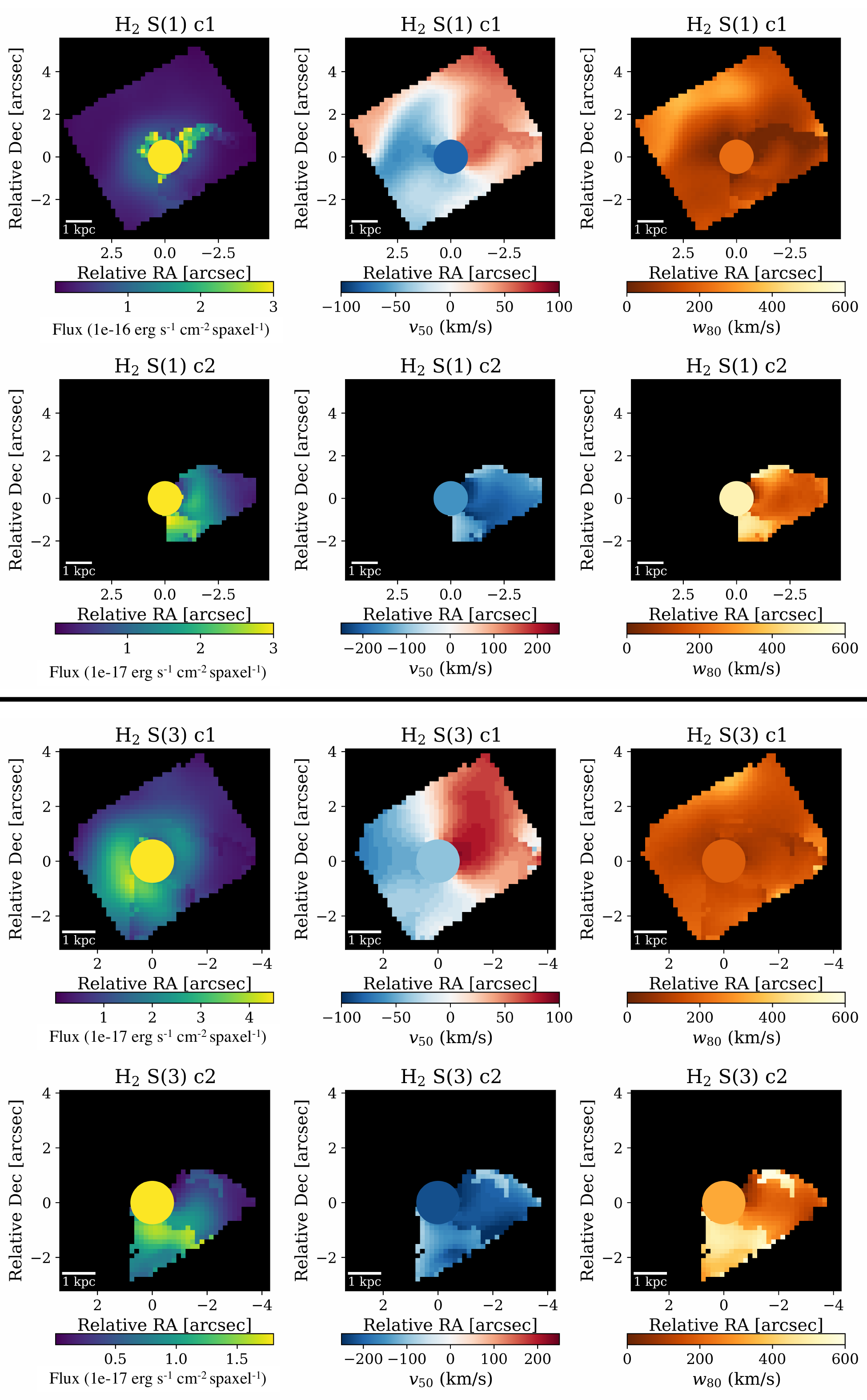}
    \caption{Maps of the two strongest warm molecular gas tracers in F10565+2448: H$_2$ 0$-$0 S(1) (top two rows) and H$_2$ 0$-$0 S(3) (bottom two rows). For each line, up to two Gaussian components are used in the fits, and they are distinguished in the panels by c1 and c2. In all panels, north is to the top and east to the left. These fits utilize the PSF-matched cubes, so the PSF sizes for each map should be the same despite H$_2$ 0$-$0 S(1) and H$_2$ 0$-$0 S(3) lying in different MIRI/MRS channels. The pixel scale is 0.20 arcsec px$^{-1}$. The central circles represent the nuclear extracted spectrum ($\sim$ 0.75\arcsec) where colors are accurate to the nuclear line parameters (see Table \ref{tab:nuc_lines}). Left column: Maps of the line fluxes in erg s$^{-1}$ cm$^{-2}$ spaxel$^{-1}$. Middle column: Median velocities (v$_{50}$) in \kms. Right column: 80th-percentile line widths ($w_{80}$). Component 1 traces a low-velocity, nearly face-on disk. Component 2 reveals an asymmetric, blueshifted outflow with $v_{50}$ of up to $-$280 km s$^{-1}$.
    \label{fig:H2_maps}}
\end{figure*}

\begin{figure*}
    \centering
    \includegraphics[width=\textwidth]{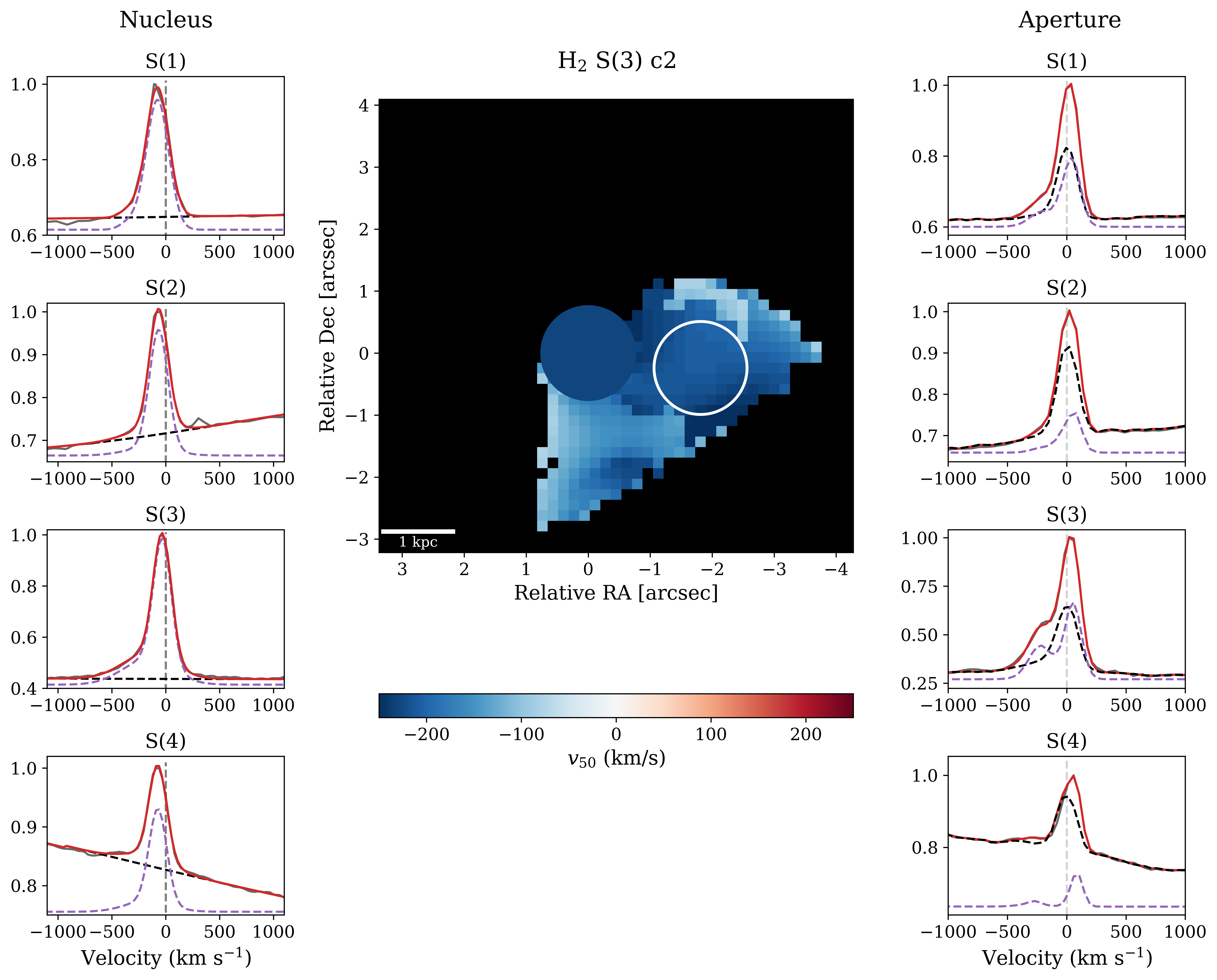}
    \caption{Comparisons of the line profiles of H$_2$ 0$-$0 S(1), S(2), S(3), and S(4). The central map showcases the $v_{50}$ of the secondary component of H$_2$ 0$-$0 S(3), using the PSF-matched cube. Right: line fits for S(1) through S(4) of the extranuclear extracted aperture marked in the central map by a white circle. Left: line fits for S(1) through S(4) for the nuclear extracted region, marked in the central map by the large central circle colored according to the nuclear S(3) $v_{50}$. The units of the vertical axes are relative flux (normalized to 1).  
    The gray solid lines in the left and right panels represent the data; black dashed lines show either the PSF-scaled spectra (right panels) or continuum fits (left panels); purple dashed lines display the line fits with two Gaussian components arbitrarily offset in the y-axis for visibility; and solid red lines trace the total fit. In the nuclear region, the outflow appears as a slightly blueshifted wing. Further from the center, the relative intensity of the outflow component grows into easily discernible split peaks. 
    \label{fig:all_H2}}
\end{figure*}

The kinematics of the warm molecular gas as a function of projected distance from the galaxy center are presented in Figure \ref{fig:pv_F10565} for the H$_2$ 0$-$0 S(1) transition. The black circles represent the rotating disk component, and the blue circles represent the outflow component. The outflow is kinematically distinct from the disk with a faster blueshifted $v_{50}$ and slightly higher $w_{80}$, supporting the outflowing nature of the gas. The asymmetric nature of the disk could be due to the turbulent nature of the merger. The nuclear component, marked by a blue star, is about the same velocity as the extended emission, which could imply that the outflow coasts at a steady velocity as it traverses the host galaxy out to 4+ kpc.

\begin{figure}
    \centering
    \includegraphics[width=\columnwidth]{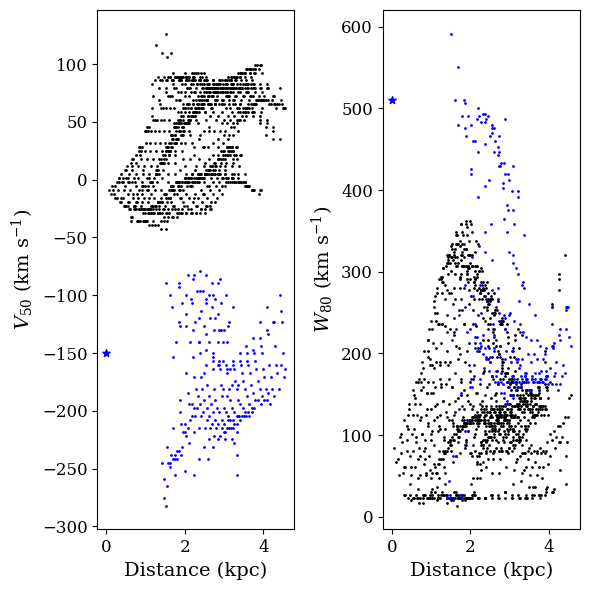}
    \caption{``Bird flock'' position-velocity diagram: Kinematics of the warm-H$_2$ gas derived from H$_2$ 0$-$0 S(1). Left: Median velocities ($v_{50}$) as a function of projected distance from the galaxy center. Black circles represent the rotating disk component, and blue circles represent the outflow component. The blue star marks the nuclear outflow derived from H$_2$ 0$-$0 S(1). Right: The $w_{80}$ of H$_2$ 0-0 S(1) as a function of projected distance from the center using the same symbols as the left panel. 
    \label{fig:pv_F10565}}
\end{figure}


Representing the warm ionized gas, the full fitted cubes for \neiii\ are shown in Figure \ref{fig:NeIII}. Similarly to the H$_2$ lines, we see a low velocity rotating disk. We do not detect an extranuclear blueshifted outflow as seen in the warm molecular gas. The central circles in Figure \ref{fig:NeIII} are colored according to the narrow component of the nuclear \neiii\ fit, showing a slight blueshift and relatively high $w_{80}$. For the nuclear fit parameters of both the disk and outflowing \neiii, refer to Table \ref{tab:nuc_lines}.

\begin{figure*}
    \centering
    \includegraphics[width=0.95\textwidth]{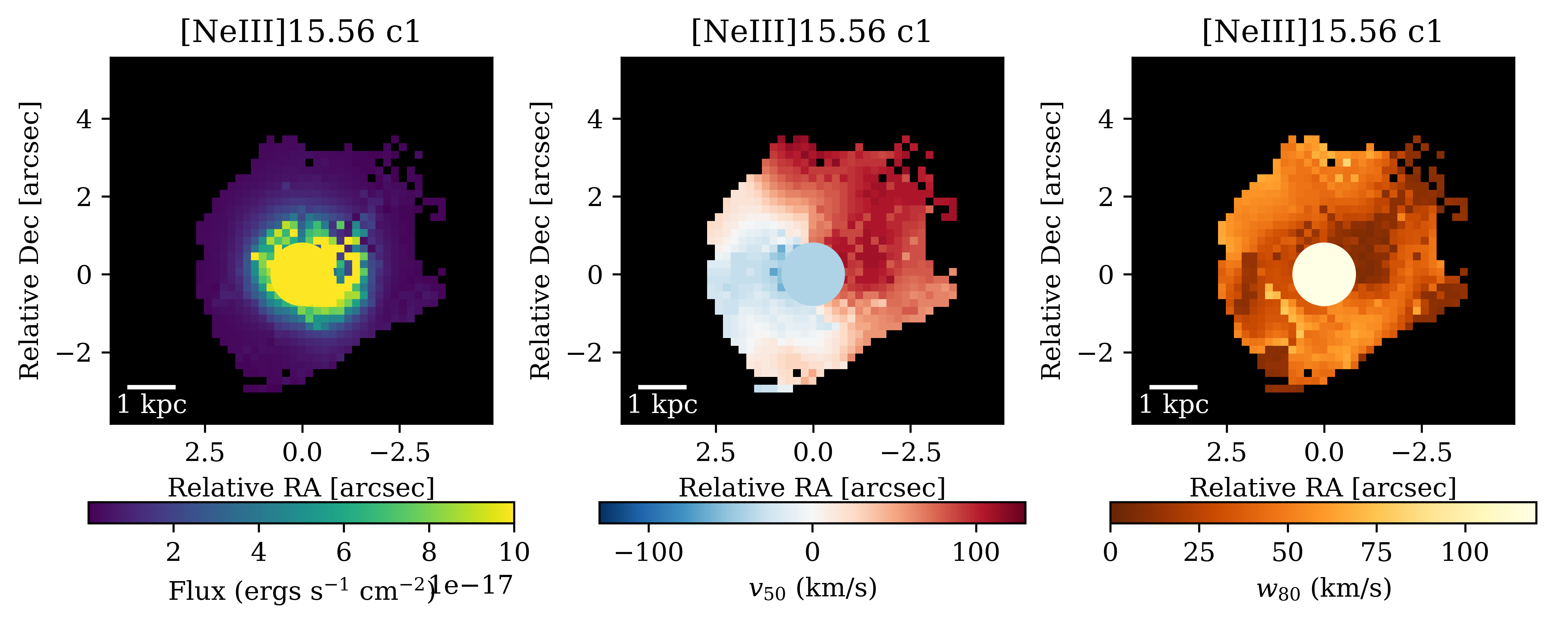}
    \caption{Results from Gaussian fits to the \neiii\ line profiles in F10565+2448. In all panels, north is to the top and east to the left. The pixel scale is 0.20 arcsec px$^{-1}$. The central circles represent the nuclear extracted spectrum ($\sim$ 0.75\arcsec) where colors are accurate to the \neiii\ low-velocity narrow component parameters (see Table \ref{tab:nuc_lines}). Left to right: Maps of the line fluxes, median velocities ($v_{50}$), and 80th-percentile line widths ($w_{80}$). The maps reveal a nearly face-on disk rotating with a maximum $|v_{50}|$ of 150 \kms. The extended blueshifted outflow seen in the warm molecular gas is not detected here.
    \label{fig:NeIII}}
\end{figure*}



\section{Discussion}
\label{sec:discussion}
\subsection{Warm Molecular Gas}

Our results have revealed the existence of a blueshifted warm-molecular outflow to the southwest with a flux-weighted mean projected velocity of $\sim -150$ \kms. These findings are consistent with the blueshifted outflow in neutral and ionized gas phases \citep{shih2010, Rupke2013}, although we find significantly slower velocities compared to the cold molecular gas phase \citep[$\pm (300-600)$ \kms;][]{cicone2014}, which also shows redshifted emission. From the noise around the H$_2$ 0$-$ S(1) line in the north-eastern part of the galaxy, we estimate an upper limit on the flux of the redshifted side of the outflow to be $\sim1/3$ that of the blueshifted side. Assuming the blueshifted and unobserved redshifted sides  have equal fluxes, we utilize the \citet{gordon2023} extinction curve to estimate an $A_V$ value of 35. $A_V$ values upwards of 90 have been observed in ULIRG nuclei \citep[e.g.,][]{spoon2004}, so it is possible for this source to be obscuring the redshifted side of the outflow in all phases except the longest wavelength CO observations.

Following the methods in \citet{Liu2020} derived from \citet{veilleux2020}, we use the S(1) disk component $v_{50}$ measurements to derive a conservative lower limit on the circular velocity
$v_{\mathrm{circ}}=\sqrt{v_{\mathrm{H}_{2}}^2 + 2\sigma_{\mathrm{H}_{2}}^2} \sim 100 \mathrm{\ km\ s}^{-1}$
and the escape velocity $v_{\mathrm{esc}} \simeq 3v_{\mathrm{circ}}=300$ \kms, noting that these velocities are not corrected for inclination. 
The stellar mass of F10565+2448 is Milky Way-like \citep[$\sim3.6\times 10^{10}$ M$_\odot$;][]{dey2024}, which, assuming a Milky Way circular velocity of 220 \kms, implies an approximate inclination of 27 deg. Assuming the outflow extends perpendicular to the disk, the inclination-corrected outflow $v_{50}$ values would not exceed $-282/\cos{27^\circ} \approx -316$ \kms, within the Milky Way's escape velocity.
Even using the projected velocities, after integrating over the outflow profiles we find that only 0.01\% of the total S(1) outflow flux exceeds the projected escape velocity of 300 \kms. This implies that the outflow is not fast enough to escape the gravitational well of the galaxy and instead will fall back as a galactic fountain.

Next we estimate the temperature and column density of the warm-molecular gas. We create excitation diagrams using the spectroscopic transitional data from \citet{roueff2019} and Equation 1 of \citet{youngblood2018}:
\begin{equation}
\label{eqn:NvJ}
    N(\nu_u,J_u) = \frac{4\pi\lambda_0}{hc}\frac{I(\nu_u,J_u\rightarrow\nu_l,J_l)}{A(\nu_u,J_u\rightarrow\nu_l,J_l)}, 
\end{equation}
where $I(\nu_u,J_u\rightarrow\nu_l,J_l)$ are the H$_2$ line intensities and $A(\nu_u,J_u\rightarrow\nu_l,J_l)$ are the Einstein A coefficients from \citet{wolniewicz1998}. To determine the temperature and column density from the excitation diagram, we implement two different methods. 

The first method uses a broken linear fit to Equation 2 of \citet{youngblood2018}: 
\begin{equation}
\label{eqn:h2temp}
    \log_{10} \frac{N(v_u, J_u)}{g(J_u)}=-\frac{1}{T\cdot \ln(10)} \frac{E(v_u, J_u)}{k_B} + \log_{10} N(0,0), 
\end{equation}
where $N(v_u, J_u)$ is the level H$_2$ column density as described in Equation \ref{eqn:NvJ}, $g(J_u)$ is the level degeneracy, $T$ is the temperature, $\frac{E(v_u, J_u)}{k_B}$ is the upper-level energy, and $N(0,0)$ is the column density in the ground $\nu=0$ $J=0$ para-H$_2$ level, which is equal to the total H$_2$ column density $N_{H_2}$ scaled by the partition function $Z(T) = 0.0247 \times T \times \left(1 - e^{-6000/T}\right)$ from \cite{roussel2007}. For the disk component, we utilize two linear fits: the first including S(1), S(2), S(3), and S(4), which are sensitive to cooler temperatures (T$_{\mathrm{cool}}$); and the second including S(5), S(6), S(7), and S(8), which are sensitive to hotter temperatures (T$_{\mathrm{warm}}$). As the outflow component only contains S(1) through S(5), we only use a single linear fit to the cooler temperature lines (S(1), S(2), S(3), S(4)). Excitation diagrams of the nuclear extracted spectrum, as well as an aperture of the extranuclear outflow region, can be seen in Figure \ref{fig:excitation}. The broken linear fits are shown in blue and red for T$_{\mathrm{cool}}$ and T$_{\mathrm{warm}}$, respectively. 

The second method is based on the H$_2$ excitation model of \citet{togi2016}. We utilize an adapted Python implementation from \citet{jones2024} to fit a power-law distribution to the excitation diagram. The model fits the lower bound on the temperature distribution $T_l$ and the power law slope with an assumed upper bound on the temperature distribution ($T_u$ = 2000 K; the model is insensitive to any values larger than 2000 K). In Figure \ref{fig:excitation}, the power law fit is shown in gray. Compared to the linear fit, the power law can better extrapolate to S(0), which is outside of the spectral range of MIRI and an indicator of cooler, more abundant gas. As such, the power law method provides consistently higher column densities and lower temperature estimates than the broken linear method. Although we could not quantify this difference consistently across varying spaxels in order to apply a global correction factor, the results from the power law fits are included in the excitation diagrams in Figure \ref{fig:excitation} for comparison. 

\begin{figure}
    \centering
    \includegraphics[width=\columnwidth]{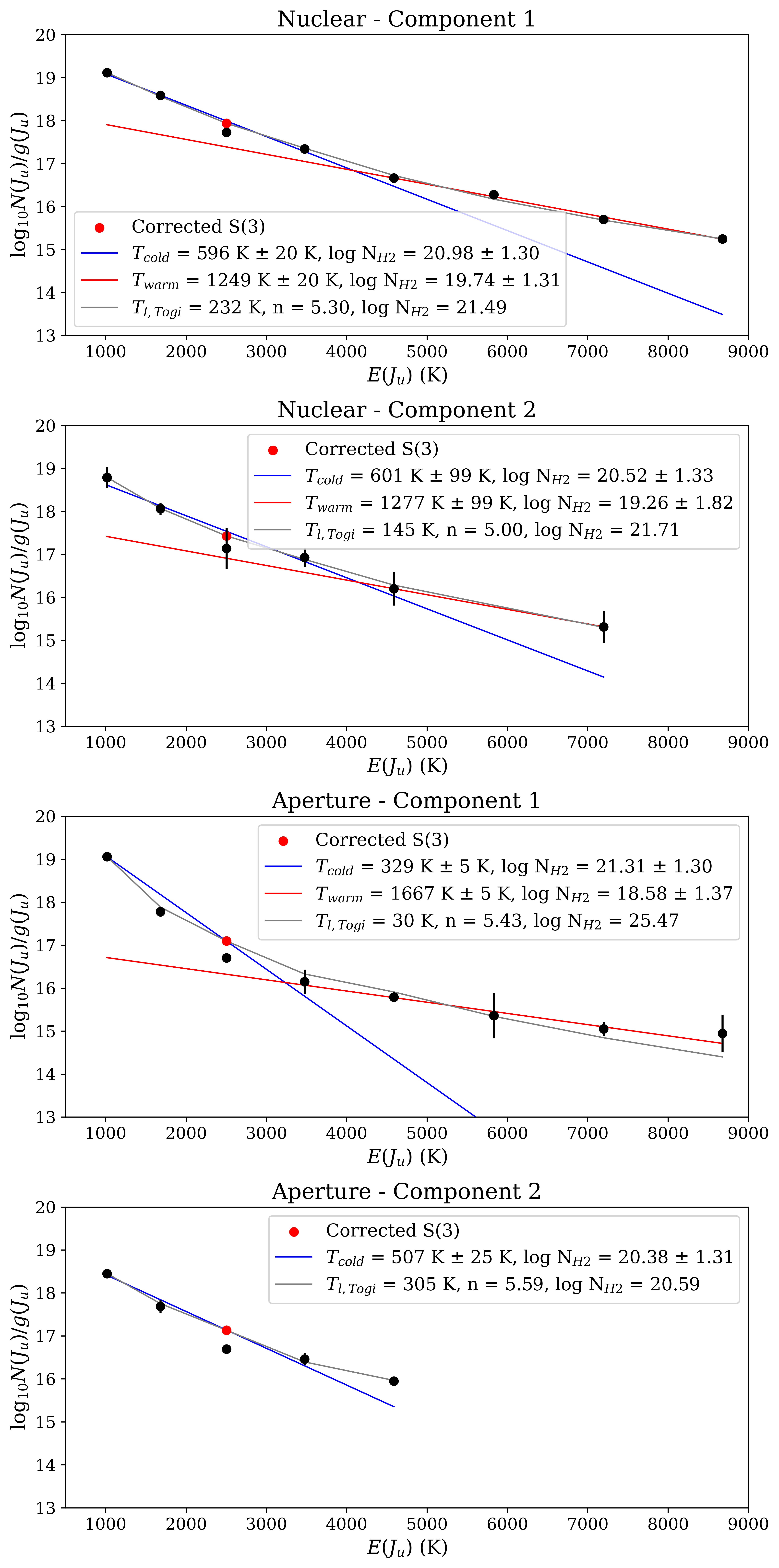}
    \caption{Example excitation diagrams based on the fluxes of the H$_2$ 0$-$0 lines. The top two panels show excitation diagrams for the nuclear extracted spectrum components 1 and 2. The bottom two panels display the excitation diagrams of components 1 and 2 for an extracted aperture in the outflow region, shown as the white circle in Figure \ref{fig:all_H2}. The blue and red lines are broken linear fits as expressed by Equation \ref{eqn:h2temp}, with the break at $E(J_u) = 4000$K. Component 2 of the outflow aperture lacks enough H$_2$ 0$-$0 lines for two linear fits, so we perform a single linear fit on only S(1) through S(4), which is shown by the blue line. Gray lines denote the power law model of \citet{togi2016}, which is used to correct for extinction in H$_2$ 0$-$0 S(3). The extinction-corrected values are shown as the red dots.
    \label{fig:excitation}}
\end{figure}

For all spaxels and the nuclear fit, we assume an ortho/para ratio of 3 for the statistical weights of the molecular transitions. Although smaller ortho/para ratios have been observed in some cases \citep[e.g.][]{habart2005}, we generally find that lower values cause discontinuities between the ortho and para transitions, and thus we adopt a value of 3 as a conservative upper limit.

In both the nuclear extracted spectrum and extranuclear regions, we see a significant drop in the S(3) line relative to other nearby lines. We attribute the drop to extinction from the silicate absorption feature from $\tau_{9.7}$ and correct for this by fitting the \citet{togi2016} method with all lines excluding S(3), correcting S(3) to the \citet{togi2016} fit, and refitting with the broken linear method. The corrected S(3) value is shown as a red dot in Figure \ref{fig:excitation}. 

While fitting the full cubes, we find that the power law method is very sensitive to the errors in the data. The power law fit highly weights the H$_2$ S(1) line, so even slight changes in flux caused by nearby fringing or residual noise can lead to higher column densities and lower temperatures than would be expected. Additionally, the lower signal S(4) $-$ S(7) lines often have larger error bars that lead to less reliable fits as well. We find that the simple linear fit  is more robust in response to fitting errors and leads to smoother temperature and column density maps than the power law fits. Thus, for the following results, we opt for the linear fits.    

Maps showing the temperature, column density, and number of H$_2$ lines detected for the disk and outflow components are shown in Figure \ref{fig:temp_dens}. We require a minimum of four lines for each fitted spaxel's excitation diagram. The central circles represent the nuclear extracted region with sizes of the PSF FWHM (0.75\arcsec) and are colored according to the results from the nuclear excitation diagrams shown in the top two panels of Figure \ref{fig:excitation}. 

The maps of the disk (top row) reveal a few cool, dense clumps to the north of the nucleus, which roughly align with the known star cluster to the north east of the nucleus \citep[see Figure 2 in][]{Rupke2013}. Aside from the clumps, the majority of the disk lies at a temperature between $\sim 300-350$ K and displays a slight radial gradient in column density, decreasing from $\sim 10^{20}$ cm$^{-2}$ around the nucleus to $\sim 10^{19}$ cm$^{-2}$ towards the edges. 

The outflow maps (bottom row) show interesting structure: areas of higher temperature ($\sim 450$K) and lower column density ($\sim 10^{18}$) around the north-western edge and directly south of the nucleus. These areas align with the regions of higher $w_{80}$ and lower $v_{50}$ seen in Figure \ref{fig:H2_maps}. This could represent the turbulent edges of the outflow as it hits the interstellar medium and begins to slow down; we discuss the potential of a shock front in Section \ref{sec:shock}. The orientation of the outflow, which is directed away from the high-density regions of the disk, may indicate that the outflow preferentially propagates along directions of minimal resistance rather than isotropically.

In the nuclear aperture, the outflow and disk are similar temperatures ($\sim 600$K), while in the outflow-centered aperture, the outflow component is slightly hotter ($\sim 500$K) than the disk ($\sim 330$K).

\begin{figure*}
    \centering
    \includegraphics[width=\textwidth]{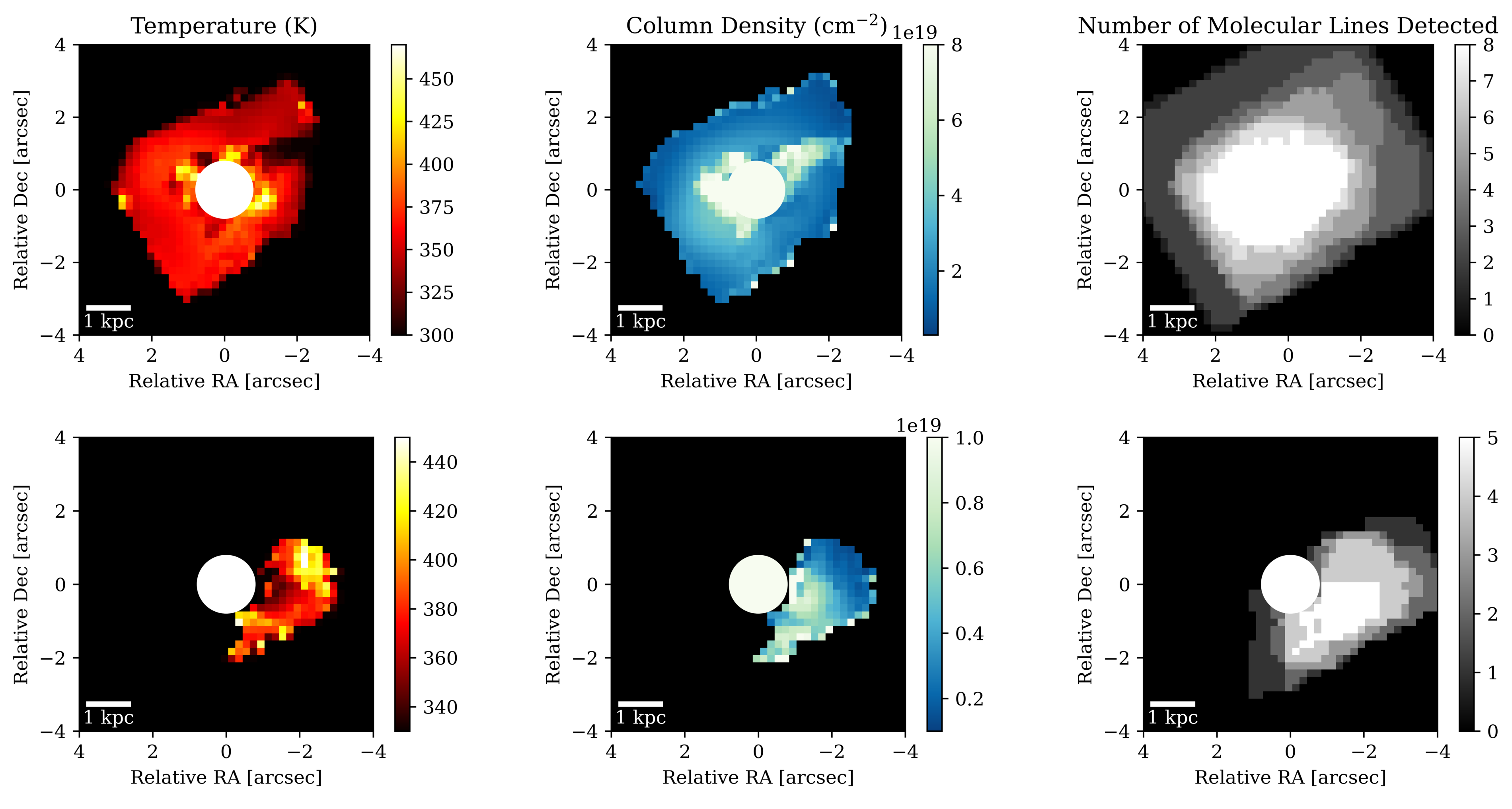}
    \caption{Maps showing the temperature (left), column density (middle), and number of H$_2$ 0$-$0 lines detected (right) for the disk component (top row) and outflow component (bottom row). The disk contains a few cool, dense clumps to the north of the nucleus, which may align with the known star cluster to the north east of the nucleus \citep[see Figure 2 in][]{Rupke2013}. The outflow maps reveal areas of higher temperature ($\sim 450$K) and lower column density ($\sim 10^{18}$) around the north-western edge and directly south of the nucleus.
    \label{fig:temp_dens}}
\end{figure*}

We calculate the total mass of warm molecular gas using the column density results from both the nuclear extracted region and the extended emission. 
Summing all spaxels in the resolved emission as well as the nuclear extracted region leads to a total warm molecular mass of $(3^{+3}_{-1}) \times 10^7$ M$_\odot$ in the disk and $(7^{+12}_{-4})\times 10^6$ M$_\odot$ in the outflow. The resolved emission contributes 26\% and 5\% of the mass in the disk and outflow, respectively. 

Comparing the warm molecular gas to the cold molecular gas, the warm molecular outflow mass is a mere 3\% of the resolved CO-derived cold molecular gas mass \citep{Fluetsch2019}, or 2\% of the unresolved OH-derived cold molecular gas mass \citep{gonzalez2017}. The warm molecular gas is an even smaller fraction of the neutral gas phase derived from resolved Na I D emission: 0.6\% \citep{Rupke2013}. Similar to other warm molecular outflows detected in ULIRGs with \jwst\ \citep[e.g.,][]{dan2025, Alonso2024}, we observe minuscule warm molecular to cold molecular gas fractions. The detection of a warm molecular outflow should also not be taken as the norm, as there are several non-detections in ULIRGs observed with \jwst\ \citep[e.g.,][]{Buiten2025, Seebeck2026, Kader2026}.

From the outflow mass, we derive the mass outflow rate
\begin{equation}
\label{eqn:massoutflow}
    \dot M_\mathrm{out} = \frac{M_\mathrm{out} v_{\mathrm{out}}}{R_\mathrm{out}},
\end{equation}
momentum outflow rate, 
\begin{equation}
\label{eqn:momentumoutflow}
    \dot P_\mathrm{out} = \dot M_\mathrm{out} v_{\mathrm{out}},
\end{equation}
and outflow power, 
\begin{equation}
\label{eqn:energyoutflow}
    \dot E_\mathrm{out} = \frac{1}{2} \dot M_\mathrm{out} v_{\mathrm{out}}^2,
\end{equation}
where $v_{\mathrm{out}}$ is the mean outflow velocity taken from H$_2$ 0$-$0 S(1) and $R_{\mathrm{out}}$ is either the length of one spaxel (0.171 kpc) for the extended emission or half of the PSF FWHM ($\sim$0.32 kpc) for the nuclear extracted region. This results in total values of $\dot M_\mathrm{out} = 4^{+6}_{-2} \mathrm{\ M}_\odot \mathrm{yr}^{-1}$, $\dot P_\mathrm{out} = (4^{+6}_{-2}) \times 10^{33}$ dyne, and $\dot E_\mathrm{out} = (3^{+4}_{-2}) \times 10^{40}$ erg s$^{-1}$ with the resolved emission contributing 12\%, 14\%, and 18\% to the totals, respectively.

\subsection{Warm Ionized Gas}
For the nuclear warm ionized gas outflow, we use the broad component luminosities of the neon-emitting gas to estimate the warm ionized outflowing mass, following a method similar to that in \citet{See2024}. First, we derive the amount of warm ionized gas represented by each neon species (\neii, \neiii, and \nev); then we sum them together for a total warm ionized gas mass. There are five main assumptions: 
\begin{enumerate}
    \item\ \neii, \neiii, and \nev\ fully encapsulate the outflowing emission (reasonable given that there is no detectable emission from higher ionized states, such as \nevi);
    \item $n_\mathrm{H}=10\times n_\mathrm{He}$ \citep[i.e., assuming a solar elemental abundance ratio;][]{Asplund2009};
    \item The solar neon abundance [Ne/H] = $-$3.91 \citep{Nicholls2017};
    \item Line emissivities calculated from \pyneb\ \citep{Luridiana2015} (assuming constant $T = 10^4$ K, which is appropriate for AGN or starburst photoionized gas, and constant $n_e = 10^3$ cm$^{-3}$, which is reasonably close to the ULIRG broad component mean of 500 cm$^{-3}$ \citep{Arribas2014});
    \item Electron densities lie below their critical densities so that collisional de-excitation is unimportant.
\end{enumerate}
This results in
\begin{equation}
\label{eqn:massneii}
    M_{ionized}^{\mathrm{[NeII]}} = 6.086 \times 10^8 \frac{C\:L_{44}(\neii) }{\langle n_{e,3} \rangle 10^{\mathrm{[Ne/H]}}} M_{\odot},
\end{equation}
\begin{equation}
\label{eqn:massneiii}
    M_{ionized}^{\mathrm{[NeIII]}} = 2.96 \times 10^8 \frac{C\:L_{44}(\neiii) }{\langle n_{e,3} \rangle 10^{\mathrm{[Ne/H]}}} M_{\odot},
\end{equation}
\begin{equation}
\label{eqn:massnev}
    M_{ionized}^{\mathrm{[NeV]}} = 2.99 \times 10^7 \frac{C\:L_{44}(\nev) }{\langle n_{e,3} \rangle 10^{\mathrm{[Ne/H]}}} M_{\odot},
\end{equation}
where $C \equiv \langle n_{e} \rangle^2/\langle n_{e}^2 \rangle $ is the electron density clumping factor, which we assume is of order unity, $L_{44}$ is the neon luminosity normalized to 10$^{44}$ \ergs, and $\langle n_{e,3} \rangle$ is the average electron density normalized to 10$^3\:\mathrm{cm}^{-3}$. We derive masses of $M_{ionized}^{\mathrm{[NeII]}} = (46 \pm 5)\times 10^{5}$ M$_\odot$, $M_{ionized}^{\mathrm{[NeIII]}} = (4.4 \pm 0.1)\times 10^{5}$ M$_\odot$, and $M_{ionized}^{\mathrm{[NeV]}} = (0.026 \pm 0.003)\times 10^{5}$ M$_\odot$. Using Equations \ref{eqn:massoutflow}, \ref{eqn:momentumoutflow}, and \ref{eqn:energyoutflow}, we can also calculate the mass, momentum, and energy outflow rates for the warm ionized gas. These, as well as the outflow radius $R_{\mathrm{out}}$, outflow velocity $v_{50}$, and compiled results of the neutral and cold molecular gas phases from the literature are summarized in Table \ref{tab:outflow}.

Comparing our unresolved warm ionized outflow properties derived from the mid-IR neon lines to the ones derived from the resolved H$\alpha$ emission \citep{Rupke2013}, we detect a smaller outflow mass ($50 \times 10^{5}$ vs $759 \times 10^{5}$ M$_\odot$), but relatively similar $\dot M_{\mathrm{out}}$, $\dot P_{\mathrm{out}}$, and $\dot E_{\mathrm{out}}$. Our warm ionized outflow mass and energetics are remarkably relatively similar to our warm molecular gas mass and energetics.

Overall, we see that neither the warm molecular nor the warm ionized gas phases contribute much to the outflow mass and energetics of this system. The neutral gas phase contains the majority of the mass, while the cold molecular gas phase contributes most strongly to the energetics.

\begin{deluxetable*}{c c c c c c c c c c}
 \tablecaption{Outflow Parameters
 \label{tab:outflow}}
 \tabcolsep=1.5pt
 \tablehead{\colhead{Ref} & \colhead{Gas Phase} & \colhead{Tracer} & \colhead{Component} & \colhead{$M_{\mathrm{out}}$} & \colhead{$v_{\mathrm{out}}$} & \colhead{$R_\mathrm{out}$} & \colhead{$\dot M_{\mathrm{out}}$} & \colhead{$\dot P_{\mathrm{out}}$} & \colhead{$\dot E_{\mathrm{out}}$}\\
 \colhead{} & \colhead{} & \colhead{} & \colhead{} & \colhead{($10^5$ M$_{\odot}$)} & \colhead{(\kms)} & \colhead{(kpc)} & \colhead{(M$_\odot$ yr$^{-1}$)} & \colhead{($10^{33}$ dyn)} & \colhead{($10^{41}$ \ergs)}\\
 \colhead{(1)} & \colhead{(2)} & \colhead{(3)} & \colhead{(4)} & \colhead{(5)} & \colhead{(6)} & \colhead{(7)} & \colhead{(8)} & \colhead{(9)} & \colhead{(10)}\\}
 \startdata
    (1) & Warm Ionized & \neii\ 12.81 & Unresolved & $46^{+5}_{-5}$ & $-150^{+10}_{-10}$ & 0.64 & $2.6^{+0.5}_{-0.5}$ & $2.4^{+0.6}_{-0.6}$ & $0.18^{+0.06}_{-0.06}$\\
    (1) & & \neiii\ 15.56 & Unresolved & $4.4^{+0.1}_{-0.1}$ & $-260^{+10}_{-10}$ & 0.64 & $0.39^{+0.03}_{-0.03}$ & $0.65^{+0.07}_{-0.07}$ & $0.08^{+0.01}_{-0.01}$\\
    (1) & & \nev\ 14.32 & Unresolved & $0.026^{+0.003}_{-0.003}$ & $-520^{+10}_{-10}$ & 0.64 & $0.0049^{+0.0006}_{-0.0006}$ & $0.016^{+0.003}_{-0.003}$ & $0.0042^{+0.0008}_{-0.0008}$\\
    (1) & & & Total & $50^{+5}_{-5}$ & & & $3.0^{+0.5}_{-0.5}$ & $3.1^{+0.6}_{-0.6}$ & $0.27^{+0.06}_{-0.06}$ \\
    (2) & & H$\alpha$ & Resolved & 759 & $-$133 & 2.0 & 1.45 & 1.277 & 0.85\\
    (1) & Warm Molecular & H$_2$ & Resolved & $3.9^{+0.5}_{-0.5}$ & $-$150 & 1.64 & $0.43^{+0.06}_{-0.06}$ & $0.51^{+0.07}_{-0.07}$ & $0.051^{+0.007}_{-0.007}$ \\
    (1) &  & H$_2$ & Unresolved & $68^{+121}_{-44}$ & $-$150 & 0.32 & $3^{+6}_{-2}$ & $3^{+5}_{-2}$ & $0.2^{+0.4}_{-0.2}$ \\
    (1) & & & Total & $72^{+121}_{-44}$ & & & $4^{+6}_{-2}$ & $4^{+6}_{-2}$ & $0.3^{+0.4}_{-0.2}$ \\
    (2) & Neutral & Na I D & Resolved & 12589 & $-$218  & 2.0 & 65 & 114 & 26 \\
    (3) & & HI & Resolved & & $-148^{+62}_{-63}$ & 1.36 & 140 & & 89 \\
    (4) & Cold Molecular & CO & Resolved & 2344 & 450 & 1.1 & 100 & 284 & 64 \\
    (5) & & OH & Unresolved & $3200^{+2800}_{-1000}$ & 250/420  & 0.480/0.400 & $250^{+150}_{-81}$ & $570^{+130}_{-200}$ & $110^{+40}_{-40}$\\
 \enddata
 \tablecomments{Meaning of the columns: (1) Reference, (2) Gas phase of the outflow, (3) emission line tracer used to derive the mass, (4) component of the outflow, either resolved or unresolved, (5) mass of the outflowing gas phase, (6) typical $v_{50}$ velocity of the outflow,  (7) radius of the outflow (PSF FWHM in the cases of unresolved components), (8) mass outflow rate, (9) momentum outflow rate, and (10) energy outflow rate. Note: For this work, we utilize half of the PSF FWHM (0.375\arcsec) for the unresolved $R_\mathrm{out}$ and the average of the H$_2$ 0$-$0 S(1) line for the warm molecular $v_{\mathrm{out}}$. For the OH fits, two outflow components were used, so both are reported for $v_{\mathrm{out}}$ and R$_{\mathrm{out}}$ with $v_{\mathrm{out}}$ being the average velocity of OH119, OH79, OH84, and OH65.}
  \tablerefs{(1) This paper, (2) \citet{Rupke2013}, (3) \citet{su2023}, (4) \citet{Fluetsch2019}, (5) \citet{gonzalez2017}}
\end{deluxetable*}

\subsection{PAH}
Ratios of the clipped fluxes of various PAH complexes can help determine the size and ionization of the PAH species. To better visualize how the PAH ratios change across the galaxy, in Figure \ref{fig:pah_plasma} we plot selected PAH ratios against each other with color maps representing distance from the center of F10565+2448 and grids from \citet{draine2021} that show varying size distributions, ionization levels, and starlight intensity values. We also include the models from \citet{hensley2023} that do not have varying size distributions nor ionization levels but reach $\log(U)$ values as low as $-3$. We find that $\log(U)$ negligibly affects the ratios below $-$1, so the \citet{hensley2023} models are displayed as a single gray circle in each plot.
To reduce scatter, we apply a signal-to-noise ratio (SNR) cut of 10\% of the maximum SNR for each grid. Our PAH 6.2/7.7 ratio is lower than the grid values, which could be explained by larger PAH grain sizes than the Draine grids employed. Outside of that, we observe tentative trends of ionization and grain size initially decreasing with increasing radius up to 1 kpc, beyond which they both increase out to 3 kpc. 

\begin{figure}
    \centering
    \includegraphics[width=\columnwidth]{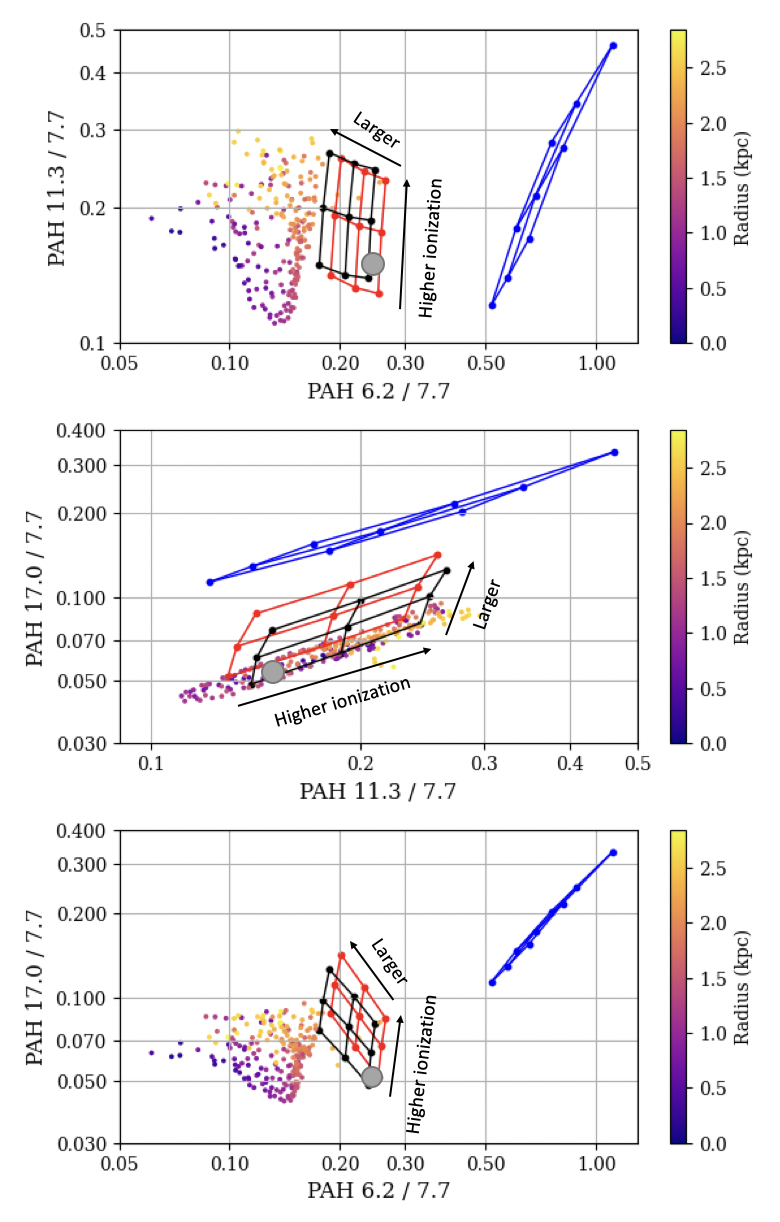}
    \caption{Selected PAH band ratios plotted against each other with colormaps representing distance from the center of F10565+2448. Upper Left: PAH 11.3/7.7 vs PAH 6.2/7.7. Upper Right: PAH 17.0/7.7 vs PAH 11.3/7.7. Lower Left: PAH 17.0/7.7 vs PAH 6.2/7.7. Grids from \citet{draine2021} with various starlight intensity ($U$) values are overplotted. Black represents $\log(U)=0$, or Milky Way level. Red represents $\log(U)=3$, or 1000 times Milky Way. Blue represents $\log(U)=6$, or $10^6$ times Milky Way. The ionization levels are based off the standard ionization, $f_{\mathrm{ion}}$, from \citet{draine2007}, where the higher and lower ionization levels correspond to factor-of-two shifts in the PAH radius for which $f_{\mathrm{ion}} = 0.5$. 
    The PAH size distributions are described by Eq. 15 in \citet{draine2021}. Gray circles represent PAH ratios of the models from \citet{hensley2023} with $\log(U)=-1$ through $-3$ (the PAH band ratios for $\log(U) < -1$ are indistinguishable from those at $\log(U) = -1$), obtained with the same clipping method as the grids. Our PAH 6.2/7.7 ratio is lower than the grid values, which may indicate that the PAH grain sizes are larger than the largest population applied in the Draine grids.
    \label{fig:pah_plasma}}
\end{figure}

To get an overall sense of the PAH in relation to other galaxies, in Figure \ref{fig:rigopoulou} we plot ratios of the nuclear PAH fluxes from \cafe\ (see Table \ref{tab:PAH}) against PAH model grids from \citet{Rigopoulou2021}. For context, a sample of Seyfert galaxies from \citet{Garcia2022} as well as a sample of Seyfert and star-forming galaxies from \citet{garcia2022b} and \citet{garcia2024} are also plotted. F10565+2448 has a larger PAH 6.2/7.7 ratio than the \citet{Garcia2022} Seyfert sample, which could be indicative of smaller grain sizes. 

\begin{figure}
    \centering
    \includegraphics[width=\columnwidth]{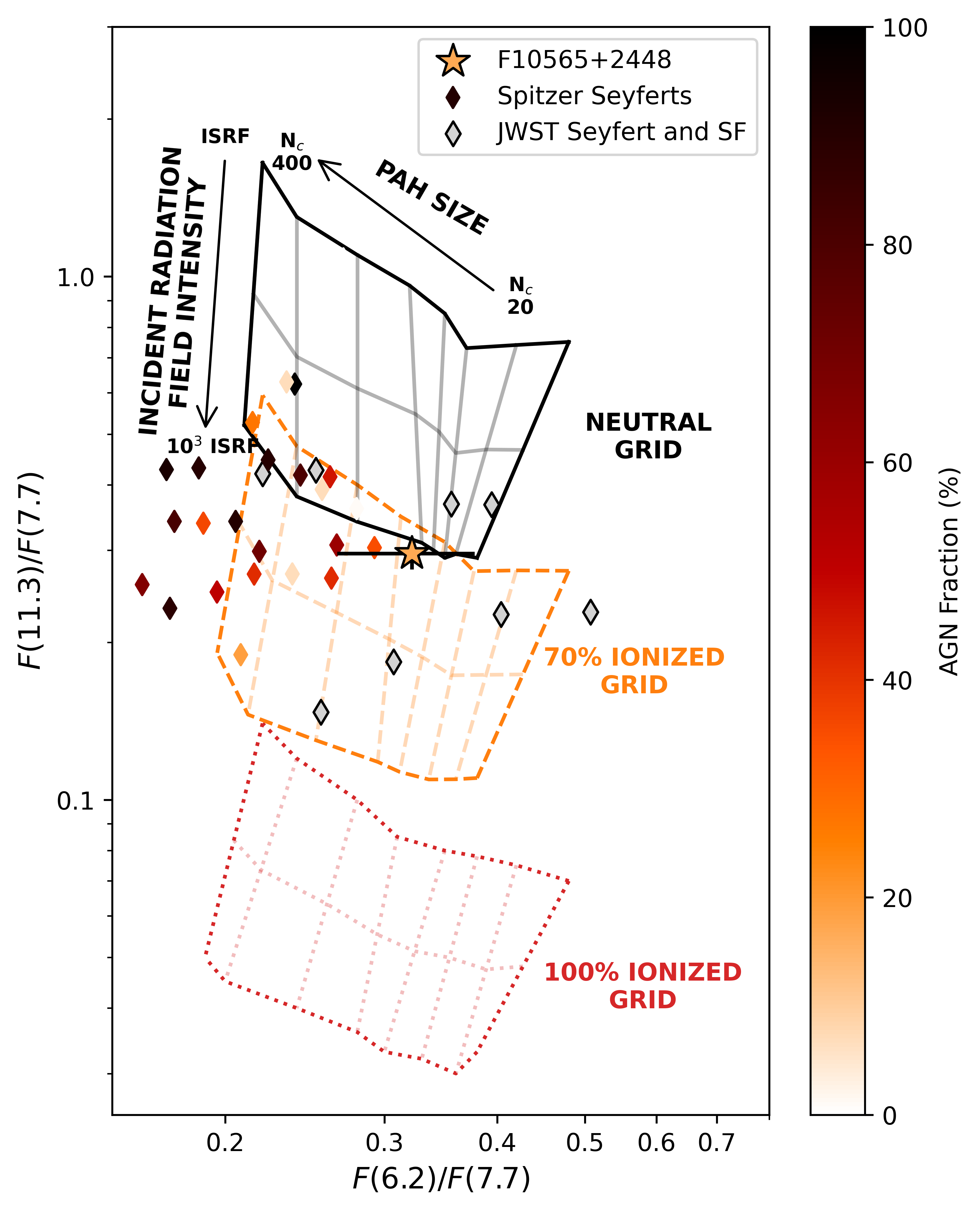}
    \caption{PAH diagnostic plot of 11.3/7.7 vs 6.2/7.7. The star indicates F10565+2448 and uses the extinction-corrected nuclear PAH fluxes from Table \ref{tab:PAH}. Colored diamonds display \spitzer\ observations of Seyfert galaxies from \citet{Garcia2022}, while gray diamonds show \jwst\ observations of Seyfert and star-forming galaxies from \citet{garcia2022b} and \citet{garcia2024}. Colors of the non-gray points indicate the mid-IR AGN fraction. The PAH model grids are from \citet{Rigopoulou2021} and show different PAH grain sizes (number of carbon atoms; $20 < N_c < 400$), incident starlight intensity levels in terms of the Interstellar Radiation Field (1 $\times$ ISRF to $10^3$ $\times$ ISRF), and ionization levels (0\% in black, 70\% in orange, and 100\% in red).
    F10565+2448 has a higher PAH 6.2/7.7 ratio compared to the Seyfert galaxies. This could indicate relatively similar ionization and/or incident radiation field intensity with smaller PAH grain sizes.   
    \label{fig:rigopoulou}}
\end{figure}

PAH is often used to estimate the star formation rate (SFR) in galaxies. Here we derive the SFR from our nuclear PAH fluxes (see Table \ref{tab:PAH}) using two methods: Equation 11 from \citet{Shipley2016} and Equation 2 from \citet{Diamond2012}. \citet{Shipley2016} gives options for deriving the SFR from PAH 6.2, 7.7, 11.3, or a combination of all three. Using their methods, we obtain $\log_{10}$SFR$_{6.2} = 1.5$, $\log_{10}$SFR$_{7.7} = 1.3$, $\log_{10}$SFR$_{11.3} = 1.4$, and $\log_{10}$SFR$_{6.2+7.7+11.3} = 1.4$. \citet{Diamond2012} has options for PAH 11.3 as well as \neii. For those we derive $\log_{10}$SFR$_{11.3} = 1.6$ and $\log_{10}$SFR$_{\neii} = 1.9$. Of the PAH-based SFR methods, we derive a mean SFR of $27\pm 10$ M$_\odot$ yr$^{-1}$, where the error represents the standard deviation across the various PAH-based methods. The literature value of $\log_{10}$SFR $= 2.176$ \citep[derived from \iras\ 8-1000$\mu$m;][]{Rupke2013} is about than an order of magnitude higher. Part of this discrepancy may come from the difference in apertures between \iras\ and \jwst; however, using existing MIRI imaging we estimate that a larger aperture would increase the PAH fluxes by a factor of $\sim 1.2$, which fails to fully explain the gap. The discrepancy suggests that SFR metrics based on PAH may need to be recalibrated in objects with powerful AGN (e.g., ULIRGs). 

\subsection{Shocks}
\label{sec:shock}
Enhanced near-IR \feii-to-hydrogen recombination line ratios have historically been used to identify shocked regions in galaxies, typically associated with supernova remnants or radio jets \citep[e.g.,][]{Alonso1997, rodriguez2005, colina2015}. More recently, this diagnostic has been extended to the mid-IR, with \citet{Alonso2025} compiling MIRI/MRS \feii\ 5.34 µm/Pf$\alpha$ ratios for Seyfert nuclei, low-luminosity AGN, and circumnuclear shocked regions, and showing that shocks preferentially occupy the regime $\log_{10}$(\feii\ 5.34 µm/Pf$\alpha$) $\gtrsim 0.7$. To assess whether shocks contribute to the excitation in our target, we constructed a map of $\log_{10}$(\feii\ 5.34 µm/Pf$\alpha$), shown in Figure \ref{fig:shock}. We highlight with white circles two localized enhancements with spaxels that reach values consistent with shock excitation, or at least Seyfert-like values ($>0.4$).

The features lie along the southern and north-western edges of the H$_2$ outflow, which are regions of the outflow where the kinematics show reduced $v_{50}$ and elevated $w_{80}$. The \feii\ line widths are also broader in those areas ($w_{80} \sim 100 - 200$ km s$^{-1}$) vs to the north or east ($w_{80} \sim 20 - 80$ km s$^{-1}$), indicative of more turbulent gas that could be associated with the shock front of the outflow.

\begin{figure}
    \centering
    \includegraphics[width=\columnwidth]{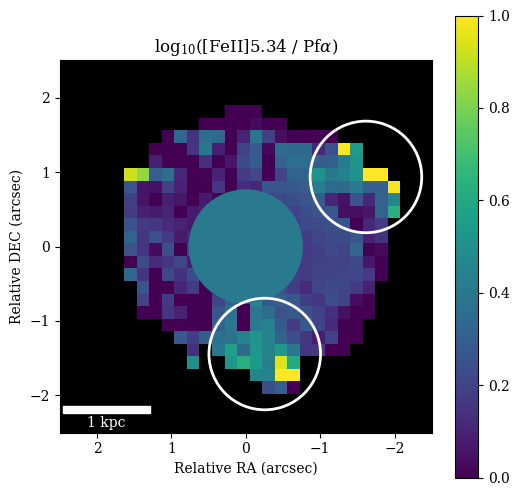}
    \caption{Map of $\log_{10}$(\feii\ 5.34$\mu$m / Pfund $\alpha$), with shocked regions $\gtrsim 0.7$ and Seyfert-like regions $\gtrsim 0.4$. We identify two regions potentially impacted by shocks: regions south and north-west of the nucleus that align with the hotter, less dense edges of the H$_2$ outflow.
    \label{fig:shock}}
\end{figure}

\subsection{Origin of the Outflow}

Initially, due to the fairly low AGN fraction (17\%), we might assume that the outflow could be driven by the nuclear starburst. The kinetic power that supernovae release is $\dot E_{\mathrm{out}} = 7.0 \times 10^{41}$ (SFR/M$_\odot$ yr$^{-1}$) erg s$^{-1}$ $\sim 10^{44} $ erg s$^{-1}$ \citep{veilleux2005}. Estimating the total kinetic power of the outflow as $\sim 10^{43}$ erg s$^{-1}$, the fraction of supernovae kinetic power converted to outflow kinetic power needs to be at least 10\%. This fraction is just within the range of hydrodynamical simulations, which report 10$-$100\% \citep[e.g.][]{strickland2000}, or even $>$30\% for the extreme case of M82 \citep{strickland2009}. Instead of the kinetic power, we can also compare the mass outflow rate. \citet{veilleux2005} estimates the mass outflow rate driven by SNe as $\dot M = 0.26\left(\frac{\mathrm{SFR}}{\mathrm{M_{\odot}\ yr}^{-1}}\right)\mathrm{\ M_{\odot}\ yr}^{-1}$. For an SFR of 150 $\mathrm{\ M_{\odot}\ yr}^{-1}$, this gives us $\dot M = 39 \mathrm{\ M_{\odot}\ yr}^{-1}$, which is about an order of magnitude smaller than the total mass outflow rate of this system. This would imply a mass-loading factor of $\sim 10$, which is higher than normal for starburst-driven outflows.
Additionally, we detect \nev\ 14.32 $\mu$m emission, which is rarely detected in pure starburst systems due to its high ionization potential that O-type stars struggle to create. In fact, our measured \nev/\neii ratio of 0.036 is similar to those of other AGN-starburst composite systems \citep[e.g.;][]{Satyapal2008, Hermosa2025}. Energetically, the outflow could be driven completely by a starburst, but the implied large mass-loading factor, as well as the detection of \nev\, motivate us to consider several other AGN-related sources of origin. 

AGN-driven outflows are often divided into energy-driven and momentum-driven regimes. In the energy-driven case, the wind cools inefficiently and expands as a hot bubble, transferring up to $\sim$5\% of the AGN bolometric luminosity to the outflow \citep{zubovas2014}. The outflow’s kinetic power corresponds to $\sim$1.5\% of the AGN radiation power, indicating that an energy-driven AGN wind could supply the required mechanical energy. 
In contrast, a momentum-conserving flow driven by radiation pressure would require $\dot P_{\rm out} \simeq L_{\rm Edd}/c$. Even assuming Eddington-limited accretion, $L_{\rm Edd}/c \sim 10^{34}$ dyn, which is two orders of magnitude below the observed $\dot P_{\rm out} \sim 10^{36}$ dyn. The radiation pressure of the AGN is insufficient, so there must be other sources of momentum driving the outflow.  

AGN-powered jets could also contribute to the observed outflow momentum. Hydrodynamical simulations show that jets can accelerate clouds up to high velocities with adequate jet power: $\frac{L_{\mathrm{jet}}}{L_{\mathrm{Edd}}} > 10^{-4}$ \citep{Wagner2012}. 5 GHz VLBI images show slight extended emission to the southwest in the same direction of the outflow, which could be indicative of a jet in this system. 
\citet{su2023} estimate the jet power based on SFR-corrected NRAO VLA Sky Survey (NVSS) 1.4 GHz flux density measurements: $L_{\mathrm{jet}} \sim 5.3 \times 10^{42}$ erg s$^{-1}$. Although $L_{\mathrm{jet}}$ is about a factor of 2 smaller than the total kinetic power of the outflow, uncertainties in both the jet power measurement and the outflow kinetic power could reconcile this difference. For a jet with $L_{\mathrm{jet}} \sim 10^{43}$ erg s$^{-1}$, \citet{Wagner2012} predicts the maximum radial velocity of accelerated clouds to be $150-300$ \kms, depending on hot-phase density and cloud volume filling factor. Most velocity measurements of the multiphase outflow in F10565+2448 lie within this range, so a jet may power at least some of the outflow. 

Overall, F10565+2448 proves itself to be an excellent example of an AGN-starburst composite system. Both an AGN and nuclear starburst are likely required to explain the observed outflow energetics. 

\section{Conclusion}
\label{sec:conclusion}

We analyzed the MIRI/MRS IFS data of the ULIRG F10565+2448 with a previously-known powerful cold molecular outflow. These new \jwst\ data provide the first spatially resolved mid-infrared view of the outflow and host galaxy. We used \qtdfit\ to separate the quasar light from the host galaxy and reveal the extended warm molecular and warm ionized gas, as well as \cafe\ to analyze the nuclear emission. These data complete the picture of this multi-phase outflow, helping us to better understand the nature of feedback in AGN-starburst composite systems. The main results of our analysis include: 

\begin{enumerate}    
    \item F10565+2448 shows evidence of a nuclear, possibly decelerating warm ionized gas outflow with median observed velocity $-150 < v_{50} < -540$ \kms\ and Neon-derived outflowing mass of $5\times 10^6$ M$_\odot$. From this, we derive a mass outflow rate of 3 M$_{\odot}$ yr$^{-1}$, which is within a factor of two of the rate derived from H$\alpha$ \citep[1.45 M$_{\odot}$ yr$^{-1}$;][]{Rupke2013}. 

    \item F10565+2448 also shows evidence of a nuclear and extranuclear warm molecular gas outflow with median observed velocities of up to $-220$ and $-280$ \kms\ respectively. We derive a total warm molecular outflowing mass of $7\times 10^6$ M$_\odot$ and total mass outflow rate of 4 M$_{\odot}$ yr$^{-1}$. These are the first resolved measurements of warm molecular gas in this system. The cold molecular and neutral gas phases remain dominant with 1$-$2 orders of magnitude higher masses and mass outflow rates.
    
    \item The extranuclear warm molecular gas reveals several cold, dense clumps to the north of the nucleus. The one to the north-east aligns with a known star cluster, so we may be detecting the warm molecular gas surrounding this cluster. The warm molecular outflow is located away from these higher density areas, which could be evidence of the outflow preferentially moving in the direction of least resistance (vs being completely isotropic). 

    \item We utilize the \feii\ 5.34/Pf$\alpha$ shock diagnostic to identify regions around the edges of the outflow which may be impacted by shocks. 

    \item PAH ratio analysis in F10565+2448 reveals trends of ionization and grain size that first decrease with radius up to 1 kpc before increasing up to 3 kpc.

    \item We discuss the energy and momentum budget of the outflow and determine that both the starburst and AGN likely contribute. We discuss the plausibility of an energy-driven and/or jet-powered outflow, while ruling out a purely radiation-pressure-driven outflow. 

\end{enumerate}

\begin{acknowledgments}

K.Y.D., J.S., and S.V.\ acknowledge partial financial support by NASA for this research through STScI grants No.\  JWST-ERS-01335, JWST-GO-01865, JWST-GO-02547, JWST GO-03869, and JWST GO-05627. M.P.S. acknowledges support under grants RYC2021-033094-I, CNS2023-145506, and PID2023-146667NB-I00 funded by MCIN/AEI/10.13039/501100011033 and the European Union NextGenerationEU/PRTR. I.G.B. is supported by the Programa Atracci\'on de Talento Investigador ``C\'esar Nombela'' via grant 2023-T1/TEC-29030 funded by the Community of Madrid. We thank the anonymous referee for their helpful comments, which greatly improved the draft during the revision process.

\end{acknowledgments}

\begin{contribution}
KYD performed the analysis and wrote the manuscript. 
JS assisted with the analysis and edited the manuscript.
SV led the design and implementation of this program, obtained the funding, supervised the analysis, and edited the manuscript. 
EGA, IGB, DL, MM, MPS, and FT provided comments on the original proposal and manuscript. DR and WL provided comments on the original proposal and are developers of \qtdfit. 
ES provided comments on the original proposal. 


\end{contribution}

\facilities{\jwst\ (MIRI/MRS)}

\software{
\qtdfit\ (\citealt{q3d2014}, \citealt{q3d2021}), 
\cafe\ (\citealt{Diaz2025}), 
\astropy\ (\citealt{astropy:2013}, \citealt{astropy:2018}, \citealt{astropy:2022}), 
\matplotlib\ \citep{matplotlib2007},
\numpy\ \citep{numpy2020},
\scipy\ \citep{SciPy2020}, \pyneb\ \citep{Luridiana2015}.
We utilized GitHub CoPilot to assist in writing parts of the code for the analysis, accessed through the Visual Studio Code extension GitHub Copilot Chat (ver. 0.35.0). We utilized ChatGPT-5 to reword a few sentences in the manuscript for clarity. 
}


\bibliography{sample701}{}

@ARTICLE{Rupke2013,
       author = {{Rupke}, David S.~N. and {Veilleux}, Sylvain},
        title = "{The Multiphase Structure and Power Sources of Galactic Winds in Major Mergers}",
      journal = {\apj},
         year = 2013,
        month = may,
       volume = {768},
       number = {1},
          eid = {75},
        pages = {75},
          doi = {10.1088/0004-637X/768/1/75},
archivePrefix = {arXiv},
       eprint = {1303.6866},
 primaryClass = {astro-ph.CO},
       adsurl = {https://ui.adsabs.harvard.edu/abs/2013ApJ...768...75R}
}

@ARTICLE{su2023,
       author = {{Su}, Renzhi and {Mahony}, Elizabeth K. and {Gu}, Minfeng and {Sadler}, Elaine M. and {Curran}, S.~J. and {Allison}, James R. and {Yoon}, Hyein and {Aditya}, J.~N.~H.~S. and {Chandola}, Yogesh and {Chen}, Yongjun and {Moss}, Vanessa A. and {Wu}, Zhongzu and {Shao}, Xi and {Liu}, Xiang and {Glowacki}, Marcin and {Whiting}, Matthew T. and {Weng}, Simon},
        title = "{Does a radio jet drive the massive multiphase outflow in the ultra-luminous infrared galaxy IRAS 10565 + 2448?}",
      journal = {\mnras},
         year = 2023,
        month = apr,
       volume = {520},
       number = {4},
        pages = {5712-5723},
          doi = {10.1093/mnras/stad370},
archivePrefix = {arXiv},
       eprint = {2302.00943},
 primaryClass = {astro-ph.GA},
       adsurl = {https://ui.adsabs.harvard.edu/abs/2023MNRAS.520.5712S}
}

@ARTICLE{Fluetsch2019,
       author = {{Fluetsch}, A. and {Maiolino}, R. and {Carniani}, S. and {Marconi}, A. and {Cicone}, C. and {Bourne}, M.~A. and {Costa}, T. and {Fabian}, A.~C. and {Ishibashi}, W. and {Venturi}, G.},
        title = "{Cold molecular outflows in the local Universe and their feedback effect on galaxies}",
      journal = {\mnras},
         year = 2019,
        month = mar,
       volume = {483},
       number = {4},
        pages = {4586-4614},
          doi = {10.1093/mnras/sty3449},
archivePrefix = {arXiv},
       eprint = {1805.05352},
 primaryClass = {astro-ph.GA},
       adsurl = {https://ui.adsabs.harvard.edu/abs/2019MNRAS.483.4586F}
}

@ARTICLE{downes1998,
       author = {{Downes}, D. and {Solomon}, P.~M.},
        title = "{Rotating Nuclear Rings and Extreme Starbursts in Ultraluminous Galaxies}",
      journal = {\apj},
         year = 1998,
        month = nov,
       volume = {507},
       number = {2},
        pages = {615-654},
          doi = {10.1086/306339},
archivePrefix = {arXiv},
       eprint = {astro-ph/9806377},
 primaryClass = {astro-ph},
       adsurl = {https://ui.adsabs.harvard.edu/abs/1998ApJ...507..615D}
}

@ARTICLE{Spinoglio1992,
       author = {{Spinoglio}, Luigi and {Malkan}, Matthew A.},
        title = "{Infrared Line Diagnostics of Active Galactic Nuclei}",
      journal = {\apj},
         year = 1992,
        month = nov,
       volume = {399},
        pages = {504},
          doi = {10.1086/171943},
       adsurl = {https://ui.adsabs.harvard.edu/abs/1992ApJ...399..504S}
}

@ARTICLE{lutz2020,
       author = {{Lutz}, D. and {Sturm}, E. and {Janssen}, A. and {Veilleux}, S. and {Aalto}, S. and {Cicone}, C. and {Contursi}, A. and {Davies}, R.~I. and {Feruglio}, C. and {Fischer}, J. and et al.},
        title = "{Molecular outflows in local galaxies: Method comparison and a role of intermittent AGN driving}",
      journal = {\aap},
         year = 2020,
        month = jan,
       volume = {633},
          eid = {A134},
        pages = {A134},
          doi = {10.1051/0004-6361/201936803},
archivePrefix = {arXiv},
       eprint = {1911.05608},
 primaryClass = {astro-ph.GA},
       adsurl = {https://ui.adsabs.harvard.edu/abs/2020A&A...633A.134L}
}

@ARTICLE{jones2023,
       author = {{Jones}, O.~C. and {{\'A}lvarez-M{\'a}rquez}, J. and {Sloan}, G.~C. and {Kavanagh}, P.~J. and {Argyriou}, I. and {Law}, D.~R. and {Labiano}, A. and {Patapis}, P. and {Mueller}, Michael and {Larson}, Kirsten L. and {Bright}, Stacey N. and {Klaassen}, P.~D. and {Fox}, O.~D. and {Gasman}, Danny and {Geers}, V.~C. and {Glauser}, Adrian M. and {Guillard}, Pierre and {Nayak}, Omnarayani and {Noriega-Crespo}, A. and {Ressler}, Michael E. and {Sargent}, B. and {Temim}, T. and {Vandenbussche}, B. and {Garc{\'\i}a Mar{\'\i}n}, Macarena},
        title = "{Observations of the planetary nebula SMP LMC 058 with the JWST MIRI medium resolution spectrometer}",
      journal = {\mnras},
         year = 2023,
        month = aug,
       volume = {523},
       number = {2},
        pages = {2519-2529},
          doi = {10.1093/mnras/stad1609},
archivePrefix = {arXiv},
       eprint = {2301.13233},
 primaryClass = {astro-ph.IM},
       adsurl = {https://ui.adsabs.harvard.edu/abs/2023MNRAS.523.2519J}
}

@ARTICLE{veilleux2013,
       author = {{Veilleux}, S. and {Mel{\'e}ndez}, M. and {Sturm}, E. and {Gracia-Carpio}, J. and {Fischer}, J. and {Gonz{\'a}lez-Alfonso}, E. and {Contursi}, A. and {Lutz}, D. and {Poglitsch}, A. and {Davies}, R. and {Genzel}, R. and {Tacconi}, L. and {de Jong}, J.~A. and {Sternberg}, A. and {Netzer}, H. and {Hailey-Dunsheath}, S. and {Verma}, A. and {Rupke}, D.~S.~N. and {Maiolino}, R. and {Teng}, S.~H. and {Polisensky}, E.},
        title = "{Fast Molecular Outflows in Luminous Galaxy Mergers: Evidence for Quasar Feedback from Herschel}",
      journal = {\apj},
         year = 2013,
        month = oct,
       volume = {776},
       number = {1},
          eid = {27},
        pages = {27},
          doi = {10.1088/0004-637X/776/1/27},
archivePrefix = {arXiv},
       eprint = {1308.3139},
 primaryClass = {astro-ph.CO},
       adsurl = {https://ui.adsabs.harvard.edu/abs/2013ApJ...776...27V}
}

@ARTICLE{Genzel1998,
       author = {{Genzel}, R. and {Lutz}, D. and {Sturm}, E. and {Egami}, E. and {Kunze}, D. and {Moorwood}, A.~F.~M. and {Rigopoulou}, D. and {Spoon}, H.~W.~W. and {Sternberg}, A. and {Tacconi-Garman}, L.~E. and {Tacconi}, L. and {Thatte}, N.},
        title = "{What Powers Ultraluminous IRAS Galaxies?}",
      journal = {\apj},
         year = 1998,
        month = may,
       volume = {498},
       number = {2},
        pages = {579-605},
          doi = {10.1086/305576},
archivePrefix = {arXiv},
       eprint = {astro-ph/9711255},
 primaryClass = {astro-ph},
       adsurl = {https://ui.adsabs.harvard.edu/abs/1998ApJ...498..579G}
}

@ARTICLE{Satyapal2008,
       author = {{Satyapal}, S. and {Vega}, D. and {Dudik}, R.~P. and {Abel}, N.~P. and {Heckman}, T.},
        title = "{Spitzer Uncovers Active Galactic Nuclei Missed by Optical Surveys in Seven Late-Type Galaxies}",
      journal = {\apj},
         year = 2008,
        month = apr,
       volume = {677},
       number = {2},
        pages = {926-942},
          doi = {10.1086/529014},
archivePrefix = {arXiv},
       eprint = {0801.2759},
 primaryClass = {astro-ph},
       adsurl = {https://ui.adsabs.harvard.edu/abs/2008ApJ...677..926S}
}

@ARTICLE{Hermosa2025,
       author = {{Hermosa Mu{\~n}oz}, L. and {Alonso-Herrero}, A. and {Labiano}, A. and {Guillard}, P. and {Pantoni}, L. and {Buiten}, V. and {Dicken}, D. and {Baes}, M. and {B{\"o}ker}, T. and {Colina}, L. and {Donnan}, F. and {Garc{\'\i}a-Bernete}, I. and {{\"O}stlin}, G. and {van der Werf}, P. and {Ward}, M.~J. and {Brandl}, B.~R. and {Walter}, F. and {Wright}, G. and {G{\"u}del}, M. and {Henning}, Th. and {Lagage}, P.-O. and {Ray}, T.},
        title = "{MICONIC: Dual active galactic nuclei, star formation, and ionised gas outflows in NGC 6240 seen with MIRI/JWST}",
      journal = {\aap},
         year = 2025,
        month = jan,
       volume = {693},
          eid = {A321},
        pages = {A321},
          doi = {10.1051/0004-6361/202452437},
archivePrefix = {arXiv},
       eprint = {2412.14707},
 primaryClass = {astro-ph.GA},
       adsurl = {https://ui.adsabs.harvard.edu/abs/2025A&A...693A.321H}
}

@ARTICLE{garcia2024,
       author = {{Garc{\'\i}a-Bernete}, I. and {Rigopoulou}, D. and {Donnan}, F.~R. and {Alonso-Herrero}, A. and {Pereira-Santaella}, M. and {Shimizu}, T. and {Davies}, R. and {Roche}, P.~F. and {Garc{\'\i}a-Burillo}, S. and {Labiano}, A. and {Hermosa Mu{\~n}oz}, L. and {Zhang}, L. and {Audibert}, A. and {Bellocchi}, E. and {Bunker}, A. and {Combes}, F. and {Delaney}, D. and {Esparza-Arredondo}, D. and {Gandhi}, P. and {Gonz{\'a}lez-Mart{\'\i}n}, O. and {H{\"o}nig}, S.~F. and {Imanishi}, M. and {Hicks}, E.~K.~S. and {Fuller}, L. and {Leist}, M. and {Levenson}, N.~A. and {Lopez-Rodriguez}, E. and {Packham}, C. and {Ramos Almeida}, C. and {Ricci}, C. and {Stalevski}, M. and {Villar Mart{\'\i}n}, M. and {Ward}, M.~J.},
        title = "{The Galaxy Activity, Torus, and Outflow Survey (GATOS): V. Unveiling PAH survival and resilience in the circumnuclear regions of AGNs with JWST}",
      journal = {\aap},
         year = 2024,
        month = nov,
       volume = {691},
          eid = {A162},
        pages = {A162},
          doi = {10.1051/0004-6361/202450086},
archivePrefix = {arXiv},
       eprint = {2409.05686},
 primaryClass = {astro-ph.GA},
       adsurl = {https://ui.adsabs.harvard.edu/abs/2024A&A...691A.162G}
}

@ARTICLE{garcia2022b,
       author = {{Garc{\'\i}a-Bernete}, I. and {Rigopoulou}, D. and {Alonso-Herrero}, A. and {Donnan}, F.~R. and {Roche}, P.~F. and {Pereira-Santaella}, M. and {Labiano}, A. and {Peralta de Arriba}, L. and {Izumi}, T. and {Ramos Almeida}, C. and {Shimizu}, T. and {H{\"o}nig}, S. and {Garc{\'\i}a-Burillo}, S. and {Rosario}, D.~J. and {Ward}, M.~J. and {Bellocchi}, E. and {Hicks}, E.~K.~S. and {Fuller}, L. and {Packham}, C.},
        title = "{A high angular resolution view of the PAH emission in Seyfert galaxies using JWST/MRS data}",
      journal = {\aap},
         year = 2022,
        month = oct,
       volume = {666},
          eid = {L5},
        pages = {L5},
          doi = {10.1051/0004-6361/202244806},
archivePrefix = {arXiv},
       eprint = {2208.11620},
 primaryClass = {astro-ph.GA},
       adsurl = {https://ui.adsabs.harvard.edu/abs/2022A&A...666L...5G}
}

@ARTICLE{sanders2003,
       author = {{Sanders}, D.~B. and {Mazzarella}, J.~M. and {Kim}, D. -C. and {Surace}, J.~A. and {Soifer}, B.~T.},
        title = "{The IRAS Revised Bright Galaxy Sample}",
      journal = {\aj},
         year = 2003,
        month = oct,
       volume = {126},
       number = {4},
        pages = {1607-1664},
          doi = {10.1086/376841},
archivePrefix = {arXiv},
       eprint = {astro-ph/0306263},
 primaryClass = {astro-ph},
       adsurl = {https://ui.adsabs.harvard.edu/abs/2003AJ....126.1607S}
}

@ARTICLE{veilleux2009_agn_fraction,
       author = {{Veilleux}, S. and {Rupke}, D.~S.~N. and {Kim}, D. -C. and {Genzel}, R. and {Sturm}, E. and {Lutz}, D. and {Contursi}, A. and {Schweitzer}, M. and {Tacconi}, L.~J. and {Netzer}, H. and {Sternberg}, A. and {Mihos}, J.~C. and {Baker}, A.~J. and {Mazzarella}, J.~M. and {Lord}, S. and {Sanders}, D.~B. and {Stockton}, A. and {Joseph}, R.~D. and {Barnes}, J.~E.},
        title = "{Spitzer Quasar and Ulirg Evolution Study (QUEST). IV. Comparison of 1 Jy Ultraluminous Infrared Galaxies with Palomar-Green Quasars}",
      journal = {\apjs},
         year = 2009,
        month = jun,
       volume = {182},
       number = {2},
        pages = {628-666},
          doi = {10.1088/0067-0049/182/2/628},
archivePrefix = {arXiv},
       eprint = {0905.1577},
 primaryClass = {astro-ph.CO},
       adsurl = {https://ui.adsabs.harvard.edu/abs/2009ApJS..182..628V}
}

@ARTICLE{mirabel1988,
       author = {{Mirabel}, I.~F. and {Sanders}, D.~B.},
        title = "{21 Centimeter Survey of Luminous Infrared Galaxies}",
      journal = {\apj},
         year = 1988,
        month = dec,
       volume = {335},
        pages = {104},
          doi = {10.1086/166909},
       adsurl = {https://ui.adsabs.harvard.edu/abs/1988ApJ...335..104M}
}

@ARTICLE{shih2010,
       author = {{Shih}, Hsin-Yi and {Rupke}, David S.~N.},
        title = "{The Complex Structure of the Multi-phase Galactic Wind in a Starburst Merger}",
      journal = {\apj},
         year = 2010,
        month = dec,
       volume = {724},
       number = {2},
        pages = {1430-1440},
          doi = {10.1088/0004-637X/724/2/1430},
archivePrefix = {arXiv},
       eprint = {1009.6020},
 primaryClass = {astro-ph.CO},
       adsurl = {https://ui.adsabs.harvard.edu/abs/2010ApJ...724.1430S}
}

@ARTICLE{glenn2001,
       author = {{Glenn}, Jason and {Hunter}, Todd R.},
        title = "{A Comparison of Tracers of Cool Gas in Galaxies and the $^{12}$CO/$^{13}$CO Luminosity Ratio in Luminous Infrared Galaxies}",
      journal = {\apjs},
         year = 2001,
        month = aug,
       volume = {135},
       number = {2},
        pages = {177-182},
          doi = {10.1086/321791},
       adsurl = {https://ui.adsabs.harvard.edu/abs/2001ApJS..135..177G}
}

@ARTICLE{wilson2008,
       author = {{Wilson}, Christine D. and {Petitpas}, Glen R. and {Iono}, Daisuke and {Baker}, Andrew J. and {Peck}, Alison B. and {Krips}, Melanie and {Warren}, Bradley and {Golding}, Jennifer and {Atkinson}, Adam and {Armus}, Lee and {Cox}, T.~J. and {Ho}, Paul and {Juvela}, Mika and {Matsushita}, Satoki and {Mihos}, J. Christopher and {Pihlstrom}, Ylva and {Yun}, Min S.},
        title = "{Luminous Infrared Galaxies with the Submillimeter Array. I. Survey Overview and the Central Gas to Dust Ratio}",
      journal = {\apjs},
         year = 2008,
        month = oct,
       volume = {178},
       number = {2},
        pages = {189-224},
          doi = {10.1086/590910},
archivePrefix = {arXiv},
       eprint = {0806.3002},
 primaryClass = {astro-ph},
       adsurl = {https://ui.adsabs.harvard.edu/abs/2008ApJS..178..189W}
}

@article{cicone2014,
	title = {Massive molecular outflows and evidence for {AGN} feedback from {CO} observations},
	volume = {562},
	issn = {0004-6361},
	url = {https://ui.adsabs.harvard.edu/abs/2014A&A...562A..21C},
	doi = {10.1051/0004-6361/201322464},
	urldate = {2024-05-28},
	journal = {Astronomy and Astrophysics},
	author = {Cicone, C. and Maiolino, R. and Sturm, E. and Graciá-Carpio, J. and Feruglio, C. and Neri, R. and Aalto, S. and Davies, R. and Fiore, F. and Fischer, J. and García-Burillo, S. and González-Alfonso, E. and Hailey-Dunsheath, S. and Piconcelli, E. and Veilleux, S.},
	month = feb,
	year = {2014},
	note = {ADS Bibcode: 2014A\&A...562A..21C},
	pages = {A21},
}

@ARTICLE{wright2023,
       author = {{Wright}, Gillian S. and {Rieke}, George H. and {Glasse}, Alistair and {Ressler}, Michael and {Garc{\'\i}a Mar{\'\i}n}, Macarena and {Aguilar}, Jonathan and {Alberts}, Stacey and {{\'A}lvarez-M{\'a}rquez}, Javier and {Argyriou}, Ioannis and {Banks}, Kimberly and {Baudoz}, Pierre and {Boccaletti}, Anthony and {Bouchet}, Patrice and {Bouwman}, Jeroen and {Brandl}, Bernard R. and {Breda}, David and {Bright}, Stacey and {Cale}, Steven and {Colina}, Luis and {Cossou}, Christophe and {Coulais}, Alain and {Cracraft}, Misty and {De Meester}, Wim and {Dicken}, Daniel and {Engesser}, Michael and {Etxaluze}, Mireya and {Fox}, Ori D. and {Friedman}, Scott and {Fu}, Henry and {Gasman}, Danny and {G{\'a}sp{\'a}r}, Andr{\'a}s and {Gastaud}, Ren{\'e} and {Geers}, Vincent and {Glauser}, Adrian Michael and {Gordon}, Karl D. and {Greene}, Thomas and {Greve}, Thomas R. and {Grundy}, Timothy and {G{\"u}del}, Manuel and {Guillard}, Pierre and {Haderlein}, Peter and {Hashimoto}, Ryan and {Henning}, Thomas and {Hines}, Dean and {Holler}, Bryan and {Detre}, {\"O}rs Hunor and {Jahromi}, Amir and {James}, Bryan and {Jones}, Olivia C. and {Justtanont}, Kay and {Kavanagh}, Patrick and {Kendrew}, Sarah and {Klaassen}, Pamela and {Krause}, Oliver and {Labiano}, Alvaro and {Lagage}, Pierre-Olivier and {Lambros}, Scott and {Larson}, Kirsten and {Law}, David and {Lee}, David and {Libralato}, Mattia and {Lorenzo Alverez}, Jose and {Meixner}, Margaret and {Morrison}, Jane and {Mueller}, Migo and {Murray}, Katherine and {Mycroft}, Matthew and {Myers}, Richard and {Nayak}, Omnarayani and {Naylor}, Bret and {Nickson}, Bryony and {Noriega-Crespo}, Alberto and {{\"O}stlin}, G{\"o}ran and {O'Sullivan}, Brian and {Ottens}, Richard and {Patapis}, Polychronis and {Penanen}, Konstantin and {Pietraszkiewicz}, Martin and {Ray}, Tom and {Regan}, Michael and {Roteliuk}, Anthony and {Royer}, Pierre and {Samara-Ratna}, Piyal and {Samuelson}, Bridget and {Sargent}, Beth A. and {Scheithauer}, Silvia and {Schneider}, Analyn and {Schreiber}, J{\"u}rgen and {Shaughnessy}, Bryan and {Sheehan}, Evan and {Shivaei}, Irene and {Sloan}, G.~C. and {Tamas}, Laszlo and {Teague}, Kelly and {Temim}, Tea and {Tikkanen}, Tuomo and {Tustain}, Samuel and {van Dishoeck}, Ewine F. and {Vandenbussche}, Bart and {Weilert}, Mark and {Whitehouse}, Paul and {Wolff}, Schuyler},
        title = "{The Mid-infrared Instrument for JWST and Its In-flight Performance}",
      journal = {\pasp},
         year = 2023,
        month = apr,
       volume = {135},
       number = {1046},
          eid = {048003},
        pages = {048003},
          doi = {10.1088/1538-3873/acbe66},
       adsurl = {https://ui.adsabs.harvard.edu/abs/2023PASP..135d8003W}
}

@ARTICLE{argyriou2023,
       author = {{Argyriou}, Ioannis and {Glasse}, Alistair and {Law}, David R. and {Labiano}, Alvaro and {{\'A}lvarez-M{\'a}rquez}, Javier and {Patapis}, Polychronis and {Kavanagh}, Patrick J. and {Gasman}, Danny and {Mueller}, Michael and {Larson}, Kirsten and {Vandenbussche}, Bart and {Glauser}, Adrian M. and {Royer}, Pierre and {Dicken}, Daniel and {Harkett}, Jake and {Sargent}, Beth A. and {Engesser}, Michael and {Jones}, Olivia C. and {Kendrew}, Sarah and {Noriega-Crespo}, Alberto and {Brandl}, Bernhard and {Rieke}, George H. and {Wright}, Gillian S. and {Lee}, David and {Wells}, Martyn},
        title = "{JWST MIRI flight performance: The Medium-Resolution Spectrometer}",
      journal = {\aap},
         year = 2023,
        month = jul,
       volume = {675},
          eid = {A111},
        pages = {A111},
          doi = {10.1051/0004-6361/202346489},
archivePrefix = {arXiv},
       eprint = {2303.13469},
 primaryClass = {astro-ph.IM},
       adsurl = {https://ui.adsabs.harvard.edu/abs/2023A&A...675A.111A}
}

@MISC{Bus2024,
  title     = "{JWST} Calibration Pipeline",
  author    = "Bushouse, Howard and Eisenhamer, Jonathan and Dencheva, Nadia
               and Davies, James and Greenfield, Perry and Morrison, Jane and
               Hodge, Phil and Simon, Bernie and Grumm, David and Droettboom,
               Michael and Slavich, Edward and Sosey, Megan and Pauly, Tyler
               and Miller, Todd and Jedrzejewski, Robert and Hack, Warren and
               Davis, David and Crawford, Steven and Law, David and Gordon,
               Karl and Regan, Michael and Cara, Mihai and MacDonald, Ken and
               Bradley, Larry and Shanahan, Clare and Jamieson, William and
               Teodoro, Mairan and Williams, Thomas and Pena-Guerrero, Maria",
  publisher = "Zenodo",
  year      =  2024
}

@MISC{Bus2022,
       author = {{Bushouse}, Howard and {Eisenhamer}, Jonathan and {Dencheva}, Nadia and {Davies}, James and {Greenfield}, Perry and {Morrison}, Jane and {Hodge}, Phil and {Simon}, Bernie and {Grumm}, David and {Droettboom}, Michael and {Slavich}, Edward and {Sosey}, Megan and {Pauly}, Tyler and {Miller}, Todd and {Jedrzejewski}, Robert and {Hack}, Warren and {Davis}, David and {Crawford}, Steven and {Law}, David and {Gordon}, Karl and {Regan}, Michael and {Cara}, Mihai and {MacDonald}, Ken and {Bradley}, Larry and {Shanahan}, Clare and {Jamieson}, William and {Teodoro}, Mairan and {Williams}, Thomas},
        title = "{JWST Calibration Pipeline}",
         year = 2022,
        month = oct,
          eid = {10.5281/zenodo.7325378},
          doi = {10.5281/zenodo.7325378},
      version = {1.8.2},
    publisher = {Zenodo},
       adsurl = {https://ui.adsabs.harvard.edu/abs/2022zndo...7325378B}
}

@ARTICLE{alam2015,
       author = {{Alam}, Shadab and {Albareti}, Franco D. and {Allende Prieto}, Carlos and {Anders}, F. and {Anderson}, Scott F. and {Anderton}, Timothy and {Andrews}, Brett H. and {Armengaud}, Eric and {Aubourg}, {\'E}ric and {Bailey}, Stephen and {Basu}, Sarbani and {Bautista}, Julian E. and {Beaton}, Rachael L. and {Beers}, Timothy C. and {Bender}, Chad F. and {Berlind}, Andreas A. and {Beutler}, Florian and {Bhardwaj}, Vaishali and {Bird}, Jonathan C. and {Bizyaev}, Dmitry and {Blake}, Cullen H. and {Blanton}, Michael R. and {Blomqvist}, Michael and {Bochanski}, John J. and {Bolton}, Adam S. and {Bovy}, Jo and {Shelden Bradley}, A. and {Brandt}, W.~N. and {Brauer}, D.~E. and {Brinkmann}, J. and {Brown}, Peter J. and {Brownstein}, Joel R. and {Burden}, Angela and {Burtin}, Etienne and {Busca}, Nicol{\'a}s G. and {Cai}, Zheng and {Capozzi}, Diego and {Carnero Rosell}, Aurelio and {Carr}, Michael A. and {Carrera}, Ricardo and {Chambers}, K.~C. and {Chaplin}, William James and {Chen}, Yen-Chi and {Chiappini}, Cristina and {Chojnowski}, S. Drew and {Chuang}, Chia-Hsun and {Clerc}, Nicolas and {Comparat}, Johan and {Covey}, Kevin and {Croft}, Rupert A.~C. and {Cuesta}, Antonio J. and {Cunha}, Katia and {da Costa}, Luiz N. and {Da Rio}, Nicola and {Davenport}, James R.~A. and {Dawson}, Kyle S. and {De Lee}, Nathan and {Delubac}, Timoth{\'e}e and {Deshpande}, Rohit and {Dhital}, Saurav and {Dutra-Ferreira}, Let{\'\i}cia and {Dwelly}, Tom and {Ealet}, Anne and {Ebelke}, Garrett L. and {Edmondson}, Edward M. and {Eisenstein}, Daniel J. and {Ellsworth}, Tristan and {Elsworth}, Yvonne and {Epstein}, Courtney R. and {Eracleous}, Michael and {Escoffier}, Stephanie and {Esposito}, Massimiliano and {Evans}, Michael L. and {Fan}, Xiaohui and {Fern{\'a}ndez-Alvar}, Emma and {Feuillet}, Diane and {Filiz Ak}, Nurten and {Finley}, Hayley and {Finoguenov}, Alexis and {Flaherty}, Kevin and {Fleming}, Scott W. and {Font-Ribera}, Andreu and {Foster}, Jonathan and {Frinchaboy}, Peter M. and {Galbraith-Frew}, J.~G. and {Garc{\'\i}a}, Rafael A. and {Garc{\'\i}a-Hern{\'a}ndez}, D.~A. and {Garc{\'\i}a P{\'e}rez}, Ana E. and {Gaulme}, Patrick and {Ge}, Jian and {G{\'e}nova-Santos}, R. and {Georgakakis}, A. and {Ghezzi}, Luan and {Gillespie}, Bruce A. and {Girardi}, L{\'e}o and {Goddard}, Daniel and {Gontcho}, Satya Gontcho A. and {Gonz{\'a}lez Hern{\'a}ndez}, Jonay I. and {Grebel}, Eva K. and {Green}, Paul J. and {Grieb}, Jan Niklas and {Grieves}, Nolan and {Gunn}, James E. and {Guo}, Hong and {Harding}, Paul and {Hasselquist}, Sten and {Hawley}, Suzanne L. and {Hayden}, Michael and {Hearty}, Fred R. and {Hekker}, Saskia and {Ho}, Shirley and {Hogg}, David W. and {Holley-Bockelmann}, Kelly and {Holtzman}, Jon A. and {Honscheid}, Klaus and {Huber}, Daniel and {Huehnerhoff}, Joseph and {Ivans}, Inese I. and {Jiang}, Linhua and {Johnson}, Jennifer A. and {Kinemuchi}, Karen and {Kirkby}, David and {Kitaura}, Francisco and {Klaene}, Mark A. and {Knapp}, Gillian R. and {Kneib}, Jean-Paul and {Koenig}, Xavier P. and {Lam}, Charles R. and {Lan}, Ting-Wen and {Lang}, Dustin and {Laurent}, Pierre and {Le Goff}, Jean-Marc and {Leauthaud}, Alexie and {Lee}, Khee-Gan and {Lee}, Young Sun and {Licquia}, Timothy C. and {Liu}, Jian and {Long}, Daniel C. and {L{\'o}pez-Corredoira}, Mart{\'\i}n and {Lorenzo-Oliveira}, Diego and {Lucatello}, Sara and {Lundgren}, Britt and {Lupton}, Robert H. and {Mack}, III, Claude E. and {Mahadevan}, Suvrath and {Maia}, Marcio A.~G. and {Majewski}, Steven R. and {Malanushenko}, Elena and {Malanushenko}, Viktor and {Manchado}, A. and {Manera}, Marc and {Mao}, Qingqing and {Maraston}, Claudia and {Marchwinski}, Robert C. and {Margala}, Daniel and {Martell}, Sarah L. and {Martig}, Marie and {Masters}, Karen L. and {Mathur}, Savita and {McBride}, Cameron K. and {McGehee}, Peregrine M. and {McGreer}, Ian D. and {McMahon}, Richard G. and {M{\'e}nard}, Brice and {Menzel}, Marie-Luise and {Merloni}, Andrea and {M{\'e}sz{\'a}ros}, Szabolcs and {Miller}, Adam A. and {Miralda-Escud{\'e}}, Jordi and {Miyatake}, Hironao and {Montero-Dorta}, Antonio D. and {More}, Surhud and {Morganson}, Eric and {Morice-Atkinson}, Xan and {Morrison}, Heather L. and {Mosser}, Ben{\^o}it and {Muna}, Demitri and {Myers}, Adam D. and {Nandra}, Kirpal and {Newman}, Jeffrey A. and {Neyrinck}, Mark and {Nguyen}, Duy Cuong and {Nichol}, Robert C. and {Nidever}, David L. and {Noterdaeme}, Pasquier and {Nuza}, Sebasti{\'a}n E. and {O'Connell}, Julia E. and {O'Connell}, Robert W. and {O'Connell}, Ross and {Ogando}, Ricardo L.~C. and {Olmstead}, Matthew D. and {Oravetz}, Audrey E. and {Oravetz}, Daniel J. and {Osumi}, Keisuke and {Owen}, Russell and {Padgett}, Deborah L. and {Padmanabhan}, Nikhil and {Paegert}, Martin and {Palanque-Delabrouille}, Nathalie and {Pan}, Kaike},
        title = "{The Eleventh and Twelfth Data Releases of the Sloan Digital Sky Survey: Final Data from SDSS-III}",
      journal = {\apjs},
         year = 2015,
        month = jul,
       volume = {219},
       number = {1},
          eid = {12},
        pages = {12},
          doi = {10.1088/0067-0049/219/1/12},
archivePrefix = {arXiv},
       eprint = {1501.00963},
 primaryClass = {astro-ph.IM},
       adsurl = {https://ui.adsabs.harvard.edu/abs/2015ApJS..219...12A}
}

@ARTICLE{Arg2023,
       author = {{Argyriou}, Ioannis and {Glasse}, Alistair and {Law}, David R. and {Labiano}, Alvaro and {{\'A}lvarez-M{\'a}rquez}, Javier and {Patapis}, Polychronis and {Kavanagh}, Patrick J. and {Gasman}, Danny and {Mueller}, Michael and {Larson}, Kirsten and {Vandenbussche}, Bart and {Glauser}, Adrian M. and {Royer}, Pierre and {Dicken}, Daniel and {Harkett}, Jake and {Sargent}, Beth A. and {Engesser}, Michael and {Jones}, Olivia C. and {Kendrew}, Sarah and {Noriega-Crespo}, Alberto and {Brandl}, Bernhard and {Rieke}, George H. and {Wright}, Gillian S. and {Lee}, David and {Wells}, Martyn},
        title = "{JWST MIRI flight performance: The Medium-Resolution Spectrometer}",
      journal = {\aap},
         year = 2023,
        month = jul,
       volume = {675},
          eid = {A111},
        pages = {A111},
          doi = {10.1051/0004-6361/202346489},
archivePrefix = {arXiv},
       eprint = {2303.13469},
 primaryClass = {astro-ph.IM},
       adsurl = {https://ui.adsabs.harvard.edu/abs/2023A&A...675A.111A}
}

@ARTICLE{See2024,
       author = {{Seebeck}, Jerome and {Veilleux}, Sylvain and {Liu}, Weizhe and {Rupke}, David S.~N. and {Vayner}, Andrey and {Wylezalek}, Dominika and {Zakamska}, Nadia L. and {Bertemes}, Caroline},
        title = "{Combined JWST{\textendash}MUSE Integral Field Spectroscopy of the Most Luminous Quasar in the Local Universe, PDS 456}",
      journal = {\apj},
         year = 2024,
        month = dec,
       volume = {976},
       number = {2},
          eid = {240},
        pages = {240},
          doi = {10.3847/1538-4357/ad815e},
archivePrefix = {arXiv},
       eprint = {2409.18086},
 primaryClass = {astro-ph.GA},
       adsurl = {https://ui.adsabs.harvard.edu/abs/2024ApJ...976..240S}
}

@ARTICLE{gordon2023,
       author = {{Gordon}, Karl D. and {Clayton}, Geoffrey C. and {Decleir}, Marjorie and {Fitzpatrick}, E.~L. and {Massa}, Derck and {Misselt}, Karl A. and {Tollerud}, Erik J.},
        title = "{One Relation for All Wavelengths: The Far-ultraviolet to Mid-infrared Milky Way Spectroscopic R(V)-dependent Dust Extinction Relationship}",
      journal = {\apj},
         year = 2023,
        month = jun,
       volume = {950},
       number = {2},
          eid = {86},
        pages = {86},
          doi = {10.3847/1538-4357/accb59},
archivePrefix = {arXiv},
       eprint = {2304.01991},
 primaryClass = {astro-ph.GA},
       adsurl = {https://ui.adsabs.harvard.edu/abs/2023ApJ...950...86G}
}

@ARTICLE{roueff2019,
       author = {{Roueff}, E. and {Abgrall}, H. and {Czachorowski}, P. and {Pachucki}, K. and {Puchalski}, M. and {Komasa}, J.},
        title = "{The full infrared spectrum of molecular hydrogen}",
      journal = {\aap},
         year = 2019,
        month = oct,
       volume = {630},
          eid = {A58},
        pages = {A58},
          doi = {10.1051/0004-6361/201936249},
archivePrefix = {arXiv},
       eprint = {1909.11585},
 primaryClass = {physics.atom-ph},
       adsurl = {https://ui.adsabs.harvard.edu/abs/2019A&A...630A..58R}
}

@ARTICLE{youngblood2018,
       author = {{Youngblood}, Allison and {France}, Kevin and {Ginsburg}, Adam and {Hoadley}, Keri and {Bally}, John},
        title = "{The Orion Fingers: H$_{2}$ Temperatures and Excitation in an Explosive Outflow}",
      journal = {\apj},
         year = 2018,
        month = apr,
       volume = {857},
       number = {1},
          eid = {7},
        pages = {7},
          doi = {10.3847/1538-4357/aab4f4},
archivePrefix = {arXiv},
       eprint = {1803.01903},
 primaryClass = {astro-ph.GA},
       adsurl = {https://ui.adsabs.harvard.edu/abs/2018ApJ...857....7Y}
}

@ARTICLE{roussel2007,
       author = {{Roussel}, H. and {Helou}, G. and {Hollenbach}, D.~J. and {Draine}, B.~T. and {Smith}, J.~D. and {Armus}, L. and {Schinnerer}, E. and {Walter}, F. and {Engelbracht}, C.~W. and {Thornley}, M.~D. and {Kennicutt}, R.~C. and {Calzetti}, D. and {Dale}, D.~A. and {Murphy}, E.~J. and {Bot}, C.},
        title = "{Warm Molecular Hydrogen in the Spitzer SINGS Galaxy Sample}",
      journal = {\apj},
         year = 2007,
        month = nov,
       volume = {669},
       number = {2},
        pages = {959-981},
          doi = {10.1086/521667},
archivePrefix = {arXiv},
       eprint = {0707.0395},
 primaryClass = {astro-ph},
       adsurl = {https://ui.adsabs.harvard.edu/abs/2007ApJ...669..959R}
}

@ARTICLE{wolniewicz1998,
       author = {{Wolniewicz}, L. and {Simbotin}, I. and {Dalgarno}, A.},
        title = "{Quadrupole Transition Probabilities for the Excited Rovibrational States of H$_{2}$}",
      journal = {\apjs},
         year = 1998,
        month = apr,
       volume = {115},
       number = {2},
        pages = {293-313},
          doi = {10.1086/313091},
       adsurl = {https://ui.adsabs.harvard.edu/abs/1998ApJS..115..293W}
}

@ARTICLE{togi2016,
       author = {{Togi}, Aditya and {Smith}, J.~D.~T.},
        title = "{Lighting the Dark Molecular Gas: H$_{2}$ as a Direct Tracer}",
      journal = {\apj},
         year = 2016,
        month = oct,
       volume = {830},
       number = {1},
          eid = {18},
        pages = {18},
          doi = {10.3847/0004-637X/830/1/18},
archivePrefix = {arXiv},
       eprint = {1607.08036},
 primaryClass = {astro-ph.GA},
       adsurl = {https://ui.adsabs.harvard.edu/abs/2016ApJ...830...18T}
}

@ARTICLE{jones2024,
       author = {{Jones}, Logan H. and {Hernandez}, Svea and {Smith}, Linda J. and {Togi}, Aditya and {Diaz-Santos}, Tanio and {Aloisi}, Alessandra and {Blair}, William and {Hirschauer}, Alec S. and {Hunt}, Leslie K. and {James}, Bethan L. and {Kumari}, Nimisha and {Lebouteiller}, Vianney and {Mingozzi}, Matilde and {Ramambason}, Lise},
        title = "{A JWST/MIRI View of the ISM in M83: I. Resolved Molecular Hydrogen Properties, Star Formation, and Feedback}",
      journal = {arXiv e-prints},
         year = 2024,
        month = oct,
          eid = {arXiv:2410.09020},
        pages = {arXiv:2410.09020},
          doi = {10.48550/arXiv.2410.09020},
archivePrefix = {arXiv},
       eprint = {2410.09020},
 primaryClass = {astro-ph.GA},
       adsurl = {https://ui.adsabs.harvard.edu/abs/2024arXiv241009020J}
}

@ARTICLE{habart2005,
       author = {{Habart}, Emilie and {Walmsley}, Malcolm and {Verstraete}, Laurent and {Cazaux}, Stephanie and {Maiolino}, Roberto and {Cox}, Pierre and {Boulanger}, Francois and {Pineau des For{\^e}ts}, Guillaume},
        title = "{Molecular Hydrogen}",
      journal = {\ssr},
         year = 2005,
        month = aug,
       volume = {119},
       number = {1-4},
        pages = {71-91},
          doi = {10.1007/s11214-005-8062-1},
       adsurl = {https://ui.adsabs.harvard.edu/abs/2005SSRv..119...71H}
}

@ARTICLE{Nicholls2017,
       author = {{Nicholls}, David C. and {Sutherland}, Ralph S. and {Dopita}, Michael A. and {Kewley}, Lisa J. and {Groves}, Brent A.},
        title = "{Abundance scaling in stars, nebulae and galaxies}",
      journal = {\mnras},
         year = 2017,
        month = apr,
       volume = {466},
       number = {4},
        pages = {4403-4422},
          doi = {10.1093/mnras/stw3235},
archivePrefix = {arXiv},
       eprint = {1612.03546},
 primaryClass = {astro-ph.GA},
       adsurl = {https://ui.adsabs.harvard.edu/abs/2017MNRAS.466.4403N}
}

@ARTICLE{Luridiana2015,
       author = {{Luridiana}, V. and {Morisset}, C. and {Shaw}, R.~A.},
        title = "{PyNeb: a new tool for analyzing emission lines. I. Code description and validation of results}",
      journal = {\aap},
         year = 2015,
        month = jan,
       volume = {573},
          eid = {A42},
        pages = {A42},
          doi = {10.1051/0004-6361/201323152},
archivePrefix = {arXiv},
       eprint = {1410.6662},
 primaryClass = {astro-ph.IM},
       adsurl = {https://ui.adsabs.harvard.edu/abs/2015A&A...573A..42L}
}

@ARTICLE{Arribas2014,
       author = {{Arribas}, S. and {Colina}, L. and {Bellocchi}, E. and {Maiolino}, R. and {Villar-Mart{\'\i}n}, M.},
        title = "{Ionized gas outflows and global kinematics of low-z luminous star-forming galaxies}",
      journal = {\aap},
         year = 2014,
        month = aug,
       volume = {568},
          eid = {A14},
        pages = {A14},
          doi = {10.1051/0004-6361/201323324},
archivePrefix = {arXiv},
       eprint = {1404.1082},
 primaryClass = {astro-ph.GA},
       adsurl = {https://ui.adsabs.harvard.edu/abs/2014A&A...568A..14A}
}

@ARTICLE{gonzalez2017,
       author = {{Gonz{\'a}lez-Alfonso}, E. and {Fischer}, J. and {Spoon}, H.~W.~W. and {Stewart}, K.~P. and {Ashby}, M.~L.~N. and {Veilleux}, S. and {Smith}, H.~A. and {Sturm}, E. and {Farrah}, D. and {Falstad}, N. and {Mel{\'e}ndez}, M. and {Graci{\'a}-Carpio}, J. and {Janssen}, A.~W. and {Lebouteiller}, V.},
        title = "{Molecular Outflows in Local ULIRGs: Energetics from Multitransition OH Analysis}",
      journal = {\apj},
         year = 2017,
        month = feb,
       volume = {836},
       number = {1},
          eid = {11},
        pages = {11},
          doi = {10.3847/1538-4357/836/1/11},
archivePrefix = {arXiv},
       eprint = {1612.08181},
 primaryClass = {astro-ph.GA},
       adsurl = {https://ui.adsabs.harvard.edu/abs/2017ApJ...836...11G}
}

@ARTICLE{dan2025,
       author = {{Dan}, Kylie Yui and {Seebeck}, Jerome and {Veilleux}, Sylvain and {Rupke}, David and {Gonzalez-Alfonso}, Eduardo and {Garcia-Bernete}, Ismael and {Liu}, Weizhe and {Lutz}, Dieter and {Melendez}, Marcio and {Pereira Santaella}, Miguel and {Sturm}, Eckhard and {Tombesi}, Francesco},
        title = "{JWST Discovery of a Very Fast Biconical Outflow of Warm Molecular Gas in the Nearby Ultraluminous Infrared Galaxy F08572+3915 NW}",
      journal = {\apj},
         year = 2025,
        month = jan,
       volume = {979},
       number = {1},
          eid = {68},
        pages = {68},
          doi = {10.3847/1538-4357/ad9a50},
archivePrefix = {arXiv},
       eprint = {2412.05859},
 primaryClass = {astro-ph.GA},
       adsurl = {https://ui.adsabs.harvard.edu/abs/2025ApJ...979...68D}
}

@ARTICLE{draine2021,
       author = {{Draine}, B.~T. and {Li}, Aigen and {Hensley}, Brandon S. and {Hunt}, L.~K. and {Sandstrom}, K. and {Smith}, J. -D.~T.},
        title = "{Excitation of Polycyclic Aromatic Hydrocarbon Emission: Dependence on Size Distribution, Ionization, and Starlight Spectrum and Intensity}",
      journal = {\apj},
         year = 2021,
        month = aug,
       volume = {917},
       number = {1},
          eid = {3},
        pages = {3},
          doi = {10.3847/1538-4357/abff51},
archivePrefix = {arXiv},
       eprint = {2011.07046},
 primaryClass = {astro-ph.GA},
       adsurl = {https://ui.adsabs.harvard.edu/abs/2021ApJ...917....3D}
}

@ARTICLE{Shipley2016,
       author = {{Shipley}, Heath V. and {Papovich}, Casey and {Rieke}, George H. and {Brown}, Michael J.~I. and {Moustakas}, John},
        title = "{A New Star Formation Rate Calibration from Polycyclic Aromatic Hydrocarbon Emission Features and Application to High-redshift Galaxies}",
      journal = {\apj},
         year = 2016,
        month = feb,
       volume = {818},
       number = {1},
          eid = {60},
        pages = {60},
          doi = {10.3847/0004-637X/818/1/60},
archivePrefix = {arXiv},
       eprint = {1601.01698},
 primaryClass = {astro-ph.GA},
       adsurl = {https://ui.adsabs.harvard.edu/abs/2016ApJ...818...60S}
}

@ARTICLE{Diamond2012,
       author = {{Diamond-Stanic}, Aleksandar M. and {Rieke}, George H.},
        title = "{The Relationship between Black Hole Growth and Star Formation in Seyfert Galaxies}",
      journal = {\apj},
         year = 2012,
        month = feb,
       volume = {746},
       number = {2},
          eid = {168},
        pages = {168},
          doi = {10.1088/0004-637X/746/2/168},
archivePrefix = {arXiv},
       eprint = {1106.3565},
 primaryClass = {astro-ph.CO},
       adsurl = {https://ui.adsabs.harvard.edu/abs/2012ApJ...746..168D}
}

@ARTICLE{putte2025,
       author = {{Van De Putte}, Dries and {Peeters}, Els and {Gordon}, Karl D. and {Smith}, John-David T. and {Lai}, Thomas S.-Y. and {Maragkoudakis}, Alexandros and {Schefter}, Bethany and {Sidhu}, Ameek and {Doshi}, Dhruvil and {Bern{\'e}}, Olivier and {Cami}, Jan and {Boersma}, Christiaan and {Dartois}, Emmanuel and {Habart}, Emilie and {Onaka}, Takashi and {Tielens}, Alexander G.~G.~M.},
        title = "{PDRs4All: XVI. Tracing aromatic infrared band characteristics in photodissociation region spectra with PAHFIT in the JWST era}",
      journal = {\aap},
         year = 2025,
        month = sep,
       volume = {701},
          eid = {A111},
        pages = {A111},
          doi = {10.1051/0004-6361/202554991},
archivePrefix = {arXiv},
       eprint = {2507.05848},
 primaryClass = {astro-ph.GA},
       adsurl = {https://ui.adsabs.harvard.edu/abs/2025A&A...701A.111V}
}

@ARTICLE{Liu2020,
       author = {{Liu}, Weizhe and {Veilleux}, Sylvain and {Canalizo}, Gabriela and {Rupke}, David S.~N. and {Manzano-King}, Christina M. and {Bohn}, Thomas and {U}, Vivian},
        title = "{Integral Field Spectroscopy of Fast Outflows in Dwarf Galaxies with AGNs}",
      journal = {\apj},
         year = 2020,
        month = dec,
       volume = {905},
       number = {2},
          eid = {166},
        pages = {166},
          doi = {10.3847/1538-4357/abc269},
archivePrefix = {arXiv},
       eprint = {2010.09008},
 primaryClass = {astro-ph.GA},
       adsurl = {https://ui.adsabs.harvard.edu/abs/2020ApJ...905..166L}
}

@ARTICLE{veilleux2020,
       author = {{Veilleux}, Sylvain and {Maiolino}, Roberto and {Bolatto}, Alberto D. and {Aalto}, Susanne},
        title = "{Cool outflows in galaxies and their implications}",
      journal = {\aapr},
         year = 2020,
        month = apr,
       volume = {28},
       number = {1},
          eid = {2},
        pages = {2},
          doi = {10.1007/s00159-019-0121-9},
archivePrefix = {arXiv},
       eprint = {2002.07765},
 primaryClass = {astro-ph.GA},
       adsurl = {https://ui.adsabs.harvard.edu/abs/2020A&ARv..28....2V}
}

@MISC{Diaz2025,
author = {{Diaz-Santos}, Tanio and {Lai}, Thomas S. -Y. and {Finnerty}, Luke and {Privon}, George and {Bonfini}, Paolo and {Larson}, Kirsten and {Marshall}, Jason and {Armus}, Lee and {Charmandaris}, Vassilis},
title = "{CAFE: Continuum And Feature Extraction tool}",
howpublished = {Astrophysics Source Code Library, record ascl:2501.001},
year = 2025,
month = jan,
eid = {ascl:2501.001},
adsurl = {https://ui.adsabs.harvard.edu/abs/2025ascl.soft01001D}
}

@ARTICLE{Wylezalek2022,
       author = {{Wylezalek}, Dominika and {Vayner}, Andrey and {Rupke}, David S.~N. and {Zakamska}, Nadia L. and {Veilleux}, Sylvain and {Ishikawa}, Yuzo and {Bertemes}, Caroline and {Liu}, Weizhe and {Barrera-Ballesteros}, Jorge K. and {Chen}, Hsiao-Wen and {Goulding}, Andy D. and {Greene}, Jenny E. and {Hainline}, Kevin N. and {Hamann}, Fred and {Heckman}, Timothy and {Johnson}, Sean D. and {Lutz}, Dieter and {L{\"u}tzgendorf}, Nora and {Mainieri}, Vincenzo and {Maiolino}, Roberto and {Nesvadba}, Nicole P.~H. and {Ogle}, Patrick and {Sturm}, Eckhard},
        title = "{First Results from the JWST Early Release Science Program Q3D: Turbulent Times in the Life of a z   3 Extremely Red Quasar Revealed by NIRSpec IFU}",
      journal = {\apjl},
         year = 2022,
        month = nov,
       volume = {940},
       number = {1},
          eid = {L7},
        pages = {L7},
          doi = {10.3847/2041-8213/ac98c3},
archivePrefix = {arXiv},
       eprint = {2210.10074},
 primaryClass = {astro-ph.GA},
       adsurl = {https://ui.adsabs.harvard.edu/abs/2022ApJ...940L...7W}
}

@ARTICLE{Rupke2023,
       author = {{Rupke}, David S.~N. and {Wylezalek}, Dominika and {Zakamska}, Nadia L. and {Veilleux}, Sylvain and {Bertemes}, Caroline and {Ishikawa}, Yuzo and {Liu}, Weizhe and {Sankar}, Swetha and {Vayner}, Andrey and {Grace Lim}, Hui Xian and {McCrory}, Ryan and {Murphree}, Grey and {Whitesell}, Lillian and {Shen}, Lu and {Liu}, Guilin and {Barrera-Ballesteros}, Jorge K. and {Chen}, Hsiao-Wen and {Diachenko}, Nadiia and {Goulding}, Andy D. and {Greene}, Jenny E. and {Hainline}, Kevin N. and {Hamann}, Fred and {Heckman}, Timothy and {Johnson}, Sean D. and {Lutz}, Dieter and {L{\"u}tzgendorf}, Nora and {Mainieri}, Vincenzo and {Nesvadba}, Nicole P.~H. and {Ogle}, Patrick and {Sturm}, Eckhard},
        title = "{First Results from the JWST Early Release Science Program Q3D: Benchmark Comparison of Optical and Mid-infrared Tracers of a Dusty, Ionized Red Quasar Wind at z = 0.435}",
      journal = {\apjl},
         year = 2023,
        month = aug,
       volume = {953},
       number = {2},
          eid = {L26},
        pages = {L26},
          doi = {10.3847/2041-8213/aced85},
archivePrefix = {arXiv},
       eprint = {2306.12541},
 primaryClass = {astro-ph.GA},
       adsurl = {https://ui.adsabs.harvard.edu/abs/2023ApJ...953L..26R}
}

@ARTICLE{Veilleux2023,
       author = {{Veilleux}, Sylvain and {Liu}, Weizhe and {Vayner}, Andrey and {Wylezalek}, Dominika and {Rupke}, David S.~N. and {Zakamska}, Nadia L. and {Ishikawa}, Yuzo and {Bertemes}, Caroline and {Barrera-Ballesteros}, Jorge K. and {Chen}, Hsiao-Wen and {Diachenko}, Nadiia and {Goulding}, Andy D. and {Greene}, Jenny E. and {Hainline}, Kevin N. and {Hamann}, Fred and {Heckman}, Timothy and {Johnson}, Sean D. and {Grace Lim}, Hui Xian and {Lutz}, Dieter and {L{\"u}tzgendorf}, Nora and {Mainieri}, Vincenzo and {Maiolino}, Roberto and {McCrory}, Ryan and {Murphree}, Grey and {Nesvadba}, Nicole P.~H. and {Ogle}, Patrick and {Sankar}, Swetha and {Sturm}, Eckhard and {Whitesell}, Lillian},
        title = "{First Results from the JWST Early Release Science Program Q3D: The Warm Ionized Gas Outflow in z   1.6 Quasar XID 2028 and Its Impact on the Host Galaxy}",
      journal = {\apj},
         year = 2023,
        month = aug,
       volume = {953},
       number = {1},
          eid = {56},
        pages = {56},
          doi = {10.3847/1538-4357/ace10f},
archivePrefix = {arXiv},
       eprint = {2303.08952},
 primaryClass = {astro-ph.GA},
       adsurl = {https://ui.adsabs.harvard.edu/abs/2023ApJ...953...56V}
}

@ARTICLE{Vayner2023,
       author = {{Vayner}, Andrey and {Zakamska}, Nadia L. and {Ishikawa}, Yuzo and {Sankar}, Swetha and {Wylezalek}, Dominika and {Rupke}, David S.~N. and {Veilleux}, Sylvain and {Bertemes}, Caroline and {Barrera-Ballesteros}, Jorge K. and {Chen}, Hsiao-Wen and {Diachenko}, Nadiia and {Goulding}, Andy D. and {Greene}, Jenny E. and {Hainline}, Kevin N. and {Hamann}, Fred and {Heckman}, Timothy and {Johnson}, Sean D. and {Grace Lim}, Hui Xian and {Liu}, Weizhe and {Lutz}, Dieter and {L{\"u}tzgendorf}, Nora and {Mainieri}, Vincenzo and {McCrory}, Ryan and {Murphree}, Grey and {Nesvadba}, Nicole P.~H. and {Ogle}, Patrick and {Sturm}, Eckhard and {Whitesell}, Lillian},
        title = "{First Results from the JWST Early Release Science Program Q3D: Ionization Cone, Clumpy Star Formation, and Shocks in a z = 3 Extremely Red Quasar Host}",
      journal = {\apj},
         year = 2023,
        month = oct,
       volume = {955},
       number = {2},
          eid = {92},
        pages = {92},
          doi = {10.3847/1538-4357/ace784},
archivePrefix = {arXiv},
       eprint = {2303.06970},
 primaryClass = {astro-ph.GA},
       adsurl = {https://ui.adsabs.harvard.edu/abs/2023ApJ...955...92V}
}

@MISC{q3d2014,
author = {{Rupke}, David S.~N.},
title = "{IFSFIT: Spectral Fitting for Integral Field Spectrographs}",
year = 2014,
month = sep,
eid = {ascl:1409.005},
pages = {ascl:1409.005},
archivePrefix = {ascl},
eprint = {1409.005},
adsurl = {https://ui.adsabs.harvard.edu/abs/2014ascl.soft09005R}
}

@MISC{q3d2021,
author = {{Rupke}, D.~S.~N. and {Schweitzer}, M. and {Viola}, V. and {Lutz}, D. and {Sturm}, E. and {Spoon}, H. and {Veilleux}, S. and {Kim}, D. -C.},
title = "{QUESTFIT: Fitter for mid-infrared galaxy spectra}",
year = 2021,
month = dec,
eid = {ascl:2112.002},
pages = {ascl:2112.002},
archivePrefix = {ascl},
eprint = {2112.002},
adsurl = {https://ui.adsabs.harvard.edu/abs/2021ascl.soft12002R}
}

@article{astropy:2013,
Adsurl = {http://adsabs.harvard.edu/abs/2013A%26A...558A..33A},
Archiveprefix = {arXiv},
Author = {{Astropy Collaboration} and {Robitaille}, T.~P. and {Tollerud}, E.~J. and {Greenfield}, P. and {Droettboom}, M. and {Bray}, E. and {Aldcroft}, T. and {Davis}, M. and {Ginsburg}, A. and {Price-Whelan}, A.~M. and {Kerzendorf}, W.~E. and {Conley}, A. and {Crighton}, N. and {Barbary}, K. and {Muna}, D. and {Ferguson}, H. and {Grollier}, F. and {Parikh}, M.~M. and {Nair}, P.~H. and {Unther}, H.~M. and {Deil}, C. and {Woillez}, J. and {Conseil}, S. and {Kramer}, R. and {Turner}, J.~E.~H. and {Singer}, L. and {Fox}, R. and {Weaver}, B.~A. and {Zabalza}, V. and {Edwards}, Z.~I. and {Azalee Bostroem}, K. and {Burke}, D.~J. and {Casey}, A.~R. and {Crawford}, S.~M. and {Dencheva}, N. and {Ely}, J. and {Jenness}, T. and {Labrie}, K. and {Lim}, P.~L. and {Pierfederici}, F. and {Pontzen}, A. and {Ptak}, A. and {Refsdal}, B. and {Servillat}, M. and {Streicher}, O.},
Doi = {10.1051/0004-6361/201322068},
Eid = {A33},
Eprint = {1307.6212},
Journal = {\aap},
Month = oct,
Pages = {A33},
Primaryclass = {astro-ph.IM},
Title = {{Astropy: A community Python package for astronomy}},
Volume = 558,
Year = 2013}

@ARTICLE{astropy:2018,
       author = {{Astropy Collaboration} and {Price-Whelan}, A.~M. and
         {Sip{\H{o}}cz}, B.~M. and {G{\"u}nther}, H.~M. and {Lim}, P.~L. and
         {Crawford}, S.~M. and {Conseil}, S. and {Shupe}, D.~L. and
         {Craig}, M.~W. and {Dencheva}, N. and {Ginsburg}, A. and {Vand
        erPlas}, J.~T. and {Bradley}, L.~D. and {P{\'e}rez-Su{\'a}rez}, D. and
         {de Val-Borro}, M. and {Aldcroft}, T.~L. and {Cruz}, K.~L. and
         {Robitaille}, T.~P. and {Tollerud}, E.~J. and {Ardelean}, C. and
         {Babej}, T. and {Bach}, Y.~P. and {Bachetti}, M. and {Bakanov}, A.~V. and
         {Bamford}, S.~P. and {Barentsen}, G. and {Barmby}, P. and
         {Baumbach}, A. and {Berry}, K.~L. and {Biscani}, F. and {Boquien}, M. and
         {Bostroem}, K.~A. and {Bouma}, L.~G. and {Brammer}, G.~B. and
         {Bray}, E.~M. and {Breytenbach}, H. and {Buddelmeijer}, H. and
         {Burke}, D.~J. and {Calderone}, G. and {Cano Rodr{\'\i}guez}, J.~L. and
         {Cara}, M. and {Cardoso}, J.~V.~M. and {Cheedella}, S. and {Copin}, Y. and
         {Corrales}, L. and {Crichton}, D. and {D'Avella}, D. and {Deil}, C. and
         {Depagne}, {\'E}. and {Dietrich}, J.~P. and {Donath}, A. and
         {Droettboom}, M. and {Earl}, N. and {Erben}, T. and {Fabbro}, S. and
         {Ferreira}, L.~A. and {Finethy}, T. and {Fox}, R.~T. and
         {Garrison}, L.~H. and {Gibbons}, S.~L.~J. and {Goldstein}, D.~A. and
         {Gommers}, R. and {Greco}, J.~P. and {Greenfield}, P. and
         {Groener}, A.~M. and {Grollier}, F. and {Hagen}, A. and {Hirst}, P. and
         {Homeier}, D. and {Horton}, A.~J. and {Hosseinzadeh}, G. and {Hu}, L. and
         {Hunkeler}, J.~S. and {Ivezi{\'c}}, {\v{Z}}. and {Jain}, A. and
         {Jenness}, T. and {Kanarek}, G. and {Kendrew}, S. and {Kern}, N.~S. and
         {Kerzendorf}, W.~E. and {Khvalko}, A. and {King}, J. and {Kirkby}, D. and
         {Kulkarni}, A.~M. and {Kumar}, A. and {Lee}, A. and {Lenz}, D. and
         {Littlefair}, S.~P. and {Ma}, Z. and {Macleod}, D.~M. and
         {Mastropietro}, M. and {McCully}, C. and {Montagnac}, S. and
         {Morris}, B.~M. and {Mueller}, M. and {Mumford}, S.~J. and {Muna}, D. and
         {Murphy}, N.~A. and {Nelson}, S. and {Nguyen}, G.~H. and
         {Ninan}, J.~P. and {N{\"o}the}, M. and {Ogaz}, S. and {Oh}, S. and
         {Parejko}, J.~K. and {Parley}, N. and {Pascual}, S. and {Patil}, R. and
         {Patil}, A.~A. and {Plunkett}, A.~L. and {Prochaska}, J.~X. and
         {Rastogi}, T. and {Reddy Janga}, V. and {Sabater}, J. and
         {Sakurikar}, P. and {Seifert}, M. and {Sherbert}, L.~E. and
         {Sherwood-Taylor}, H. and {Shih}, A.~Y. and {Sick}, J. and
         {Silbiger}, M.~T. and {Singanamalla}, S. and {Singer}, L.~P. and
         {Sladen}, P.~H. and {Sooley}, K.~A. and {Sornarajah}, S. and
         {Streicher}, O. and {Teuben}, P. and {Thomas}, S.~W. and
         {Tremblay}, G.~R. and {Turner}, J.~E.~H. and {Terr{\'o}n}, V. and
         {van Kerkwijk}, M.~H. and {de la Vega}, A. and {Watkins}, L.~L. and
         {Weaver}, B.~A. and {Whitmore}, J.~B. and {Woillez}, J. and
         {Zabalza}, V. and {Astropy Contributors}},
        title = "{The Astropy Project: Building an Open-science Project and Status of the v2.0 Core Package}",
      journal = {\aj},
         year = 2018,
        month = sep,
       volume = {156},
       number = {3},
          eid = {123},
        pages = {123},
          doi = {10.3847/1538-3881/aabc4f},
archivePrefix = {arXiv},
       eprint = {1801.02634},
 primaryClass = {astro-ph.IM},
       adsurl = {https://ui.adsabs.harvard.edu/abs/2018AJ....156..123A}
}

@ARTICLE{astropy:2022,
       author = {{Astropy Collaboration} and {Price-Whelan}, Adrian M. and {Lim}, Pey Lian and {Earl}, Nicholas and {Starkman}, Nathaniel and {Bradley}, Larry and {Shupe}, David L. and {Patil}, Aarya A. and {Corrales}, Lia and {Brasseur}, C.~E. and {N{"o}the}, Maximilian and {Donath}, Axel and {Tollerud}, Erik and {Morris}, Brett M. and {Ginsburg}, Adam and {Vaher}, Eero and {Weaver}, Benjamin A. and {Tocknell}, James and {Jamieson}, William and {van Kerkwijk}, Marten H. and {Robitaille}, Thomas P. and {Merry}, Bruce and {Bachetti}, Matteo and {G{"u}nther}, H. Moritz and {Aldcroft}, Thomas L. and {Alvarado-Montes}, Jaime A. and {Archibald}, Anne M. and {B{'o}di}, Attila and {Bapat}, Shreyas and {Barentsen}, Geert and {Baz{'a}n}, Juanjo and {Biswas}, Manish and {Boquien}, M{'e}d{'e}ric and {Burke}, D.~J. and {Cara}, Daria and {Cara}, Mihai and {Conroy}, Kyle E. and {Conseil}, Simon and {Craig}, Matthew W. and {Cross}, Robert M. and {Cruz}, Kelle L. and {D'Eugenio}, Francesco and {Dencheva}, Nadia and {Devillepoix}, Hadrien A.~R. and {Dietrich}, J{"o}rg P. and {Eigenbrot}, Arthur Davis and {Erben}, Thomas and {Ferreira}, Leonardo and {Foreman-Mackey}, Daniel and {Fox}, Ryan and {Freij}, Nabil and {Garg}, Suyog and {Geda}, Robel and {Glattly}, Lauren and {Gondhalekar}, Yash and {Gordon}, Karl D. and {Grant}, David and {Greenfield}, Perry and {Groener}, Austen M. and {Guest}, Steve and {Gurovich}, Sebastian and {Handberg}, Rasmus and {Hart}, Akeem and {Hatfield-Dodds}, Zac and {Homeier}, Derek and {Hosseinzadeh}, Griffin and {Jenness}, Tim and {Jones}, Craig K. and {Joseph}, Prajwel and {Kalmbach}, J. Bryce and {Karamehmetoglu}, Emir and {Ka{l}uszy{'n}ski}, Miko{l}aj and {Kelley}, Michael S.~P. and {Kern}, Nicholas and {Kerzendorf}, Wolfgang E. and {Koch}, Eric W. and {Kulumani}, Shankar and {Lee}, Antony and {Ly}, Chun and {Ma}, Zhiyuan and {MacBride}, Conor and {Maljaars}, Jakob M. and {Muna}, Demitri and {Murphy}, N.~A. and {Norman}, Henrik and {O'Steen}, Richard and {Oman}, Kyle A. and {Pacifici}, Camilla and {Pascual}, Sergio and {Pascual-Granado}, J. and {Patil}, Rohit R. and {Perren}, Gabriel I. and {Pickering}, Timothy E. and {Rastogi}, Tanuj and {Roulston}, Benjamin R. and {Ryan}, Daniel F. and {Rykoff}, Eli S. and {Sabater}, Jose and {Sakurikar}, Parikshit and {Salgado}, Jes{'u}s and {Sanghi}, Aniket and {Saunders}, Nicholas and {Savchenko}, Volodymyr and {Schwardt}, Ludwig and {Seifert-Eckert}, Michael and {Shih}, Albert Y. and {Jain}, Anany Shrey and {Shukla}, Gyanendra and {Sick}, Jonathan and {Simpson}, Chris and {Singanamalla}, Sudheesh and {Singer}, Leo P. and {Singhal}, Jaladh and {Sinha}, Manodeep and {Sip{H{o}}cz}, Brigitta M. and {Spitler}, Lee R. and {Stansby}, David and {Streicher}, Ole and {{{S}}umak}, Jani and {Swinbank}, John D. and {Taranu}, Dan S. and {Tewary}, Nikita and {Tremblay}, Grant R. and {Val-Borro}, Miguel de and {Van Kooten}, Samuel J. and {Vasovi{'c}}, Zlatan and {Verma}, Shresth and {de Miranda Cardoso}, Jos{'e} Vin{'i}cius and {Williams}, Peter K.~G. and {Wilson}, Tom J. and {Winkel}, Benjamin and {Wood-Vasey}, W.~M. and {Xue}, Rui and {Yoachim}, Peter and {Zhang}, Chen and {Zonca}, Andrea and {Astropy Project Contributors}},
        title = "{The Astropy Project: Sustaining and Growing a Community-oriented Open-source Project and the Latest Major Release (v5.0) of the Core Package}",
      journal = {\apj},
         year = 2022,
        month = aug,
       volume = {935},
       number = {2},
          eid = {167},
        pages = {167},
          doi = {10.3847/1538-4357/ac7c74},
archivePrefix = {arXiv},
       eprint = {2206.14220},
 primaryClass = {astro-ph.IM},
       adsurl = {https://ui.adsabs.harvard.edu/abs/2022ApJ...935..167A}
}

@Article{matplotlib2007,
  Author    = {Hunter, J. D.},
  Title     = {Matplotlib: A 2D graphics environment},
  Journal   = {Computing in Science \& Engineering},
  Volume    = {9},
  Number    = {3},
  Pages     = {90--95},
  publisher = {IEEE COMPUTER SOC},
  doi       = {10.1109/MCSE.2007.55},
  year      = 2007
}

@Article{numpy2020,
 title         = {Array programming with {NumPy}},
 author        = {Charles R. Harris and K. Jarrod Millman and St{\'{e}}fan J.
                 van der Walt and Ralf Gommers and Pauli Virtanen and David
                 Cournapeau and Eric Wieser and Julian Taylor and Sebastian
                 Berg and Nathaniel J. Smith and Robert Kern and Matti Picus
                 and Stephan Hoyer and Marten H. van Kerkwijk and Matthew
                 Brett and Allan Haldane and Jaime Fern{\'{a}}ndez del
                 R{\'{i}}o and Mark Wiebe and Pearu Peterson and Pierre
                 G{\'{e}}rard-Marchant and Kevin Sheppard and Tyler Reddy and
                 Warren Weckesser and Hameer Abbasi and Christoph Gohlke and
                 Travis E. Oliphant},
 year          = {2020},
 month         = sep,
 journal       = {Nature},
 volume        = {585},
 number        = {7825},
 pages         = {357--362},
 doi           = {10.1038/s41586-020-2649-2},
 publisher     = {Springer Science and Business Media {LLC}},
 url           = {https://doi.org/10.1038/s41586-020-2649-2}
}

@ARTICLE{SciPy2020,
  author  = {Virtanen, Pauli and Gommers, Ralf and Oliphant, Travis E. and
            Haberland, Matt and Reddy, Tyler and Cournapeau, David and
            Burovski, Evgeni and Peterson, Pearu and Weckesser, Warren and
            Bright, Jonathan and {van der Walt}, St{\'e}fan J. and
            Brett, Matthew and Wilson, Joshua and Millman, K. Jarrod and
            Mayorov, Nikolay and Nelson, Andrew R. J. and Jones, Eric and
            Kern, Robert and Larson, Eric and Carey, C J and
            Polat, {\.I}lhan and Feng, Yu and Moore, Eric W. and
            {VanderPlas}, Jake and Laxalde, Denis and Perktold, Josef and
            Cimrman, Robert and Henriksen, Ian and Quintero, E. A. and
            Harris, Charles R. and Archibald, Anne M. and
            Ribeiro, Ant{\^o}nio H. and Pedregosa, Fabian and
            {van Mulbregt}, Paul and {SciPy 1.0 Contributors}},
  title   = {{{SciPy} 1.0: Fundamental Algorithms for Scientific
            Computing in Python}},
  journal = {Nature Methods},
  year    = {2020},
  volume  = {17},
  pages   = {261--272},
  adsurl  = {https://rdcu.be/b08Wh},
  doi     = {10.1038/s41592-019-0686-2},
}

@ARTICLE{zubovas2014,
       author = {{Zubovas}, Kastytis and {Nayakshin}, Sergei},
        title = "{Energy- and momentum-conserving AGN feedback outflows}",
      journal = {\mnras},
         year = 2014,
        month = may,
       volume = {440},
       number = {3},
        pages = {2625-2635},
          doi = {10.1093/mnras/stu431},
archivePrefix = {arXiv},
       eprint = {1403.3933},
 primaryClass = {astro-ph.GA},
       adsurl = {https://ui.adsabs.harvard.edu/abs/2014MNRAS.440.2625Z}
}

@ARTICLE{veilleux2005,
       author = {{Veilleux}, Sylvain and {Cecil}, Gerald and {Bland-Hawthorn}, Joss},
        title = "{Galactic Winds}",
      journal = {\araa},
         year = 2005,
        month = sep,
       volume = {43},
       number = {1},
        pages = {769-826},
          doi = {10.1146/annurev.astro.43.072103.150610},
archivePrefix = {arXiv},
       eprint = {astro-ph/0504435},
 primaryClass = {astro-ph},
       adsurl = {https://ui.adsabs.harvard.edu/abs/2005ARA&A..43..769V}
}

@ARTICLE{strickland2000,
       author = {{Strickland}, David K. and {Stevens}, Ian R.},
        title = "{Starburst-driven galactic winds - I. Energetics and intrinsic X-ray emission}",
      journal = {\mnras},
         year = 2000,
        month = may,
       volume = {314},
       number = {3},
        pages = {511-545},
          doi = {10.1046/j.1365-8711.2000.03391.x},
archivePrefix = {arXiv},
       eprint = {astro-ph/0001395},
 primaryClass = {astro-ph},
       adsurl = {https://ui.adsabs.harvard.edu/abs/2000MNRAS.314..511S}
}

@ARTICLE{strickland2009,
       author = {{Strickland}, David K. and {Heckman}, Timothy M.},
        title = "{Supernova Feedback Efficiency and Mass Loading in the Starburst and Galactic Superwind Exemplar M82}",
      journal = {\apj},
         year = 2009,
        month = jun,
       volume = {697},
       number = {2},
        pages = {2030-2056},
          doi = {10.1088/0004-637X/697/2/2030},
archivePrefix = {arXiv},
       eprint = {0903.4175},
 primaryClass = {astro-ph.CO},
       adsurl = {https://ui.adsabs.harvard.edu/abs/2009ApJ...697.2030S}
}

@ARTICLE{Wagner2012,
       author = {{Wagner}, A.~Y. and {Bicknell}, G.~V. and {Umemura}, M.},
        title = "{Driving Outflows with Relativistic Jets and the Dependence of Active Galactic Nucleus Feedback Efficiency on Interstellar Medium Inhomogeneity}",
      journal = {\apj},
         year = 2012,
        month = oct,
       volume = {757},
       number = {2},
          eid = {136},
        pages = {136},
          doi = {10.1088/0004-637X/757/2/136},
archivePrefix = {arXiv},
       eprint = {1205.0542},
 primaryClass = {astro-ph.CO},
       adsurl = {https://ui.adsabs.harvard.edu/abs/2012ApJ...757..136W}
}

@ARTICLE{Buiten2025,
       author = {{Buiten}, Victorine A. and {van der Werf}, Paul P. and {Viti}, Serena and {Dicken}, Daniel and {Alonso Herrero}, Almudena and {Wright}, Gillian S. and {Baes}, Maarten and {B{\"o}ker}, Torsten and {Brandl}, Bernhard R. and {Colina}, Luis and {Garc{\'\i}a Mar{\'\i}n}, Macarena and {Greve}, Thomas R. and {Guillard}, Pierre and {Jones}, Olivia C. and {Hermosa Mu{\~n}oz}, Laura and {Labiano}, {\'A}lvaro and {{\"O}stlin}, G{\"o}ran and {Pantoni}, Lara and {Walter}, Fabian and {Ward}, Martin J. and {Perna}, Michele and {van Dishoeck}, Ewine F. and {Henning}, Thomas and {G{\"u}del}, Manuel and {Ray}, Thomas P.},
        title = "{The rich JWST spectrum of the western nucleus of Arp 220: Shocked hot core chemistry dominates the inner disc}",
      journal = {\aap},
         year = 2025,
        month = jul,
       volume = {699},
          eid = {A312},
        pages = {A312},
          doi = {10.1051/0004-6361/202554141},
archivePrefix = {arXiv},
       eprint = {2502.10271},
 primaryClass = {astro-ph.GA},
       adsurl = {https://ui.adsabs.harvard.edu/abs/2025A&A...699A.312B}
}

@ARTICLE{Alonso2024,
       author = {{Alonso Herrero}, A. and {Hermosa Mu{\~n}oz}, L. and {Labiano}, A. and {Guillard}, P. and {Buiten}, V.~A. and {Dicken}, D. and {van der Werf}, P. and {{\'A}lvarez-M{\'a}rquez}, J. and {B{\"o}ker}, T. and {Colina}, L. and {Eckart}, A. and {Garc{\'\i}a-Mar{\'\i}n}, M. and {Jones}, O.~C. and {Pantoni}, L. and {P{\'e}rez-Gonz{\'a}lez}, P.~G. and {Rouan}, D. and {Ward}, M.~J. and {Baes}, M. and {{\"O}stlin}, G. and {Royer}, P. and {Wright}, G.~S. and {G{\"u}del}, M. and {Henning}, Th. and {Lagage}, P.-O. and {van Dishoeck}, E.~F.},
        title = "{MICONIC: JWST/MIRI MRS observations of the nuclear and circumnuclear regions of Mrk 231}",
      journal = {\aap},
         year = 2024,
        month = oct,
       volume = {690},
          eid = {A95},
        pages = {A95},
          doi = {10.1051/0004-6361/202450071},
archivePrefix = {arXiv},
       eprint = {2407.02180},
 primaryClass = {astro-ph.GA},
       adsurl = {https://ui.adsabs.harvard.edu/abs/2024A&A...690A..95A}
}

@ARTICLE{Alonso2025,
       author = {{Alonso Herrero}, A. and {Hermosa Mu{\~n}oz}, L. and {Labiano}, A. and {Guillard}, P. and {Garc{\'\i}a-Mar{\'\i}n}, M. and {Dicken}, D. and {Garc{\'\i}a-Burillo}, S. and {Pantoni}, L. and {Buiten}, V. and {Colina}, L. and {B{\"o}ker}, T. and {Baes}, M. and {Eckart}, A. and {Evangelista}, L. and {{\"O}stlin}, G. and {Rouan}, D. and {van der Werf}, P. and {Walter}, F. and {Ward}, M.~J. and {Wright}, G. and {G{\"u}del}, M. and {Henning}, Th. and {Lagage}, P.-O.},
        title = "{MICONIC: JWST/MIRI MRS reveals a fast ionized gas outflow in the central region of Centaurus A}",
      journal = {\aap},
         year = 2025,
        month = jul,
       volume = {699},
          eid = {A334},
        pages = {A334},
          doi = {10.1051/0004-6361/202554823},
archivePrefix = {arXiv},
       eprint = {2506.15286},
 primaryClass = {astro-ph.GA},
       adsurl = {https://ui.adsabs.harvard.edu/abs/2025A&A...699A.334A}
}

@ARTICLE{colina2015,
       author = {{Colina}, Luis and {Piqueras L{\'o}pez}, Javier and {Arribas}, Santiago and {Riffel}, Rog{\'e}rio and {Riffel}, Rogemar A. and {Rodriguez-Ardila}, Alberto and {Pastoriza}, Miriani and {Storchi-Bergmann}, Thaisa and {Alonso-Herrero}, Almudena and {Sales}, Dinalva},
        title = "{Understanding the two-dimensional ionization structure in luminous infrared galaxies. A near-IR integral field spectroscopy perspective}",
      journal = {\aap},
         year = 2015,
        month = jun,
       volume = {578},
          eid = {A48},
        pages = {A48},
          doi = {10.1051/0004-6361/201425567},
archivePrefix = {arXiv},
       eprint = {1504.02724},
 primaryClass = {astro-ph.GA},
       adsurl = {https://ui.adsabs.harvard.edu/abs/2015A&A...578A..48C}
}

@ARTICLE{rodriguez2005,
       author = {{Rodr{\'\i}guez-Ardila}, A. and {Riffel}, R. and {Pastoriza}, M.~G.},
        title = "{Molecular hydrogen and [FeII] in active galactic nuclei - II. Results for Seyfert 2 galaxies}",
      journal = {\mnras},
         year = 2005,
        month = dec,
       volume = {364},
       number = {3},
        pages = {1041-1053},
          doi = {10.1111/j.1365-2966.2005.09638.x},
       adsurl = {https://ui.adsabs.harvard.edu/abs/2005MNRAS.364.1041R}
}

@ARTICLE{Alonso1997,
       author = {{Alonso-Herrero}, Almudena and {Rieke}, M.~J. and {Rieke}, G.~H. and {Ruiz}, M.},
        title = "{Using Near-Infrared [Fe II] Lines to Identify Active Galaxies}",
      journal = {\apj},
         year = 1997,
        month = jun,
       volume = {482},
       number = {2},
        pages = {747-756},
          doi = {10.1086/304184},
       adsurl = {https://ui.adsabs.harvard.edu/abs/1997ApJ...482..747A}
}

@ARTICLE{Rigopoulou2021,
       author = {{Rigopoulou}, D. and {Barale}, M. and {Clary}, D.~C. and {Shan}, X. and {Alonso-Herrero}, A. and {Garc{\'\i}a-Bernete}, I. and {Hunt}, L. and {Kerkeni}, B. and {Pereira-Santaella}, M. and {Roche}, P.~F.},
        title = "{The properties of polycyclic aromatic hydrocarbons in galaxies: constraints on PAH sizes, charge and radiation fields}",
      journal = {\mnras},
         year = 2021,
        month = jul,
       volume = {504},
       number = {4},
        pages = {5287-5300},
          doi = {10.1093/mnras/stab959},
archivePrefix = {arXiv},
       eprint = {2011.10114},
 primaryClass = {astro-ph.GA},
       adsurl = {https://ui.adsabs.harvard.edu/abs/2021MNRAS.504.5287R}
}

@ARTICLE{Garcia2022,
       author = {{Garc{\'\i}a-Bernete}, I. and {Rigopoulou}, D. and {Alonso-Herrero}, A. and {Pereira-Santaella}, M. and {Roche}, P.~F. and {Kerkeni}, B.},
        title = "{Polycyclic aromatic hydrocarbons in Seyfert and star-forming galaxies}",
      journal = {\mnras},
         year = 2022,
        month = jan,
       volume = {509},
       number = {3},
        pages = {4256-4275},
          doi = {10.1093/mnras/stab3127},
archivePrefix = {arXiv},
       eprint = {2011.10882},
 primaryClass = {astro-ph.GA},
       adsurl = {https://ui.adsabs.harvard.edu/abs/2022MNRAS.509.4256G}
}

@ARTICLE{draine2007,
       author = {{Draine}, B.~T. and {Li}, Aigen},
        title = "{Infrared Emission from Interstellar Dust. IV. The Silicate-Graphite-PAH Model in the Post-Spitzer Era}",
      journal = {\apj},
         year = 2007,
        month = mar,
       volume = {657},
       number = {2},
        pages = {810-837},
          doi = {10.1086/511055},
archivePrefix = {arXiv},
       eprint = {astro-ph/0608003},
 primaryClass = {astro-ph},
       adsurl = {https://ui.adsabs.harvard.edu/abs/2007ApJ...657..810D}
}

@ARTICLE{silk1998,
       author = {{Silk}, Joseph and {Rees}, Martin J.},
        title = "{Quasars and galaxy formation}",
      journal = {\aap},
         year = 1998,
        month = mar,
       volume = {331},
        pages = {L1-L4},
          doi = {10.48550/arXiv.astro-ph/9801013},
archivePrefix = {arXiv},
       eprint = {astro-ph/9801013},
 primaryClass = {astro-ph},
       adsurl = {https://ui.adsabs.harvard.edu/abs/1998A&A...331L...1S}
}

@ARTICLE{dekel1986,
       author = {{Dekel}, A. and {Silk}, J.},
        title = "{The Origin of Dwarf Galaxies, Cold Dark Matter, and Biased Galaxy Formation}",
      journal = {\apj},
         year = 1986,
        month = apr,
       volume = {303},
        pages = {39},
          doi = {10.1086/164050},
       adsurl = {https://ui.adsabs.harvard.edu/abs/1986ApJ...303...39D}
}

@ARTICLE{hopkins2014,
       author = {{Hopkins}, Philip F. and {Kere{\v{s}}}, Du{\v{s}}an and {O{\~n}orbe}, Jos{\'e} and {Faucher-Gigu{\`e}re}, Claude-Andr{\'e} and {Quataert}, Eliot and {Murray}, Norman and {Bullock}, James S.},
        title = "{Galaxies on FIRE (Feedback In Realistic Environments): stellar feedback explains cosmologically inefficient star formation}",
      journal = {\mnras},
         year = 2014,
        month = nov,
       volume = {445},
       number = {1},
        pages = {581-603},
          doi = {10.1093/mnras/stu1738},
archivePrefix = {arXiv},
       eprint = {1311.2073},
 primaryClass = {astro-ph.CO},
       adsurl = {https://ui.adsabs.harvard.edu/abs/2014MNRAS.445..581H}
}

@ARTICLE{somerville2015,
       author = {{Somerville}, Rachel S. and {Dav{\'e}}, Romeel},
        title = "{Physical Models of Galaxy Formation in a Cosmological Framework}",
      journal = {\araa},
         year = 2015,
        month = aug,
       volume = {53},
        pages = {51-113},
          doi = {10.1146/annurev-astro-082812-140951},
archivePrefix = {arXiv},
       eprint = {1412.2712},
 primaryClass = {astro-ph.GA},
       adsurl = {https://ui.adsabs.harvard.edu/abs/2015ARA&A..53...51S}
}

@ARTICLE{heckman2017,
       author = {{Heckman}, Timothy M. and {Thompson}, Todd A.},
        title = "{Galactic Winds and the Role Played by Massive Stars}",
      journal = {arXiv e-prints},
         year = 2017,
        month = jan,
          eid = {arXiv:1701.09062},
        pages = {arXiv:1701.09062},
          doi = {10.48550/arXiv.1701.09062},
archivePrefix = {arXiv},
       eprint = {1701.09062},
 primaryClass = {astro-ph.GA},
       adsurl = {https://ui.adsabs.harvard.edu/abs/2017arXiv170109062H}
}

@ARTICLE{murray2005,
       author = {{Murray}, Norman and {Quataert}, Eliot and {Thompson}, Todd A.},
        title = "{On the Maximum Luminosity of Galaxies and Their Central Black Holes: Feedback from Momentum-driven Winds}",
      journal = {\apj},
         year = 2005,
        month = jan,
       volume = {618},
       number = {2},
        pages = {569-585},
          doi = {10.1086/426067},
archivePrefix = {arXiv},
       eprint = {astro-ph/0406070},
 primaryClass = {astro-ph},
       adsurl = {https://ui.adsabs.harvard.edu/abs/2005ApJ...618..569M}
}

@ARTICLE{king2003,
       author = {{King}, Andrew},
        title = "{Black Holes, Galaxy Formation, and the M$_{BH}$-{\ensuremath{\sigma}} Relation}",
      journal = {\apjl},
         year = 2003,
        month = oct,
       volume = {596},
       number = {1},
        pages = {L27-L29},
          doi = {10.1086/379143},
archivePrefix = {arXiv},
       eprint = {astro-ph/0308342},
 primaryClass = {astro-ph},
       adsurl = {https://ui.adsabs.harvard.edu/abs/2003ApJ...596L..27K}
}

@ARTICLE{hopkins2012,
       author = {{Hopkins}, Philip F. and {Quataert}, Eliot and {Murray}, Norman},
        title = "{Stellar feedback in galaxies and the origin of galaxy-scale winds}",
      journal = {\mnras},
         year = 2012,
        month = apr,
       volume = {421},
       number = {4},
        pages = {3522-3537},
          doi = {10.1111/j.1365-2966.2012.20593.x},
archivePrefix = {arXiv},
       eprint = {1110.4638},
 primaryClass = {astro-ph.CO},
       adsurl = {https://ui.adsabs.harvard.edu/abs/2012MNRAS.421.3522H}
}

@ARTICLE{dimatteo2005,
       author = {{Di Matteo}, Tiziana and {Springel}, Volker and {Hernquist}, Lars},
        title = "{Energy input from quasars regulates the growth and activity of black holes and their host galaxies}",
      journal = {\nat},
         year = 2005,
        month = feb,
       volume = {433},
       number = {7026},
        pages = {604-607},
          doi = {10.1038/nature03335},
archivePrefix = {arXiv},
       eprint = {astro-ph/0502199},
 primaryClass = {astro-ph},
       adsurl = {https://ui.adsabs.harvard.edu/abs/2005Natur.433..604D}
}

@ARTICLE{king2005,
       author = {{King}, Andrew},
        title = "{The AGN-Starburst Connection, Galactic Superwinds, and M$_{BH}$-{\ensuremath{\sigma}}}",
      journal = {\apjl},
         year = 2005,
        month = dec,
       volume = {635},
       number = {2},
        pages = {L121-L123},
          doi = {10.1086/499430},
archivePrefix = {arXiv},
       eprint = {astro-ph/0511034},
 primaryClass = {astro-ph},
       adsurl = {https://ui.adsabs.harvard.edu/abs/2005ApJ...635L.121K}
}

@ARTICLE{Elmegreen1977,
       author = {{Elmegreen}, B.~G. and {Lada}, C.~J.},
        title = "{Sequential formation of subgroups in OB associations.}",
      journal = {\apj},
         year = 1977,
        month = jun,
       volume = {214},
        pages = {725-741},
          doi = {10.1086/155302},
       adsurl = {https://ui.adsabs.harvard.edu/abs/1977ApJ...214..725E}
}

@ARTICLE{krumholz2009,
       author = {{Krumholz}, Mark R. and {Matzner}, Christopher D.},
        title = "{The Dynamics of Radiation-pressure-dominated H II Regions}",
      journal = {\apj},
         year = 2009,
        month = oct,
       volume = {703},
       number = {2},
        pages = {1352-1362},
          doi = {10.1088/0004-637X/703/2/1352},
archivePrefix = {arXiv},
       eprint = {0906.4343},
 primaryClass = {astro-ph.SR},
       adsurl = {https://ui.adsabs.harvard.edu/abs/2009ApJ...703.1352K}
}

@ARTICLE{silk2013,
       author = {{Silk}, Joseph},
        title = "{Unleashing Positive Feedback: Linking the Rates of Star Formation, Supermassive Black Hole Accretion, and Outflows in Distant Galaxies}",
      journal = {\apj},
         year = 2013,
        month = aug,
       volume = {772},
       number = {2},
          eid = {112},
        pages = {112},
          doi = {10.1088/0004-637X/772/2/112},
archivePrefix = {arXiv},
       eprint = {1305.5840},
 primaryClass = {astro-ph.CO},
       adsurl = {https://ui.adsabs.harvard.edu/abs/2013ApJ...772..112S}
}

@ARTICLE{zinn2013,
       author = {{Zinn}, P.-C. and {Middelberg}, E. and {Norris}, R.~P. and {Dettmar}, R.-J.},
        title = "{Active Galactic Nucleus Feedback Works Both Ways}",
      journal = {\apj},
         year = 2013,
        month = sep,
       volume = {774},
       number = {1},
          eid = {66},
        pages = {66},
          doi = {10.1088/0004-637X/774/1/66},
archivePrefix = {arXiv},
       eprint = {1306.6468},
 primaryClass = {astro-ph.CO},
       adsurl = {https://ui.adsabs.harvard.edu/abs/2013ApJ...774...66Z}
}

@ARTICLE{Maiolino2017,
       author = {{Maiolino}, R. and {Russell}, H.~R. and {Fabian}, A.~C. and {Carniani}, S. and {Gallagher}, R. and {Cazzoli}, S. and {Arribas}, S. and {Belfiore}, F. and {Bellocchi}, E. and {Colina}, L. and {Cresci}, G. and {Ishibashi}, W. and {Marconi}, A. and {Mannucci}, F. and {Oliva}, E. and {Sturm}, E.},
        title = "{Star formation inside a galactic outflow}",
      journal = {\nat},
         year = 2017,
        month = mar,
       volume = {544},
       number = {7649},
        pages = {202-206},
          doi = {10.1038/nature21677},
archivePrefix = {arXiv},
       eprint = {1703.08587},
 primaryClass = {astro-ph.GA},
       adsurl = {https://ui.adsabs.harvard.edu/abs/2017Natur.544..202M}
}

@ARTICLE{Rupke2005,
       author = {{Rupke}, David S. and {Veilleux}, Sylvain and {Sanders}, D.~B.},
        title = "{Outflows in Infrared-Luminous Starbursts at z < 0.5. II. Analysis and Discussion}",
      journal = {\apjs},
         year = 2005,
        month = sep,
       volume = {160},
       number = {1},
        pages = {115-148},
          doi = {10.1086/432889},
archivePrefix = {arXiv},
       eprint = {astro-ph/0506611},
 primaryClass = {astro-ph},
       adsurl = {https://ui.adsabs.harvard.edu/abs/2005ApJS..160..115R}
}
\bibliographystyle{aasjournalv7}

\end{document}
